\documentclass[12pt]{article}
\usepackage{latexsym}\usepackage{epsfig,amssymb,euscript}
\usepackage{amsmath} \usepackage{bbm,slashed}
\usepackage[usenames,dvipsnames]{xcolor}
\usepackage{ulem}
\usepackage{cite}
\usepackage{subfigure}
\usepackage{caption}
\newcommand{\be}{\begin{equation}}
\newcommand{\ee}{\end{equation}}
\newcommand{\bea}{\begin{eqnarray}}
\newcommand{\eea}{\end{eqnarray}}

\newcommand{\Tr}{{\rm Tr}}

\begin{document}
\begin{titlepage}
\begin{flushright}
IPPP/13/13\\
DCPT/13/26
\end{flushright}
\bigskip
\def\thefootnote{\fnsymbol{footnote}}
\begin{center}
\vskip -10pt
{\LARGE
{\bf
Phenomenology of \\ 
General Gauge Mediation \\
\vspace{0.2in}
in light of a 125 GeV Higgs
}
}
\end{center}

\bigskip
\begin{center}
{\large
Phill Grajek$^{1,4}$,
Alberto Mariotti$^{1,2,4}$ and
Diego Redigolo$^{3,4}$} 
\end{center}

\renewcommand{\thefootnote}{\arabic{footnote}}

\begin{center}
$^1${Theoretische Natuurkunde and IIHE/ELEM\\ 
Vrije Universiteit Brussel, Pleinlaan 2, B-1050 Brussels, Belgium\\}
$^2${Institute for Particle Physics Phenomenology\\
Durham University, Durham DH1 3LE, United Kingdom\\}
$^3${Physique Th\'eorique et Math\'ematique \\
Universit\'e Libre de Bruxelles, C.P. 231, 1050 Bruxelles, Belgium\\}
$^4${International Solvay Institutes, Brussels, Belgium\\}
\end{center}

\noindent
\begin{center} {\bf Abstract} \end{center}
We explore the phenomenology of the full General Gauge Mediation parameter space in the MSSM focusing on the consequences of having a fundamental Higgs around 125 GeV. Assuming GUT-complete structure of the hidden sector, we allow for deviations from the strict definition of gauge mediated SUSY-breaking coming from mild violations of messenger-parity and from extra couplings between the Higgs multiplets and the hidden sector. Relaxing the GUT assumption, our parameter space is defined by the property of having vanishing A-terms at the messenger scale. In this extended setup we focus on the possibility of splitting the $SU(3)$ mass parameters of GGM. 
In all these scenarios we investigate the possible spectra, discussing to what extent having an Higgs mass around 125 GeV is constraining the GGM parameter space and what are the possible signatures at LHC.    
\vfill

\end{titlepage}

\setcounter{footnote}{0}


\tableofcontents

\section{Introduction}

The General Gauge Mediation (GGM) formalism \cite{Meade:2008wd, Buican:2008ws} provides a model-independent definition of the gauge mediation mechanism by requiring that the SUSY-breaking hidden sector is coupled to a supersymmetric extension of the Standard Model only through supersymmetric gauge interactions. As a consequence, the two sectors become decoupled in the limit in which we set to zero the three gauge couplings $(g_1,g_2,g_3)$\footnote{In all the explicit formulas we use the GUT normalisation for $g_1$ rather than the Standard Model normalization.} associated to the ordinary gauge group of the Standard Model ($U(1)\times SU(2) \times SU(3)$). 

Most of the explicit realizations of gauge mediation (see \cite{Giudice:1998bp} for a review and original references) are included in the GGM formalism which has the advantage of providing a model-independent identification of the parameter space of gauge mediated SUSY-breaking theories. Moreover, the GGM parameter space is completely calculable in the sense that it has been shown to be completely realizable in terms of weakly coupled messenger models \cite{Buican:2008ws, Carpenter:2008w}. Each point in the parameter space determines a spectrum of superpartners at an UV scale which, after an RG-flow evolution, corresponds to a spectrum of sparticles at the electroweak scale that characterizes the phenomenology of the associated model. 

The primary motivation of this work is to complement the already existing phenomenological studies of gauge mediation scenarios in the Minimal Supersymmetric Standard Model \cite{Carpenter:2008he, Rajaraman:2009ga, Abel:2009ve, Abel:2010vba, Dolan:2011ie, Grellscheid:2011ij, Kats:2011qh} with a complete analysis of the full GGM paramater space. Most importantly, after the announced evidence for the production of a Standard Model-like Higgs boson with mass near 125 GeV at LHC \cite{:2012gk, :2012gu}, it is of capital interest to understand more quantitatively the implications of assuming a 125 GeV Higgs in the MSSM for the full GGM parameter space, taking a step forward in the same direction of the analyses already carried out in \cite{Draper:2011aa, Ajaib:2012vc}.        

Leaving aside the Higgs sector and assuming messenger parity, the GGM parameter space is described by 6 independent parameters which account for the SUSY-breaking masses for gauginos and sfermions  at a certain UV scale that we call $M_{\text{mess}}$ in analogy with the weakly coupled models with messenger fields. We can write explicit formulas for the UV soft masses factorizing the dependence on the coupling constants and the typical loop factors:   
\begin{align}
& M_{\tilde{\lambda}_{i}}(M_{\text{mess}})=\frac{g_{i}^2(M_{\text{mess}})}{(4\pi)^2}\Lambda_{G_{i}}\ , \label{gmass}\\
& m^2_{\tilde{f}}(M_{\text{mess}})=2\sum_{i=1}^{3}q^{2}_{\tilde{f}_{i}}k_{i}\frac{g^4_{i}(M_{\text{mess}})}{(4\pi)^4}\Lambda_{S_{i}}^2\ .\label{sfmass}
\end{align}
$q^{2}_{\tilde{f}_{i}}=(Y^2,3/4,4/3)$ is the quadratic Casimir of the representation $\tilde{f}$ under the $i^{th}$ gauge group and $k_i=(3/5,1,1)$. $\Lambda_{G_{i}}$ and $\Lambda_{S_{i}}$ are model-dependent functions of the SUSY-breaking scales of the hidden sector and of the typical UV scale $M_{\text{mess}}$ that we assume to be unique for simplicity. Moreover, $M_{\text{mess}}$ sets the length of the RG-flow evolution and plays the role of an extra parameter in GGM \cite{Jaeckel:2011ma, Jaeckel:2011qj} so that we end up with 6+1 parameters\footnote{In general the $\Lambda_{G_{i}}$ parameters could be complex but we will assume them real in order to avoid potentially dangerous CP-violating extra-phases.}. The three scalar mass scales $\Lambda_{S_{i}}$ determine five soft masses for the two matter doublets and three singlets of the MSSM so that we have two mass sum rules at the messenger scale $\Tr ( Y m^2)=0$ and $\Tr ( (B-L) m^2)=0$ which are generic predictions of GGM \cite{Meade:2008wd}. 
 
In the Higgs sector, the two soft masses $m^2_{H_{u,d}}$ receive a gauge mediation contribution at the messenger scale $M_{\text{mess}}$ like all the other scalars. At leading order, this contribution is exactly as the one for the slepton doublet:
 \be
\label{higgsA}
m_{H_u}^2(M_{\text{mess}})=m_{H_d}^2(M_{\text{mess}})=
m_{\tilde{E}_L}^2(M_{\text{mess}})=
\frac{k_1}{2} \frac{g_1^4(M_{\text{mess}})}{(4 \pi)^4}\Lambda_{S_{1}}^2+
\frac{3 k_2}{2} \frac{g_2^4(M_{\text{mess}})}{(4 \pi)^4}\Lambda_{S_{2}}^2\ .
\ee
Therefore,  the Higgs soft terms are determined as functions of the parameters $\Lambda_{S_{1}}^2$ and $\Lambda_{S_{2}}^2$, the parameter space remains 6+1 dimensional and both the GGM sum rules are satisfied at the messenger scale if messenger parity is assumed.

In order to achieve a phenomenologically acceptable electroweak symmetry breaking (EWSB) in gauge mediation, we need to find a way out from the so-called $\mu/B_{\mu}$ problem \cite{Dvali:1996cu, Csaki:2008sr, Komargodski:2008ax, DeSimone:2011va}, generating a viable $\mu$ term and finding a suppression mechanism for the $B_{\mu}$ contributions. One possibility that goes under the name of Pure General Gauge Mediation \cite{Abel:2009ve, Abel:2010vba} is to keep the strict definition of GGM, having $B_{\mu}(M_{\text{mess}})\simeq 0$ at the UV scale and generating a viable EWSB by RG-flow evolution \cite{Rattazzi:1996fb}. In this scenario $B_{\mu}\ll\mu$ and, as a consequence, we generically get high values of $\tan\beta$ in the IR.

Another option is to relax the definition of GGM, allowing for direct couplings between Higgs sector and hidden sector. In this context it is possible to show that in a large class of models the new interactions generate a non-vanishing $B_{\mu}$ in the UV that can be of the same order of $\mu$ or suppressed by an $U(1)_{R}$ symmetry, but at the same time they add new contributions to the Higgs soft masses $\delta m^2_{u,d}$ and to the $A$-terms $\delta A_{u,d}$ \cite{Csaki:2008sr, Komargodski:2008ax, DeSimone:2011va}.

In our analysis we are going to use $\tan\beta$ as free input parameter instead of $B_{\mu}$ and allowing for possible extra contributions to the Higgs soft masses $\delta m^2_{u,d}$ for consistency. The two Higgs soft masses at the messenger scale become
\begin{equation}\label{NHmasses}
m_{H_{u,d}}^2(M_{\text{mess}})=m_{\tilde{E}_L}^2(M_{\text{mess}})+\delta m^2_{u,d}(M_{\text{mess}})\ .
\end{equation}
Since the contributions to the $A$-terms have been shown to be generically suppressed with respect to $\delta m^2_{u,d}$ \cite{Komargodski:2008ax}, we will ignore them in the following.
This extended scenario that we call Extended General Gauge Mediation (EGGM) has 6+1+3 parameters and its characterizing feature is the vanishing of the $A$-terms at high scale\footnote{In principle this is not always the case and some recent studies have shown how it is possible to obtain large $A$-terms in gauge mediation scenarios \cite{Chacko:2001km,Chacko:2002et,Evans:2011bea, Evans:2012hg, Kang:2012ra, Craig:2012xp, Craig:2013wga}.}. A careful discussion of the fate of the GGM sum rules in this context has been carried out in \cite{Jaeckel:2011qj}.

The GGM parameter space have been for longtime considered prohibitively large for an exhaustive survey and this is even more true if we consider its extension that accounts for the $\mu/B_{\mu}$ problem. Moreover, in most of the model building, the hidden sector and the mediation sector are assumed to have a complete GUT structure at the messenger scale in order to easily achieve gauge coupling unification so that the whole soft spectrum is fully determined by two independent parameters $\Lambda_{G}$ and $\Lambda_{S}$ which account for gaugino masses and sfermion masses. In this simplified setting that we call Constrained General Gauge Mediation (CGGM) the parameter space is restricted to 2+1+1 dimensions and it becomes 2+1+3 dimensional including extra contributions to the Higgs soft masses.

The CGGM, with or without the inclusion of extra-parameters in the Higgs sector, has been subject of an intense study in order to understand which are the regions of the parameter space which satisfies the bounds from  collider direct searches, from flavor observables and from Higgs physics \cite{Abel:2009ve, Abel:2010vba, Dolan:2011ie, Grellscheid:2011ij}. 

Considering the most general extension of the CGGM we allow also for mild violations of messenger parity in the hidden sector. The assumption of messenger parity  sets to zero possible messenger parity violating (MPV) $D_{Y}$-term contributions coming from the hidden sector dynamics. In order to define the most general GGM parameter space, we allow also for MPV effects \cite{Dimopoulos:1996ig, Argurio:2012qt} that we can parametrize with one extra SUSY-breaking scale: 
\begin{equation}
\delta m^2_{\tilde{f}}(M_{\text{mess}})=k_{1}\frac{g_{1}^2(M_{\text{mess}})}{(4\pi)^2}Y_{\tilde{f}}\Lambda_{D}^2\label{D-tad}\ .
\end{equation}
These contributions arise generically at 1-loop in GGM and violate the usual GGM sum rules separately, letting us with only one combination of the two still satisfied at the messenger scale: $4\Tr((4Y-5(B-L))m^2)= 0$. 

1-loop $D_{Y}$ terms are potentially dangerous because they could dominate over the usual gauge mediation ones \eqref{sfmass} leading, eventually, to tachyonic sfermion masses at the electroweak scale.  
However, assuming a GUT-complete structure of the hidden sector, the extra MPV contributions \eqref{D-tad} arise generically at two loops being of the same order of the usual gauge mediation effects \eqref{sfmass} \cite{Dimopoulos:1996ig, Argurio:2012qt}.

Consequently, the suppression of the $D_{Y}$ tadpoles due to the $G_{GUT}$ assumption opens up a new viable direction in the parameter space, alleviating the possible issue with tachyonic spectra at the electroweak scale. 
We will explain in detail in section \ref{2.3} how the contribution \eqref{D-tad} is modified under the assumption of GUT-completeness in the hidden sector and we will show the possible phenomenological consequences of these effects. A more careful study of the phenomenological consequences of 1-loop $D_{Y}$ tadpoles that could arise in the full GGM parameter space is left for future investigations. 

A generic result of our entire analysis is that
assuming a Higgs around 125 GeV in the MSSM is pushing most of the viable spectra of the CGGM out of the present collider searches. This result, that can be viewed as the top-down counterpart of the analysis performed in \cite{Draper:2011aa}, seems to indicate that enforcing the Higgs constraint in the standard GGM scenarios and requiring a viable spectrum for colliders force us to give up GUT-completeness at the messenger scale and explore new regions of the parameter space of GGM in which non-standard hierarchies between the soft terms in the UV are allowed.

As a byproduct, we will show to what extent the GGM framework is still an effective signature generator for LHC searches \cite{Kats:2011qh}, investigating the allowed low-energy phenomenologies which have a Higgs around 125 GeV and their corresponding collider signatures. 
In order to pursue this analysis we make use of some model-independent features of gauge mediated scenarios as organizing principles for classifying the low-energy spectrum:
\begin{enumerate}
\item The gravitino is always the lightest particle in the spectrum (LSP);
\item The nature of the next-to-lightest-superpartner (NLSP) dictates much of the phenomenology; 
\item For each NLSP type we can have different signatures at hadron colliders depending mostly on the decay length of the NLSP, on the nature of the NNLSP and on the mass scale of colored superpartners.
\end{enumerate}

In the end, we will also comment on the possibility of having tachyonic UV boundary conditions for the squark masses, provided the resulting low energy spectrum results tachyon-free. This scenario has been first suggested in \cite{Dermisek:2006ey, Essig:2007kh} as a way to minimize the fine-tuning problem in the MSSM and has received renewed attention as a mechanism to satisfy the Higgs mass constraint in gauge mediation keeping reasonably light stop masses \cite{Draper:2011aa}. We will indeed show this mechanism at work in section 6. 

We believe that our analysis can help to give an overall picture of the effects of a Higgs around  125 GeV for GGM, revealing which kind of scenarios are still permitted within a well defined UV framework in view of the current status of the LHC searches for new physics. 
\subsection{Outline and Summary of the Results}
In section 2 we give the details about our scan and review the basic phenomenology of GGM that will be relevant for our study. We first define the ranges for all the parameters of the EGGM parameter space and then discuss the different constraints imposed in our scan and their phenomenological implications. Our main focus will be on spectra which have a Higgs around 125 GeV. 

The sizeable fine-tuning induced by requiring a viable EWSB breaking and a Higgs mass around 125 is an unavoidable drawback of our approach which is compensated by having a well defined and fully calculable UV framework which is able to predict all the soft parameters of the MSSM, ensuring their flavor universality. 

One of the main motivations of this work is to understand to what extent GGM remains a powerful framework from which to obtain simplified SUSY spectra for collider searches \cite{Kats:2011qh}, and for this reason we summarize in section 2 the main physical features of the spectrum on which we will be interested in view of collider signature studies.

In sections 3, 4 and 5 we focus on the phenomenology of a restricted parameter space which is theoretically appealing because it arises naturally from the most general hidden sector with a GUT-complete structure. In section 3 the phenomenology of the CGGM case is reviewed. Requiring a Higgs mass around 125 GeV is pushing all the colored scalars at the multi-TeV scale as expected. 

The most interesting region of the parameter space for collider searches is achieved when the gaugino masses are screened with respect to the scalar ones. This scenario gives a low energy spectrum with Bino NLSP, neutral Wino NNLSP and charged Wino and gluino light enough to trigger EW and colored production. The scalar spectrum is  characterized by squark and sleptons at the multi-TeV scale very much like in Split SUSY scenarios \cite{ArkaniHamed:2004fb, Giudice:2004tc, ArkaniHamed:2004yi} with minimal splitting \cite{Wells:2003tf,  ArkaniHamed:2012gw, Arvanitaki:2012ps}. This kind of spectrum is quite generic in the gaugino screening region and does not feel the details of the UV parameters in the scalar sector,  thus we will find it substantially unmodified in sections 3, 4 and 5. In section 4 we will see that accidental cancellations in the EWSB condition can lead also to very light Higgsinos which can 
modify the experimental signature 
of this scenario. 
 
In the gaugino mediation region of CGGM, where the gaugino masses are heavier than the scalar ones, we get the standard gauge mediation spectrum with stau NLSP mostly right-handed. Allowing for extra 
contributions to the Higgs masses, we can modify this scenario obtaining a reversed hierarchy among the right-handed sleptons and a selectron NLSP, or a region of sneutrino co-NLSP with left-handed sleptons much lighter than the right-handed ones. A region of sneutrino co-NLSP can also be obtained through mild violations of messenger parity as we will see in section 5. 

The detectability of the gaugino mediation scenarios is very much suppressed by the heaviness of all the gaugino masses that is an unavoidable consequence of the assumption of GUT completeness. Motivated by this last observation, in section 6 we present scenarios where 
the supersymmetry breaking scales associated to the colored and uncolored sector
are split.

In subsection 6.1 we will see how splitting the gluino mass scale can lead to spectra with gluino NLSP in the gaugino screening region, while in the gaugino mediation region to a light stau NLSP together with quite light electroweak gauginos.
The last mechanism enhances the EW production of light sleptons and can be extended directly to all the cases with different slepton NLSP that we found in section 4. Conversely, the effect of MPV D-terms in section 5 has to be reconsidered for models which are not GUT-complete but we postpone this analysis for future works.

In subsection 6.2 we will disentangle the SUSY-breaking parameter $\Lambda_{S_{3}}$ which contributes to the squark masses. Setting the latter around 10 TeV, we get the typical gauge mediation spectra with neutralino NLSP, stau NLSP or stau-neutralino co-NLSP that can be produced in a hadronic collider via a sufficiently light gluino which has a 3-body decay channel through virtual squarks. This mechanism can be used to enhance the colored production in all the cases with different slepton NLSP that we found in section 4.

Allowing for tachyonic boundary conditions for the squarks, we briefly comment on the possibility of achieving maximal mixing scenarios in gauge mediation with zero $A$-terms. This scenario might allow for spectra with light stop masses in gauge mediation \cite{Draper:2011aa}, alleviating the fine-tuning problem \cite{Dermisek:2006ey, Essig:2007kh}. Unfortunately, in our simulation we do not find any region of the parameter space with a stop lighter than 2 TeV, signaling an unavoidable fine-tuning in gauge mediation with suppressed $A$-terms.

In section 7 we briefly discuss more general scenarios that can be achieved in the EGGM parameter space and section 8 contains our conclusions, where further perspectives of this work are discussed.  

\section{Phenomenology}

Before presenting the constraints on the parameter space and the details of the various spectra, we summarize the general strategy of our scan and discuss our assumptions and the constraints that we applied.

We choose to use the standard top-down approach whereby we first fix a certain number of boundary conditions
at the high scale $M_{\text{mess}}$ and then obtain the low energy data through RG evolution. Of course, this approach
can become computationally inefficient when the number of high energy parameters increases, since most of the
model points generated in a scan of the parameter space will not pass the low-energy constraints on the
soft spectrum.  For this reason, in order to explore the 6+1+3 dimensional EGGM parameter space, we are going to
proceed step-by-step, exploring sub-spaces of the general parameter space, according to each physically
interesting region.   

For each case we utilized a high performance computing cluster to perform a random scan
of the constrained parameter space under study.  To compute the soft spectrum from the EGGM soft terms we use a version of SoftSUSY 3.3.4 \cite{Allanach:2001kg} which we modified to accept three $\Lambda_{S_i}$ and three $\Lambda_{G_i}$, 
extra-contributions to the Higgs soft masses, and extra-contributions to each superpartner mass coming from mild messenger parity violations in the CGGM case.

Restricting to $\mu>0$ we consider the following range for the EGGM parameters:
\begin{align}
& 10^{3}\ \text{GeV}<\Lambda_{G_{i}}<10^{10} \text{ GeV}\ ,\  \vert\Lambda_{S_{i}}^2\vert <10^{20}\text{ GeV}^2\ ,\  \vert\delta m^2_{u,d}\vert <10^{20}\text{ GeV}^2\ ;\nonumber \\
& \text{Max}(\Lambda_{G_i},\Lambda_{S_i})
< M_{\text{mess}}<10^{14}\text{ GeV}\ ,\ 3<\tan\beta<70\ .\label{EGGMpar}
\end{align}    
As mentioned in the introduction, we allow also for negative values of $\Lambda_{S_i}^2$ and $\delta m^2_{u,d}$, corresponding to tachyonic squared masses in the UV. 
Restricting to GUT-complete hidden sectors with only one $\Lambda_G$ and $\Lambda_S$, we add MPV contributions parametrized by $\Lambda_D$: 
\begin{equation}
-10^{20}\ \text{GeV}^2<\Lambda_D^2<10^{20}\ \text{GeV}^2\ .
\end{equation} 

The lower bound on $M_{\text{mess}}$ is dictated by the requirement of no-tachyons in the messenger sector.
Moreover, concerning the explored region for $(M_{\text{mess}}, \Lambda_{G_i},\Lambda_{S_i})$,
we require to be in a regime where gauge mediation dominates over gravity effects. 
This translates into a bound on the gravitino
mass that we loosely estimate as
\be
\label{gravitinomass}
m_{3/2}\simeq \frac{M_{\text{mess}}\text{Max}(\Lambda_{G_i},\Lambda_{S_i}; \Lambda_{D})}{\sqrt{3}\tilde{k}
M_{Planck}}   < 10\ \text{GeV}\ .
\ee
Taking the maximum of all the UV SUSY-breaking parameters gives an estimate of the value of $F/M_{\text{mess}}$, where $F$ is the scale of SUSY-breaking felt by the messenger sector.

The gravitino mass obtained from super-Higgs mechanism \cite{Deser:1977uq} is proportional to the SUSY-breaking scale $F_0$ of the complete hidden sector that may differ from $F$ depending on how SUSY-breaking is communicated to the messenger sector. In order to parametrize this model dependent effect we define the coefficient $\tilde{k}=F/F_{0}$. In direct mediation scenarios $\tilde{k}$ would be typically of order 1, whereas we can get $\tilde{k}\ll1$ in scenarios where the SUSY-breaking effects are induced in the messenger sector through loop corrections. In the following we will always take $\tilde{k}=1$ for simplicity, but the possibility of varying this parameter should always be kept in mind. 
This completes the definition of the UV parameter space we scan over. 

The low-energy sparticle spectrum generated on the EGGM parameter space has to satisfy different constraints. First of all, SoftSUSY 3.3.4 ensures that the low-energy spectrum leads to a viable EWSB in the MSSM. The existence of a stable EWSB vacuum gives us two relations between the Higgs sector parameters 
\bea
\frac{m^{2}_{Z}}{2}&=&-\vert\mu\vert^2-\frac{(m^2_{H_{u}}+\delta m^2_{u}+\Sigma_{u})\tan^2\beta-(m^2_{H_{d}}+\delta m^2_{d}+\Sigma_{d})}{\tan^2\beta-1}\ , \label{min1}\\
\frac{2\tan\beta}{1+\tan^2\beta}&=&\frac{2B_{\mu}}{2\vert\mu\vert^2+m^2_{H_{u}}+\delta m^2_{u}+\Sigma_{u}+m^2_{H_{d}}+\delta m^2_{d}+\Sigma_{d}}=\frac{2 B_{\mu}}{m^2_{A}}>1\label{min2}\ .
\eea
$\Sigma_u$ and $\Sigma_d$ in the above formulas encode the radiative corrections to the masses of $H_{u}$ and $H_{d}$ induced by gauge interactions and top or bottom Yukawa interactions respectively. For a given value of $\tan\beta$, the $B_{\mu}$ parameter can be always expressed in terms of the CP-odd neutral scalar mass $m_{A}$ using \eqref{min2}. We will see in the following how the constraints \eqref{min1} and \eqref{min2}, which are distinctive features of the MSSM, will play a crucial role in determining some properties of the spectrum.

We then impose the experimental constraints, starting by setting the mass of the lightest Higgs boson to be at 125 GeV. 
In SoftSUSY 3.3.4 the
Higgs mass is computed at full one loop order 
plus zero-momentum two loop corrections in $\alpha_s, y_t, y_b, y_{\tau}$.
Taking into account possible
uncertainties in this
computation 
we allow for the range
\be
\label{Higgsmass}
123\ \text{GeV} \leq m_{h} \leq 127\ \text{GeV}\ .
\ee

In \cite{Draper:2011aa} was shown how the Higgs mass constraint might have strong consequences on the models of gauge mediation with vanishing $A$-terms. The main point is that, within the MSSM, the Higgs mass is bounded from above at the tree-level and at 1-loop receives contributions which depend logarithmically on the stop masses and polynomially in their mixing \cite{Martin:1997ns}:
\begin{align}
\label{Higgsone}
&m_{h\ (\text{1-loop})}^2=m_{Z}^2\left(\frac{1-\tan^2\beta}{1+\tan^2\beta}\right)^2+\frac{3m_{t}^4}{4\pi^2v^2}\left(\log\left(\frac{M_{S}^2}{m^2_{t}}\right)+\frac{X_{t}^2}{M_{S}^2}\left(1-\frac{X_{t}^2}{12M_{S}^2}\right)\right)\ ,\\
&\text{where}\ M_{S}=\sqrt{m_{\tilde{t}_{1}}m_{\tilde{t}_{2}}}\ ,\ X_{t}=A_{t}-\frac{\mu}{\tan\beta}\ ,
\end{align}
where the top mass is fixed at $m_{t}=173.5 \text{ GeV}$ and $v=2m_{W}/g_{2}\simeq 174.1 \text{ GeV}$.

A Higgs around 125 GeV is well above the MSSM tree-level upper bound of 91 GeV so that imposing the Higgs constraint would generically induce multi-TeV stop masses. The only possibility to avoid this scenario without modifying the MSSM is to maximize the $X_{t}$ contribution approaching the maximal mixing scenario where $\vert X_{t}/M_{S}\vert\simeq2$. However, it has been shown in \cite{Brummer:2012ns} that obtaining the maximal mixing scenario is particularly difficult in the EGGM parameter space in which the $A$-terms are zero at the UV scale.

We focus our analysis on the $\mu>0$ case for being in the favorable situation to get large $X_{t}$, since the $A_{t}$-term generated along the flow (always negative in the notations of \cite{Allanach:2001kg}) sums up with the $\mu$ contribution. 
In our top-down scan it will become quantitatively more clear how requiring a Higgs mass around 125 GeV puts severe constraints on the UV parameter space, reducing the regions with interesting phenomenology for colliders. 

In order to restrict our attention to the region of the parameter space which is relevant for collider searches, we set an upper bound of $10$ TeV on all the sparticle masses. This bound is large enough to include all the interesting spectra for collider physics and, at the same time, it excludes the possibility of having spectra with large mass splitting which can be difficult to simulate with SoftSUSY 3.3.4.

The choice of an upper bound of $10$ TeV on all the sparticle masses give also an indication of the maximum amount of tuning that we are allowing in our scenarios. The issue of giving a precise quantitative estimate of the fine-tuning in supersymmetric theories \cite{Barbieri:1987fn, deCarlos:1993yy} and more specifically in the context of gauge mediated SUSY-breaking theories \cite{Ciafaloni:1996zh, Abel:2009ve, Kobayashi:2009rn,Lalak:2013bqa} has been extensively discussed in the literature and we refer to that for more details on the subject.
 
Since we are assuming the MSSM as low energy theory, a Higgs mass within the allowed window \eqref{Higgsmass} and a solution for the EWSB conditions \eqref{min1} and \eqref{min2} would be always achieved  as the consequence of a certain amount of tuning of the UV parameters. Moreover, assuming GGM with zero $A$-terms as an UV completion, we are excluding the possibility of reaching the maximal mixing scenario in which the level of fine-tuning would be reduced to some extent \cite{Essig:2007kh}. 

The approach of our study follows more the spirit of \cite{ArkaniHamed:2012gw}, considering the fact that we are dropping the issue of naturalness in order to realize in a simple way other attractive features of the supersymmetric extensions of the Standard Model. In particular, we allow for a sizeable fine-tuning in the Higgs potential in order to have a fully calculable parameter space (EGGM) as an UV completion which can explain the origin of all the soft terms in the MSSM and their flavor universality. 
     
Implementing stringent constraints from direct searches in a global top-down scan of the EGGM parameter space can be very subtle, since the lower bounds on the sparticles masses from collider searches are very much dependent on the analysis, on the complete structure of the spectrum, and on the decay length of the NLSP.
Moreover, we will see that in a large portion of the parameter space the constraints from direct searches are still not competitive with the much stringent Higgs constraint \eqref{Higgsmass}, which is independent from the details of the spectrum\footnote{This is in line with the analysis of the CGGM case \cite{Grellscheid:2011ij} in which the LEP bound on the Higgs mass \cite{Schael:2006cr} was shown to be more constraining than the ATLAS and CMS direct searches on $1/\text{fb}$ of data \cite{Aad:2011ib, Chatrchyan:2011zy} in a large portion of the parameter spcae.}.

For both these reasons we will be very conservative on the lower mass bounds on the uncolored particles at the level of the simulation, implementing only the bounds from direct searches at Tevatron and LEP already used in \cite{Carpenter:2008he,Abel:2009ve} which are summarize in the Table 1. With regard to the colored superpartners, the bounds we implement at the level simulation are not updated to the current bounds from LHC searches: $m_{\tilde{g}}\geq 51\text{ GeV}$ \cite{Kaplan:2008pt}, $m_{\tilde{t}}\geq 92.6\ , m_{\tilde{b}}\geq89, m_{\tilde{u},\tilde{d},\tilde{s},\tilde{c}}\geq 97 \text{ GeV}$ \cite{Abbiendi:2002mp,Heister:2002hp}.  The reason for this choice is that we want to highlight how assuming an Higgs mass around 125 GeV already significantly constrains the colored spectrum in gauge mediation.
We will then comment case by case on how the parameter space of EGGM will be further restricted by LHC direct searches focusing on the relevant processes for each region of the parameter space \cite{Kats:2011qh}. A more quantitatively precise analysis of the consequences of LHC direct searches on the EGGM parameter space would require a MonteCarlo simulation of signal events and a direct implementation of the experimental cuts similar to what was done for the CGGM parameter space in \cite{Dolan:2011ie}. This analysis would be essential in order to identify GGM benchmark points with a Higgs at 125 GeV and we leave it for future works. 
\begin{table}
\begin{center}
\begin{tabular}{|c|c|c|c|c|}
\hline
 Superpartner & Lower Bound & Source\\ \hline
 Neutralinos &  $m_{\tilde{N}}\geq 46$ GeV & \cite{Abdallah:2003xe}\\ \hline
 Charginos   &$m_{\tilde{C}}\geq45$ GeV &\cite{Caso:1998tx} \\ \hline
Sleptons	  &$m_{\tilde{\tau}}\geq68\ ,\ m_{\tilde{e},\tilde{\mu}}\geq85$ GeV&\cite{Barate:2000tu, Barate:1999gm}\\ \hline
Sneutrinos  & $m_{\tilde{\nu}}\geq51$ GeV&\cite{Decamp:1991uy}\\ \hline
\end{tabular}
  \caption{Lower bounds on uncolored sparticle masses implemented in the simulation.}
  \end{center}
  \end{table}
  
We also require the generated spectra to satisfy the constraints coming from flavour physics. An appealing feature which is characterizing all the gauge mediated models is to predict nearly exact flavor invariance of the soft terms. However, precision measurement of B-observables  are particularly sensible to the exchange of new particles and some of these effects can even be enhanced in the large $\tan\beta$ regime. For this reason, the experimental constraints on the rare branching ratios BR($B_{s}\to\mu^+\mu^-$), BR($B\to X_{s}\gamma$) and BR($B^{\pm}\to\tau^{\pm}\nu_{\tau}$) lead to strong lower bounds on MSSM sparticle masses. 

We use SuperIso v3.3 \cite{Mahmoudi:2008tp} to calculate all the flavor observables which are listed in the Table 2. We take the constraints on the flavor observables from the updated version of the SuperIso manual imposing either the combined experimental value or the $95\%$ C.L. bound depending on the sensitivity of our setup to the specific flavor observable.
\begin{table}[ht]
\begin{center}
\begin{tabular}{|c|c|c|c|c|}
\hline
 Observable & Constraint & Source \\ \hline
 BR($B\to X_{s}\gamma$)  & $2.16\times 10^{-4}\leq\text{BR}(B\to X_{s}\gamma)\leq 4.93\times10^{-4}$ & $95\%$ C.L. \cite{Mahmoudi:2008tp,Barberio:2008fa}\\ \hline
 $\Delta_0(B\to K^{*}\gamma)$   & $-1.7\times 10^{-2}\leq\Delta_0(B\to K^{*}\gamma)\leq 8.9\times10^{-2}$    &$95\%$ C.L. \cite{Mahmoudi:2008tp, Aubert:2008af, Nakao:2004th}\\ \hline
 BR($B_{s}\to \mu^{+}\mu^{-}$)   & $\text{BR}(B_{s}\to \mu^{+}\mu^{-}\leq 4.5\times10^{-9})$   &combined exp. \cite{Aaij:2012ac} \\ \hline
 BR($B_{u}\to \tau\nu_{\tau}$)  & $0.71\times 10^{-4}\leq\text{BR}(B_{u}\to \tau\nu_{\tau})\leq 2.57\times10^{-4}$   &$95\%$ C.L. \cite{Mahmoudi:2008tp,Aubert:2007dsa}\\ \hline
R($B_{u}\to \tau\nu_{\tau}$)  & $0.56\leq\text{R}(B_{u}\to \tau\nu_{\tau})\leq 2.7$   &$95\%$ C.L. \cite{Mahmoudi:2008tp,Aubert:2007dsa}\\ \hline
BR($B\to D_{0}\tau\nu_{\tau}$) & $4.5\times 10^{-3}\leq\text{BR}(B\to D_{0}\tau\nu_{\tau})\leq 12.7\times 10^{-3}$ & combined exp. \cite{Aubert:2007dsa} \\ \hline
$\xi_{Dl\nu}$   & $0.247\leq\xi_{Dl\nu}\leq 0.585$    & combined exp. \cite{Aubert:2007dsa}\\ \hline
$\frac{BR(K\to \mu\nu)}{BR(\pi\to\mu\nu)}$   & $0.6257\leq\frac{BR(K\to \mu\nu)}{BR(\pi\to\mu\nu)}\leq 0.6459$   &$95\%$ C.L. \cite{Mahmoudi:2008tp,Antonelli:2010yf}\\ \hline
$R_{\mu23}$   & $0.992\leq R_{\mu23}\leq 1.006$    & combined exp. \cite{Antonelli:2010yf}\\ \hline
\end{tabular}
  \caption{Constraints on flavour observables}
  \end{center}
  \end{table}

This concludes the set of all the constraints that we will impose through our study.
Although we perform the scan over the complete parameter space (\ref{EGGMpar}),
we often restrict to two dimensional slices in order to present the results in useful plots.
In particular we will often
 fix $M_{\text{mess}}$ and $\tan \beta$ to the values ($10^7\ \text{GeV}, 10^{13}\ \text{GeV}$) 
 and ($10 \pm 5, 35 \pm5$) respectively.
 The two choices of $M_{\text{mess}}$ correspond to short and long
 RG running, whereas the choices of $\tan \beta$ correspond to moderate-small (compatible
 with the large Higgs mass)
and large $\tan \beta$ scenarios.

For every sub-case we will discuss the physical features that are the basics to perform a more detailed phenomenological and
LHC-oriented analysis:
\begin{itemize}
\item The restriction imposed by the Higgs bound (\ref{Higgsmass}) on the parameter space ;
\item The allowed NLSP cases ;
\item The gluino, stop, and first generation squarks masses ;
\item The allowed mass range for each NLSP case ;
\item The nature of the NNLSP and the mass splitting with the NLSP ;
\item The decay length of the NLSP (promptly decaying or long-lived) .
\end{itemize}
More details on the low-energy spectra will be discussed case by case when needed.
 
In gauge mediated theories,
the decay of the NLSP is always a two body decay to its standard model partner and the gravitino LSP.
The decay rate can be computed from the effective action for the gravitino 
(see for instance \cite{Martin:1997ns,Giudice:1998bp})
and it is given approximately by
\be
\Gamma(\tilde x \to x G)=\frac{\tilde{k}^2 m_{\tilde x}^5}{16 \pi F^2}\simeq\frac{m_{\tilde x}^5}{16 \pi (\sqrt{3}M_{Planck}m_{3/2})^2}\ , \label{decayrateNLSP}
\ee
where we neglected numerical factors due to possible NLSP mixing\footnote{This factors are particularly important for NLSPs like the neutralino $\tilde{N}_{1}$ which can have different decay channel depending their dominant components.}.
We can rewrite this as a function of the gravitino mass using the estimate (\ref{gravitinomass}).
The decay length in meters results then
\be
\label{decaylength}
L=\left(\frac{100 \ \text{GeV}}{m_{NSLP}} \right)^5 
\left(\frac{\sqrt{3} M_{Planck} m_{3/2}}{(100 \ \text{TeV})^2} \right)^2 10^{-4} \ \text{m}
\ee
and it will be used to discriminate between NLSP decaying inside or outside the detector. Substituting the estimate for the gravitino mass \eqref{gravitinomass} we see that $L$ is proportional to $M_{\text{mess}}^2$ which, therefore, controls the decay length of the NLSP in gauge mediation.

We conclude this section with a remark about cosmology.
In our analysis we did not put any constraint related to the nature of dark matter
and its abundance. In most of the region of the parameter space the
LSP (the gravitino) cannot be the dark matter, typically being too heavy. 

A strong bound on heavy stable gravitino comes from 
its contribution to the present energy density of the Universe. In the absence of a mechanism of dilution, the gravitino mass cannot be too large in order to avoid the overclosure of the Universe \cite{Moroi:1993mb,deGouvea:1997tn}.
In the short running case with $M_{\text{mess}}=10^6\ \text{GeV}$ we have $\tilde{k}m_{3/2}\simeq \text{KeV}$ that it is just at the edge of the cosmological upper bound on the gravitino mass, whereas in the long running case with $M_{\text{mess}}=10^{13}\ \text{GeV}$ $\tilde{k} m_{3/2}\simeq  \text{GeV}$. In both cases we have to assume that some mechanism is at work in order to dilute the gravitino abundance at early time, and also that a proper dark matter candidate can be identified within the particles of the hidden sector \cite{Dimopoulos:1996gy}.

From \eqref{decaylength} we see that for very large $M_{\text{mess}}$ the NLSP can become very long lived, eventually interfering with nucleosynthesis products \cite{Gherghetta:1998tq}. The damaging effects of NLSP decay products on nucleosynthesis depend on the precise calculation of the NLSP abundance at the time of freeze out which requires a full knowledge of the spectrum. However for $L\gtrsim 3\times10^{12} \text{ m}$ a fairly generic bound on $M_{\text{mess}}$ can be derived considering the effects of NLSP decay into hadronic jets during nuclesynthesis which would induce $^7\! \text{Li}$ overproduction. 
Choosing $M_{\text{mess}}=10^{13} \text{ GeV}$ could be at the border of the allowed region for most of the NLSP choices \cite{Gherghetta:1998tq}, however this problem can be easily overcome by lowering the value of $M_{\text{mess}}$ without changing our main results. 

\section{Constrained General Gauge Mediation}\label{cGGM}

We begin the analysis of the parameter space of EGGM by studying the simplest situation, 
that is the Constrained General Gauge Mediation (CGGM) case in which we have one scale for the gaugino masses and one scale for the 
scalar masses. This scenario arises naturally in hidden sectors with a GUT-complete structure which can realize  gauge and mass unification in a simple way. Assuming $\Lambda_S^2>0$ we ignore the possibility of having tachyonic sleptons and squarks in the UV theory so that we have  
\be
\Lambda_{G_i}= \Lambda_G \qquad \Lambda_{S_i}= \Lambda_S \qquad \qquad i=1,2,3 \qquad \Lambda_S^2>0\ .
\ee

This case has been extensively studied in the papers \cite{Abel:2009ve, Abel:2010vba, Dolan:2011ie, Grellscheid:2011ij}, 
in the case of vanishing $B_{\mu}$ term 
at the messenger scale.
We repeat here the analysis allowing for generic values of $B_{\mu}$,
focusing on the consequences of a 125 GeV Higgs mass
on this simplified scenario. 
We use this case to check the validity of our software and
as a guideline to analyze situations with more free parameters.

We have scanned over the entire parameter space, 
including $M_{\text{mess}}$ and $\tan\beta$. The range of $\tan\beta$ gets restricted to $5.3\leq\tan\beta\leq65$ independently of any other parameter. The absolute lower bound for $\tan\beta$ is determined by the Higgs mass constraint in agreement with \cite{Draper:2011aa}, while the upper bound follows from requiring the stau squared mass matrix to be always positive definite at the electroweak scale. The no-tachyon condition in the messenger sector and the Higgs bound set a lower bound for $M_{\text{mess}}$ which is roughly $M_{\text{mess}}\geq 10^6 \text{ GeV}$.  
 
In order to show the results, we fix the values for $M_{\text{mess}}$ and $\tan\beta$,
as explained in the previous section.
The generic dependence of the sparticle spectrum on these parameters can be easily
inferred by analyzing the four distinct cases we present.
For each one we will show several plots in the $\Lambda_G$ $\Lambda_S$ plane.

The parameters $\Lambda_G$ and $\Lambda_S$ determine the overall scale of the
gaugino and of the scalar masses at the messenger scale.
Their ratio can be related to a particular class of gauge mediated models in the UV.
When $\Lambda_G \gg \Lambda_S$ the scenario is the so-called gaugino mediation which can be explicitly realized in messenger models using a large number of
messengers \cite{Dimopoulos:1996yq}, or in quiver-like construction as in \cite{Green:2010ww,Auzzi:2010mb}.
In this scenario the gauginos are much heavier than the scalars at the mediation scale.
Along the RG flow, the gaugino induce positive squared masses to the scalars
via one loop quantum corrections. 
Depending on the length of the RG flow (so on $M_{\text{mess}}$) 
the spectrum at
the EW scale can be still hierarchical, with gauginos heavier than scalars, 
or quite democratic. 

When $\Lambda_G \ll \Lambda_S$ the scenario is the so-called gaugino screening, which can be realized in direct mediation models if an approximate R-symmetry is present at the mediation scale  \cite{ArkaniHamed:1998kj}, 
or if the gaugino masses are suppressed by further loop factors with respect to the scalar masses in semi-direct mediation models
\cite{Argurio:2009ge}. The result is that the gauginos are lighter in this scenario, independently on the RG flow length.

Finally, in the region where $\Lambda_G \simeq \Lambda_S$, the resulting phenomenology
is essentially the one of Minimal Gauge Mediation.
The Minimal Gauge Mediation scenario \cite{Dimopoulos:1996yq}, with one 
pair of vector-like messenger in the $5+\bar{5}$ of $SU(5)$ which is coupled to a spurion field $X=M_{\text{mess}}+\theta^2 F$,
coincides with our CGGM with
$\Lambda_G=\Lambda_S$ only in the case in which the SUSY-breaking scale $\sqrt{F}$ is much
smaller than $M$. 
When $\sqrt{F} \lesssim M_{\text{mess}}$ there are higher order corrections in the MGM model which typically enhance the gaugino masses \cite{Martin:1996zb}.
This situation is included in our full scan, being 
eventually mimicked by slight deviations from the relation $\Lambda_G=\Lambda_S$.

\begin{figure}[ht]
\captionsetup[subfigure]{labelformat=empty}
\begin{center}
\subfigure{
\includegraphics[width=7.5cm]{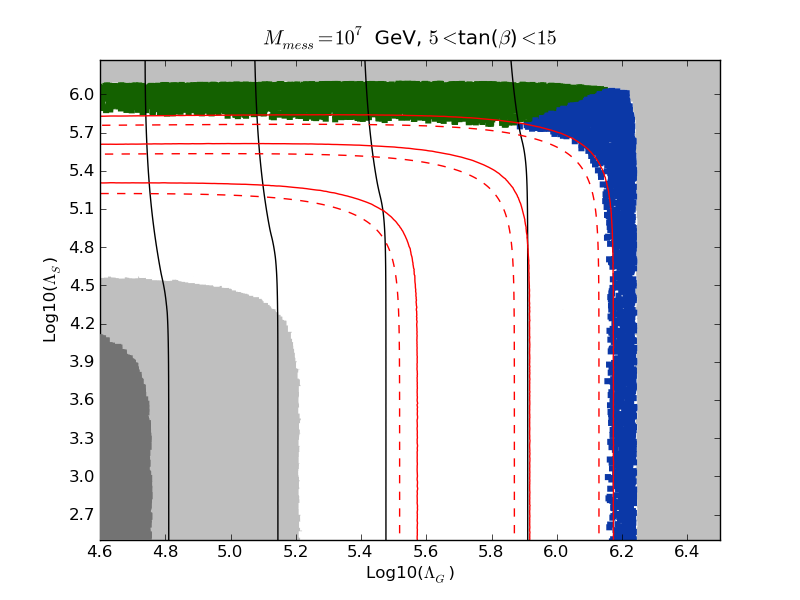}
}
\subfigure{
\includegraphics[width=7.5cm]{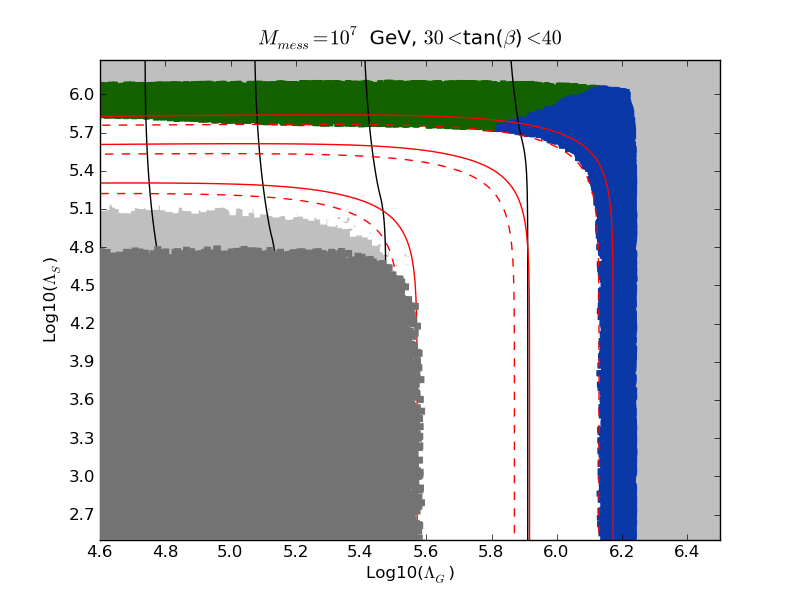}
}
\subfigure{
\includegraphics[width=7.5cm]{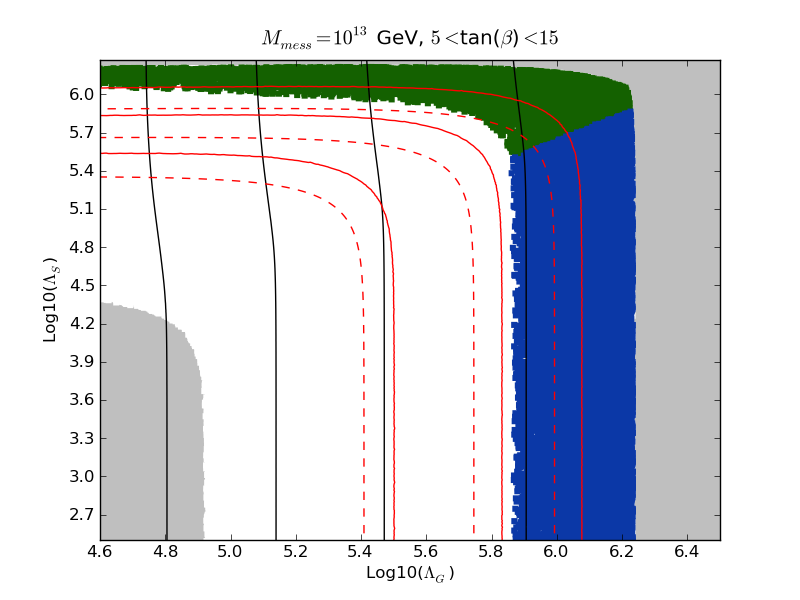}
}
\subfigure{
\includegraphics[width=7.5cm]{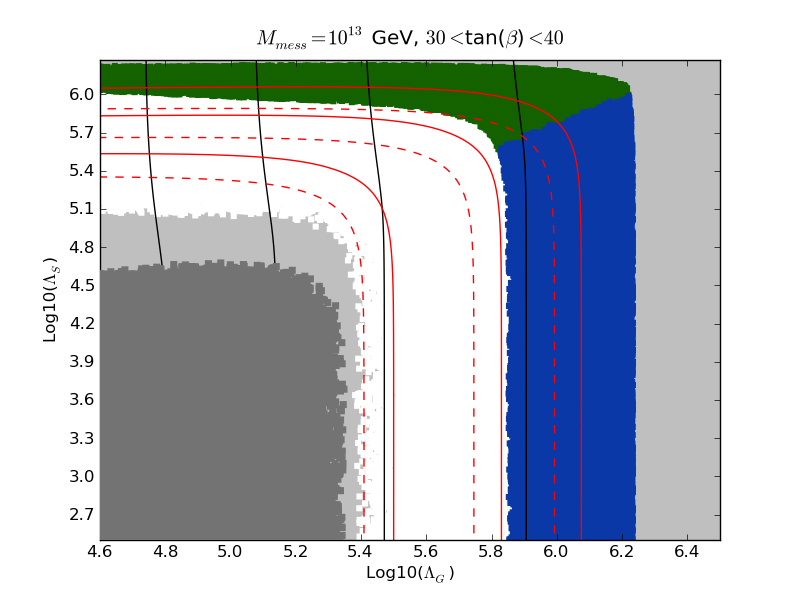}
}
\caption{
\label{CGGMa}\footnotesize
Logarithmic plot in the $\Lambda_G, \Lambda_S$ plane. Explanations of the colors is 
in the text. 
The black, red, dashed-red contour plots identify the gluino, lightest stop, first generation masses
respectively.
The scales of the contours are $500$ GeV, $(1,2,5)$ TeV for the gluino and
$(1.5,3,5)$ TeV for the stop and for the first generation squarks.
}
\end{center}
\end{figure}

In Figure \ref{CGGMa} we show the plots in logarithmic scale 
in the $(\Lambda_G,\Lambda_S)$ plane
for fixed values of $M_{\text{mess}}=10^7,10^{13}$ GeV and $\tan \beta = 10, 35 \pm 5$.
The dark grey region corresponds to points where SoftSUSY did not converge.
In the light grey region, SoftSUSY computes the spectrum but it does not satisfy the
constraints that we described in the previous section. 
The light grey region in the bottom left corner of the plots is due to collider and flavor observables constraints, the tiny light grey band in the upper part of our plots is due to the upper bound of 10 TeV on the squark masses, while the light grey band in the right part of our plots is due to the upper bound of 10 TeV on the gluino mass. 
The white region is excluded because it does not satisfy the Higgs mass constraint.
Finally, the blue and green regions are the allowed ones compatible with all the constraints.

Blue corresponds to stau NLSP, whereas green corresponds to neutralino NLSP.
As can be expected,
the stau is the NLSP in the region of gaugino mediation, whereas the neutralino is the NLSP
when there is gaugino screening. In the intermediate region, where $\Lambda_G \simeq \Lambda_S$,
both stau and neutralino can be the NLSP, depending on the length of the RG flow and on 
$\tan\beta$.

The black, red, dashed-red contour plots identify the gluino, lightest stop, first generation squark masses
respectively.
The 125 GeV Higgs is obtained through a heavy stop, essentially always above 3 TeV.
The shape of the allowed region is slightly dependent on the messenger mass and basically independent on $\tan \beta$.
For $\tan \beta>5.3$ the tree level upper bound on the Higgs mass is already saturated as can be seen from \eqref{Higgsone}. As a consequence the parameter space is only slightly enlarged for large $\tan\beta$. A more important role is played by $M_{\text{mess}}$, a large value of which helps 
to satisfy the Higgs mass bound by generating along the flow larger squark masses and sizable $A$-terms through loop corrections controlled by the gluino mass.  
As a consequence, the allowed region for the parameters $\Lambda_G,\Lambda_S$ gets larger for large $M_{\text{mess}}$ and the stop mass can be smaller to some extent.

The stop is always the lightest of the squarks, and the 
first generation squarks are as a result very heavy in 
the region of the parameter space compatible with the Higgs mass constraint.
A large mass for the squarks is obtained or via a large UV value of the 
$\Lambda_S$ parameter or induced by a large gluino mass through gaugino mediation.
This explains the shape of the contours for the stop and for the first generation squark masses.

The gluino is not directly related to the Higgs mass, and it can be 
light, provided $\Lambda_S$ is large enough.
In the region of neutralino NLSP, the gluino is the lightest of the colored sparticles,
and it is relevant for LHC phenomenology. In the region of stau NLSP, all the colored sparticles are very heavy,
essentially decoupled for what concerns collider phenomenology.

\begin{figure}[ht]
\begin{center}
\subfigure{
\includegraphics[width=7.5cm]{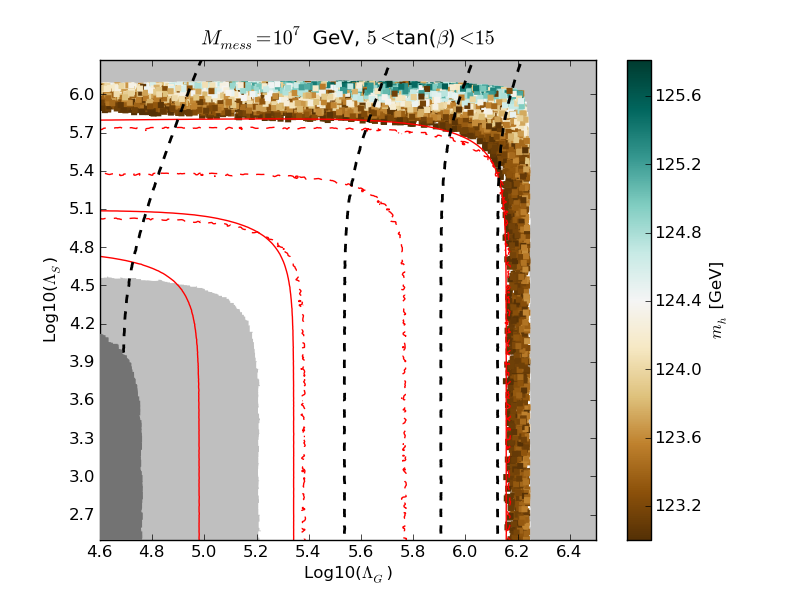}
}
\subfigure{
\includegraphics[width=7.5cm]{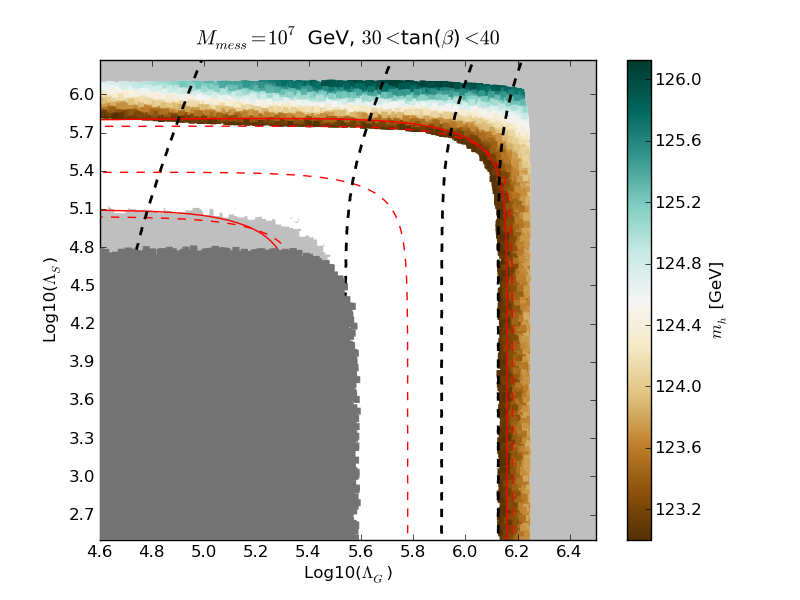}
}
\subfigure{
\includegraphics[width=7.5cm]{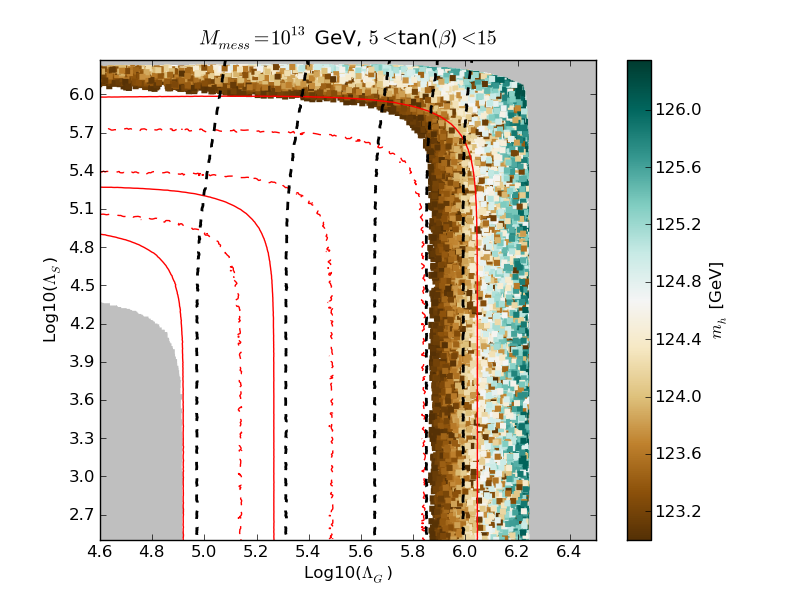}
}
\subfigure{
\includegraphics[width=7.5cm]{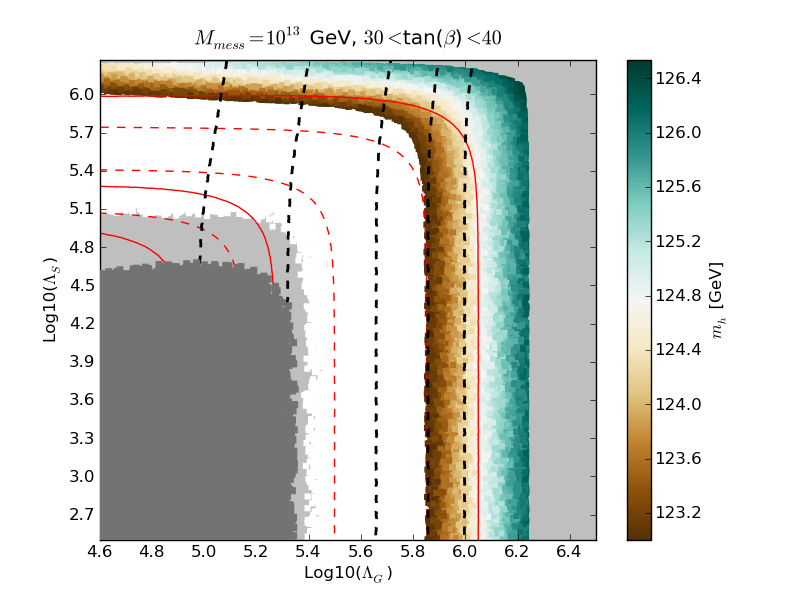}
}
\caption{
\label{CGGMb}\footnotesize
Logarithmic plot in the $\Lambda_G, \Lambda_S$ plane. 
The gradient indicates the Higgs mass on the allowed region.
The red, dashed red and dashed black contours 
indicates $M_S ,\mu$ and $A_t$ respectively.
The scales of the contours are $(0.5,1,5)$ TeV for $M_S$, 
$(0.5,1,2)$ TeV for $\mu$, and $(-0.2,-1,-2,-3)$ TeV  for $A_t$ and also $-4$ TeV in the second row.
}
\end{center}
\end{figure}
In figure \ref{CGGMb} we show the value of the Higgs mass and the
physical quantity entering in the one loop formula (\ref{Higgsone}).
The contours display the values of $\mu, A_t, M_S$.
A negative $A_t$ term is generated along the RG flow by the gluino mass,
explaining why it increases with increasing $\Lambda_G$.
Moreover, for large $M_{\text{mess}}$,
the induced $A_t$ term can be quite large thus the
allowed region extends to smaller values for the average stop mass $M_S$,
corresponding to a lightest stop mass eigenvalue around $3$ TeV.
However, since the gluino mass also induces squark masses 
at 1-loop, we cannot reach scenarios of maximal mixing ratio
$\frac{X_t}{M_s}\simeq2$ \cite{Brummer:2012ns} which is the relevant quantity in the 1-loop formula of the
Higgs mass (\ref{Higgsone}).
The heavy Higgs mass is thus obtained predominantly 
through a large value of $M_S$, 
explaining the shape of the allowed region.

The $\mu$ parameter is also quite large, 
indicating a fine-tuning in the minimization of the Higgs potential. This feature is easy to understand by expanding the EWSB condition \eqref{min1} for large $\tan\beta$:
\be
\frac{m^2_{Z}}{2}\simeq -\vert\mu\vert^2-(m^2_{H_{u}}+\Sigma_u)+\mathcal{O}\left(\frac{\Sigma_u-\Sigma_d}{\tan^2\beta}\right)\ .
\ee
At low energies $m^2_{H_{u}}+\Sigma_u$ becomes large and negative because $\Sigma_u$ is dominated by the negative contribution from top Yukawa interactions which are proportional to $m_{\tilde{t}}^2$. This effect induces a $\mu\gg m_{Z}$, essentially $\vert\mu\vert\simeq\sqrt{-(m^2_{H_{u}}+\Sigma_u)}$. In the gaugino screening region the positive value of $\mu$ arises from the compensation of the large positive contributions to $m^2_{H_{u}}$ proportional to $\Lambda_S^2$ and the large negative contributions to $\Sigma_u$ driven by the large value of the stop mass squared, controlled again by $\Lambda_S^2$. This mechanism is essentially independent on $M_{\text{mess}}$ and $\Lambda_G$. Conversely, in the gaugino mediation region, the masses of the up-Higgs and of the stop are generated by gaugino 1-loop effects, which are proportional to the Wino mass and the gluino mass respectively. For $M_{\text{mess}}=10^7$ the only scenario which satisfies the Higgs mass constraint is the one in which the gluino mass reaches the upper bound $m_{\tilde{g}}=10 \text{ TeV}$ and sets up the maximal value for the stop mass around $5 \text{ TeV}$ which gives the lower bound $\mu \geq 2 \text{ TeV}$. For $M_{\text{mess}}=10^{13}$ GeV we get viable scenarios with a lighter stop mass around 3 TeV, but this effect is compensated by the fact that top-Yukawa contribution in $\Sigma_u$ is enhanced by $\log (\frac{M_{\text{mess}}}{m_{\tilde{t}}})$ and the $\mu$ lower bound results again around 2 TeV.

A large $\mu$ term induces large masses for all the Higgsinos which are always above 2 TeV in the allowed parameter space. Similarly, the heavy Higgses $H^{0}, A^{0}, H^{\pm}$ result always above 2 TeV, being decoupled from the IR physics even for quite large $\tan\beta$. 
\begin{figure}[ht]
\begin{center}
\subfigure{
\includegraphics[width=7.5cm]{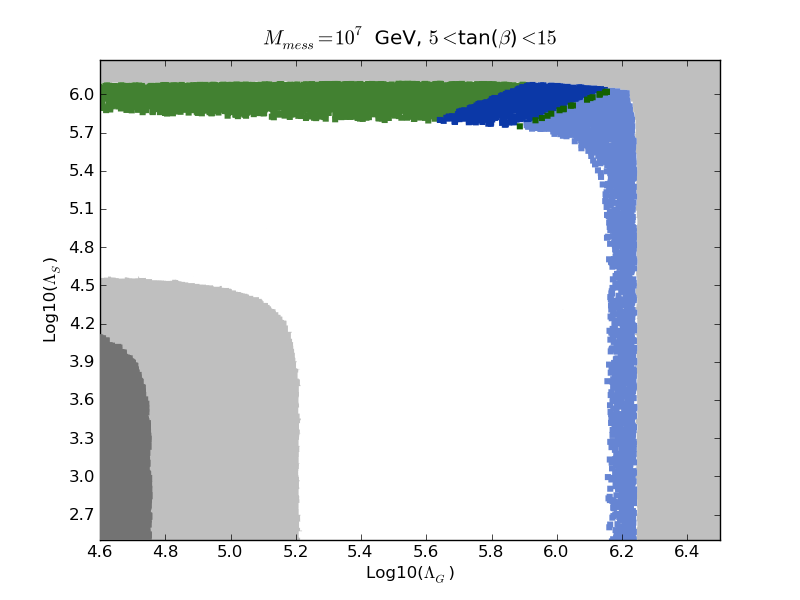}
}
\subfigure{
\includegraphics[width=7.5cm]{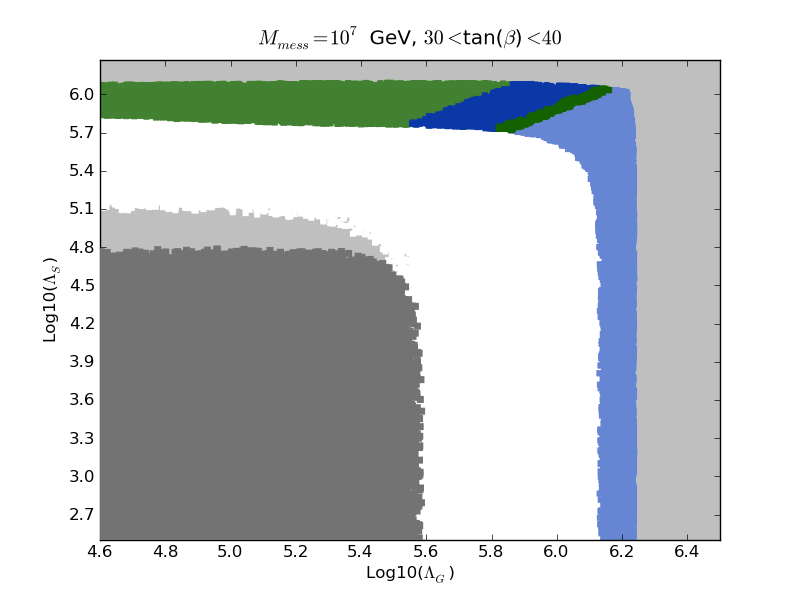}
}
\subfigure{
\includegraphics[width=7.5cm]{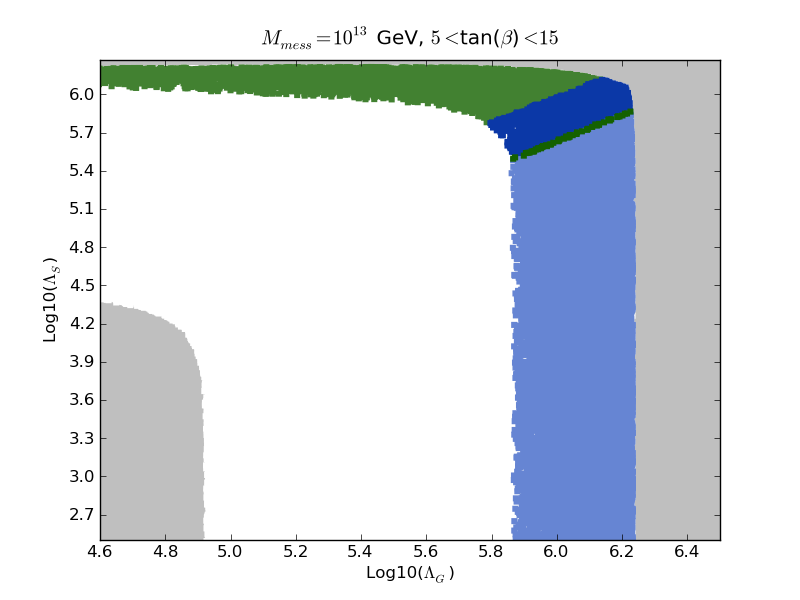}
}
\subfigure{
\includegraphics[width=7.5cm]{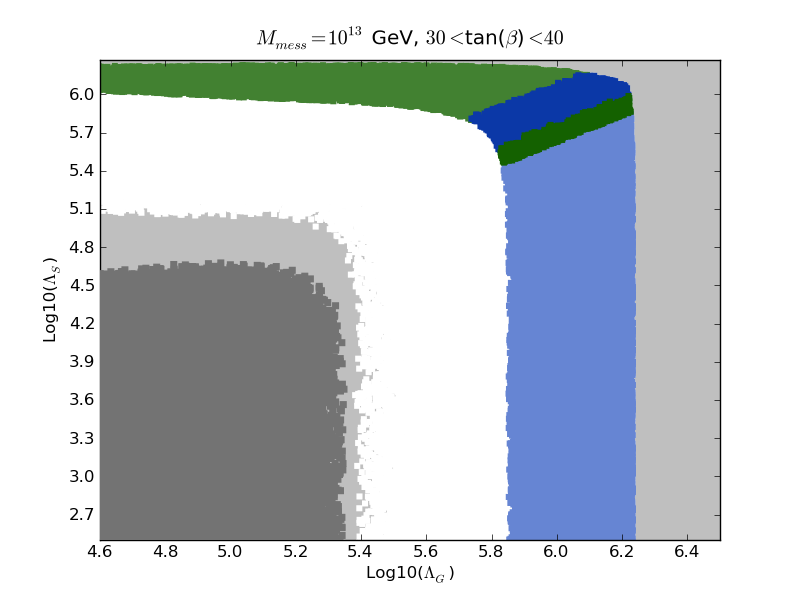}
}
\caption{
\label{CGGMc}\footnotesize
Logarithmic plot for the NNLSP in the $\Lambda_G, \Lambda_S$ plane. 
The NNLSP colors are light green for the second lightest neutralino,
blue for the stau, pale blue for the smuon, and green for the lightest neutralino.
}
\end{center}
\end{figure}
\begin{figure}[ht]
\begin{center}
\subfigure{
\includegraphics[width=7.5cm]{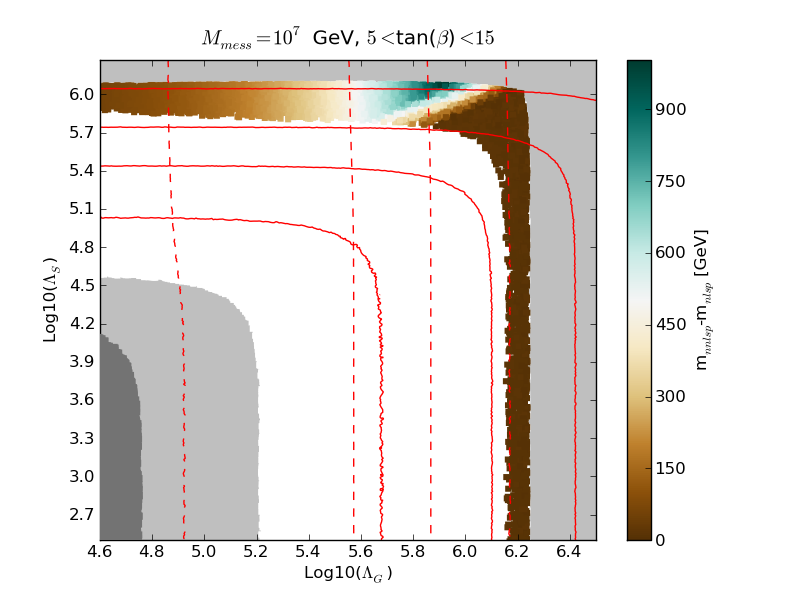}
}
\subfigure{
\includegraphics[width=7.5cm]{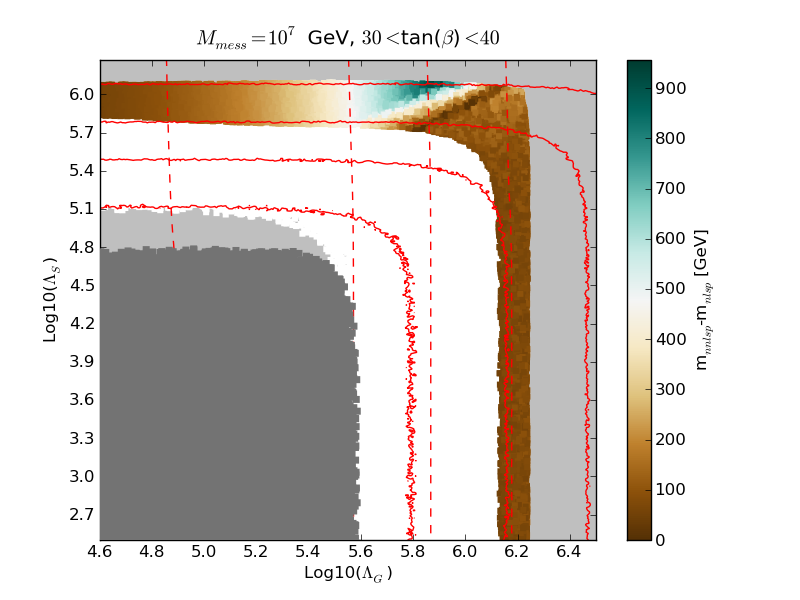}
}
\subfigure{
\includegraphics[width=7.5cm]{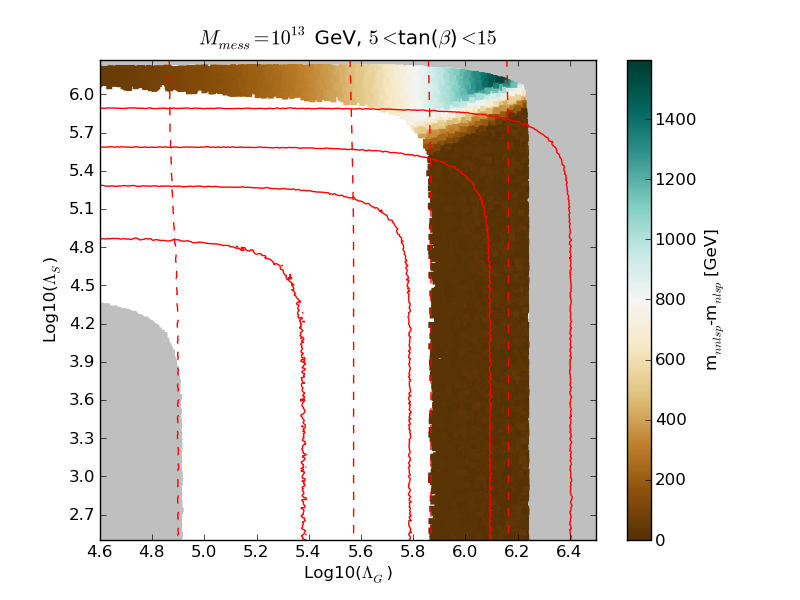}
}
\subfigure{
\includegraphics[width=7.5cm]{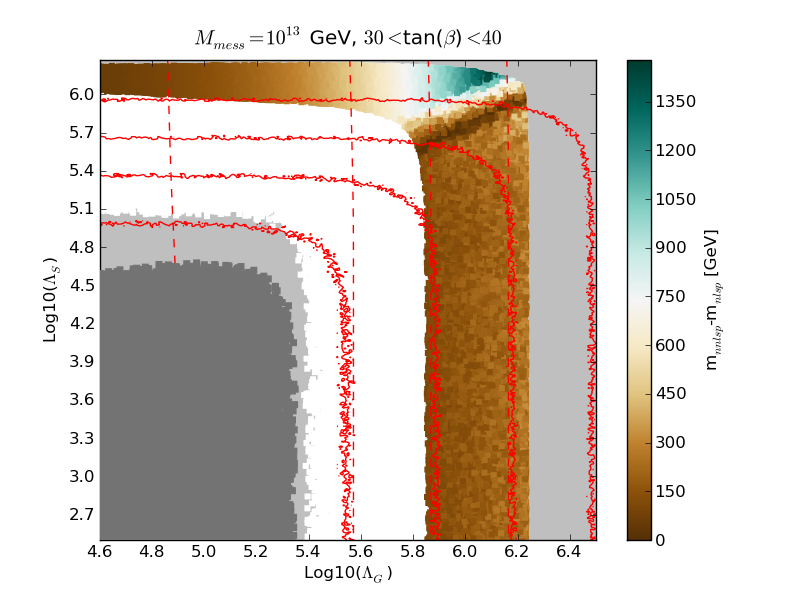}
}
\caption{
\label{CGGMd}\footnotesize
Logarithmic plot in the $\Lambda_G, \Lambda_S$ plane. 
The gradient represents the mass difference $m_{\text{NNLSP}}-m_{\text{NLSP}}$.
The solid and dashed contours identify the NLSP masses, stau and neutralino
respectively.
The scales of the contours are $(100,500)$ GeV and $(1,2)$ TeV
for the neutralino, and $(200,500)$ GeV and $(1,2)$ TeV for the stau.
}
\end{center}
\end{figure}

In figure \ref{CGGMc} we show the NNLSP species and in figure \ref{CGGMd} 
the mass difference between the 
NNLSP and the NLSP, with contours for the values of the NLSP mass,
i.e. stau and neutralino. 

In the gaugino mediation region, the stau mass is dominated by gaugino mediation contribution as can be observed by the shape of the contours and it varies from 460 GeV to 1.5 TeV for long running and from and from 480 GeV to 800 GeV for short running. 
In the long-running cases, with $M_{\text{mess}}=10^{13}$ GeV, 
the gaugino mediation region gets larger and extends to smaller values of $\Lambda_{G}$,
since the Higgs mass bound is more easily satisfied.
As a consequence, we can have lighter stau masses with respect to the short-running cases. 
At the same time, a larger values of $\tan\beta$ maximize the off diagonal contribution 
in the stau mass matrix, making even smaller the lightest mass eigenvalue $\tilde{\tau}_{1}$.
Thus the scenario with the lightest allowed stau NLSP (around 460 GeV) is realized in the
$M_{\text{mess}}=10^{13}$, $\tan \beta = 35 \pm 5$ case.
Observe that 
the dependence on $\tan\beta$ of the stau mass explains why its contours have small fluctuations in our plots,
 in which $\tan\beta$ can vary within a range. 
 
In the stau NLSP region with small $\tan\beta$, the typical scenario with co-slepton NLSP is realized, specifically the right-handed smuon is the NNLSP with the selectron almost degenerate. The mass splitting between the selectron and the smuon is always negligible, while the splitting between the third and the first two generations depends mostly on $\tan\beta$ and can be more than 50 GeV for $\tan\beta=35\pm5$, opening the possibility of having three-body decays through virtual neutralinos $\tilde{l}_{R}\rightarrow\tilde{\tau}_{1}\tau l$ \footnote{With the label $l=e,\mu$ we will often indicate the first two families which we can consider always degenerate in gauge mediation.} \cite{Ambrosanio:1997bq}.  

The lightest neutralino mass eigenvalue $m_{\tilde{N}_{1}}$\footnote{In the following we are going to follow the SoftSUSY convention for the neutralino and chargino mass eigenvalues \cite{Allanach:2001kg}, ordering them from the lightest to the heaviest.} is mostly Bino in all the allowed parameter space since $\mu\gtrsim M_{\tilde{\lambda}_{1}}$ and it varies between 50 GeV and few TeV being substantially determined by the value of $\Lambda_G$ at $M_{\text{mess}}$. 
In the neutralino NLSP case, the most common NNLSP is the neutral Wino which can also be very light in the gaugino screening region where $\mu\gtrsim M_{\tilde{\lambda}_{2}}$.
For values of $\Lambda_G \simeq \Lambda_S \simeq10^6\ \text{GeV}$ we are in the
transition region between stau and neutralino NLSP in which, however, the NLSP
mass is quite large, being around the TeV scale.

\begin{figure}[ht]
\begin{center}
\subfigure{
\includegraphics[width=7.5cm]{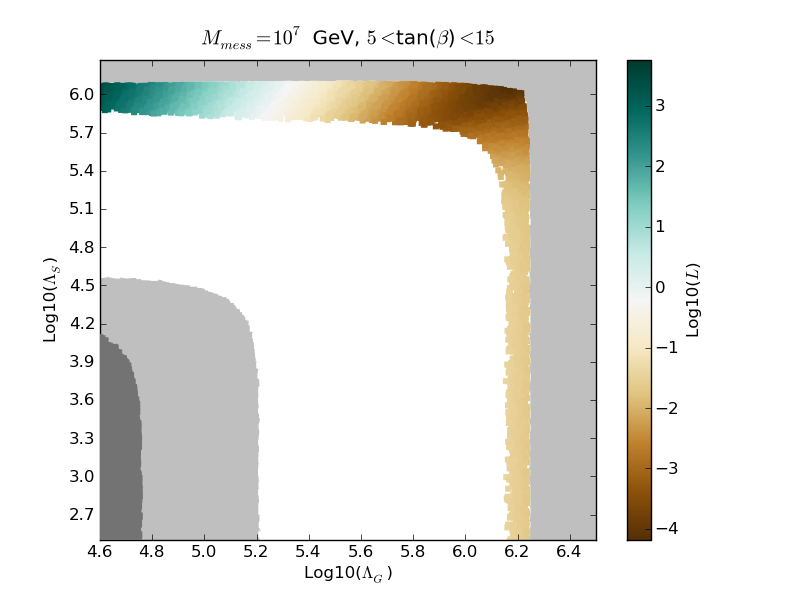}
}
\subfigure{
\includegraphics[width=7.5cm]{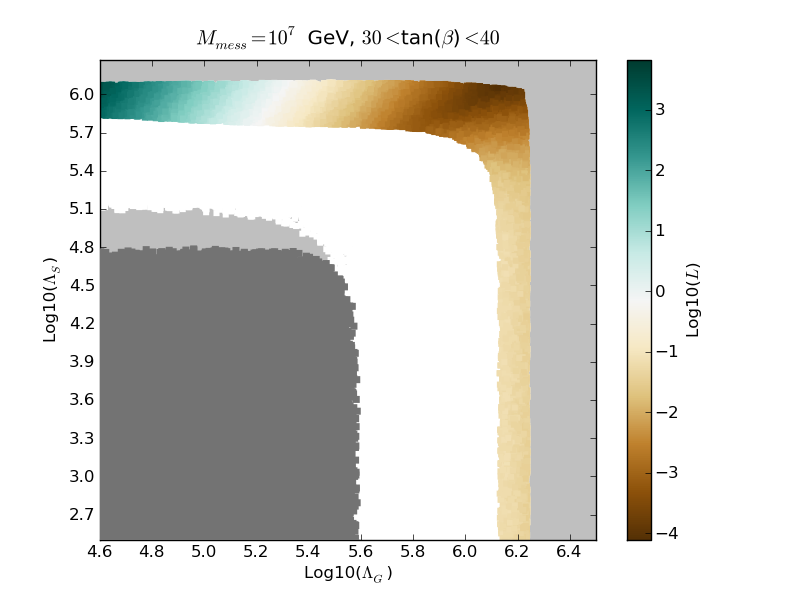}
}
\subfigure{
\includegraphics[width=7.5cm]{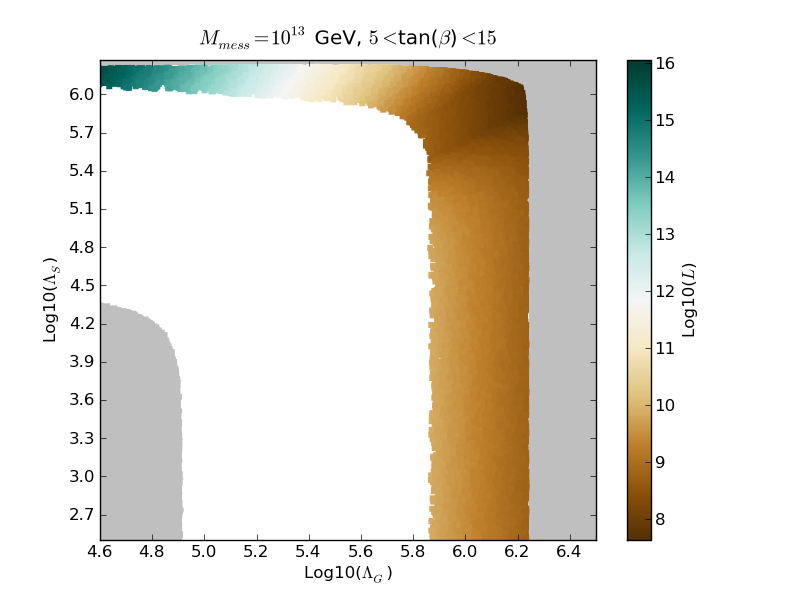}
}
\subfigure{
\includegraphics[width=7.5cm]{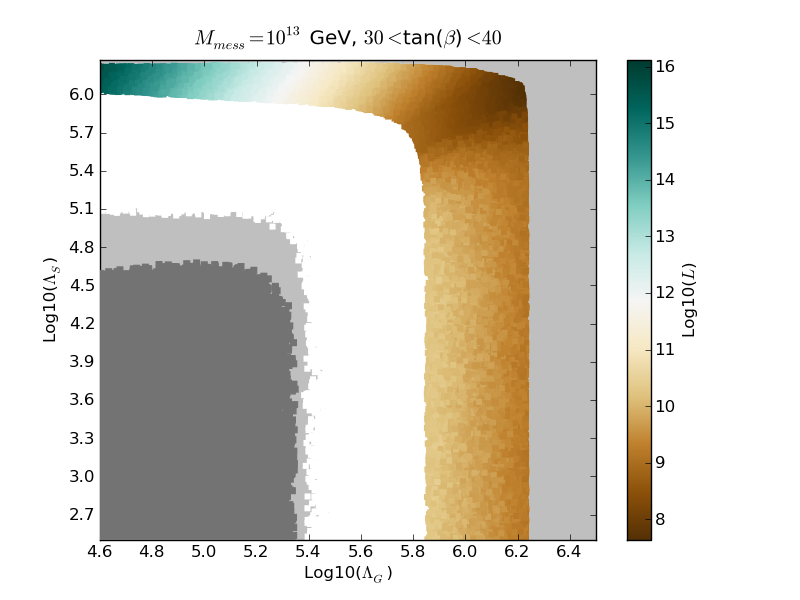}
}
\caption{
\label{CGGMe}\footnotesize
Logarithmic plot in the $\Lambda_G, \Lambda_S$ plane. 
The gradient indicates the decay length of the two body NLSP decay into 
gravitino plus NLSP partner.
Different regions have different NLSP type, and the decay length is computed accordingly.
}
\end{center}
\end{figure}

As discussed in section 2, the 
LSP in gauge mediation is always the gravitino, and the
NLSP is unstable and decays to its superpartner plus gravitino. This decay can happen inside or outside the detector, 
characterizing the collider signatures, and it is hence
essential for comparison with LHC direct searches. 
In figure \ref{CGGMe} we show the decay length of the NLSP in meters, 
computed using equation (\ref{decaylength}). 
For large messenger mass $M_{\text{mess}}=10^{13}$ GeV, the NLSP always 
decays outside the detector, while for small messenger mass $M_{\text{mess}}=10^{7}$ GeV,
the NLSP can decay inside or outside the detector depending on the value
of the UV parameters $\Lambda_G,\Lambda_S$. As we can see from the plots in Figure \ref{CGGMe}, in the short running case with $M_{\text{mess}}=10^7\text{ GeV}$ the stau is always promptly decaying, while the neutralino decay can be prompt for $m_{\tilde{N}_{1}}\gtrsim 500 \text{ GeV}$ or displaced for $m_{\tilde{N}_{1}}\lesssim 500 \text{ GeV}$. 

In general, LHC direct searches can drastically reduce the allowed parameter region, since the bound from direct searches summarized in Table 1 are very mild. In principle a dedicated analysis is needed in order to show the effects of LHC searches on the parameter space.
However, after imposing the Higgs mass constraint, the spectra that we get in CGGM parameter space are so simplified that we can already make some qualitative statement about the possible consequences of direct searches constraints. 

For short running, in the gaugino mediation region with stau NLSP, every colored particle is heavier than 5 TeV because of the Higgs bound, and hence decoupled from the IR dynamics. Moreover, the Wino mass is heavier than 2 TeV and the stau always around 500 GeV. Hence, both colored and electroweak production are very suppressed and this region will be very difficult to probe at LHC even at 14 TeV with $\mathcal{O}(100/\text{fb})$ \cite{Kats:2011qh}. In the long running case the stau decay length is around $10^9\ m$ which would imply a life-time well below $10^4 \text{ s}$. The stau and the gauginos get slightly lighter increasing $M_{\text{mess}}$ but still the colored and electroweak production at LHC are very suppressed by the fact that $m_{\tilde{g}}\gtrsim4 \text{ TeV}$, $m_{\tilde{C_{1}}}\gtrsim1.5 \text{ TeV}$ and $m_{\tilde{\tau}}\gtrsim460 \text{ GeV}$. 

Conversely, in the gaugino screening region for $\Lambda_G\lesssim0.1\Lambda_S $, we have a potentially interesting region for collider signatures in which $m_{\tilde{g}}\lesssim2\ \text{TeV}$, $m_{\tilde{C_{1}}}\lesssim600\ \text{GeV}$ and we get the Bino NLSP with $m_{\tilde{N}_{1}}\lesssim400\ \text{GeV}$ and the neutral Wino NNLSP . All the scalar spectrum is at the multi-TeV scale making this scenario very close to Split SUSY spectra with minimal splitting \cite{Wells:2003tf, ArkaniHamed:2012gw, Arvanitaki:2012ps}.
 
In the long running case with $M_{\text{mess}}=10^{13}$ we can have a very light Bino NLSP with $m_{\tilde{N}_1}\lesssim100\text{ GeV}$ and a decay length of  $10^{16}\ m$ which is already ruled out by nucleosynthesis bounds on $^7\! \text{Li}$ overproduction \cite{Gherghetta:1998tq}. This problem can be easily solved by reducing the value of $M_{\text{mess}}$. Taking $M_{\text{mess}}=10^7$ GeV, we have, automatically, a simplified spectrum for neutralino NLSP which can be produced via gluino decay or neutralino NNLSP/ chargino decay. 
The neutralino NLSP decays into a photon and a gravitino with a decay length larger than $\mathcal{O}(1) \text{ m}$. The displaced decay of the neutralino can lead to particularly interesting signatures with high transverse momentum isolated photon not be pointing to the interaction point which were studied in \cite{Kawagoe:2003jv,Meade:2010ji, Park:2011vw}. Recent CMS searches \cite{Chatrchyan:2012vxa} restrict the neutralino mass to values $m_{\tilde{N}_1}\gtrsim200\text{ GeV}$ for decay length around the meter possibly reducing the allowed region for $M_{\text{mess}}=10^7\text{ GeV}$.    

Lowering $M_{\text{mess}}$ down to its lower bound it is possible to obtain a region in which the neutralino NLSP is light and promptly decaying. In this case the allowed region can be further reduced by considering the LHC bounds on neutralino NLSP scenario given by CMS searches of $\gamma\gamma+\text{jet}+\text{MET}$ and $\gamma+\text{jets}+\text{MET}$ \cite{:2012mx,CMS-PAS-SUS-12-018}, ATLAS $\gamma\gamma+\text{MET}$ \cite{:2012afa}. The analysis performed in \cite{Kats:2011qh} based on $1/\text{fb}$ data already pushes the gluino mass at around 900 GeV for a neutralino lighter than 900 GeV and the Wino mass around 400 GeV for a neutralino lighter than 350 GeV.

\section{Extra Higgs soft terms}\label{2.2}
In the previous cases we have followed the usual strategy of considering $\tan \beta$
as a free parameter, obtaining the values of $\mu$ and $B_{\mu}$ by demanding 
EWSB with the correct gauge boson masses.
The soft parameters $m^2_{H_{u}}$ and $m^2_{H_{d}}$ were fixed to their gauge mediation contribution which coincides with the slepton doublet soft masses (\ref{higgsA}).

However, as discussed in the introduction, 
in order to explain the generation of $\mu$ and $B_{\mu}$ in gauge mediation, we should typically 
admit direct couplings of the Higgses of the MSSM with hidden sector operators.
From a model building perspective, the kind of extra-interactions that we want to take into account can be parametrized as 
\begin{equation}
W=\int d^2\theta(\lambda_{u}H_{u}\mathcal{O}_d+\lambda_{d}H_{d}\mathcal{O}_u)\ .\label{extraHiggs}
\end{equation}
These couplings always generate extra contributions to the Higgs soft masses $\delta m^2_{u,d}$ at 1-loop\footnote{We ignore the possibility of having extra interactions of the form $W=\int d^2\theta\lambda^2_{s}\mathcal{O}_{s}H_{u}H_{d}$. This kind of interactions are renormalizable only if $O_{s}$ is a fundamental field, leading to NMSSM-like scenarios which are out of our setup.}
\begin{equation}
\delta m^2_{u,d}=\frac{\vert\lambda_{u,d}\vert^2}{(4\pi)^2}\Lambda_{H_{d,u}}^2\ ,
\end{equation}
where the scales $\Lambda_{H_{d,u}}$ parametrize the contributions coming from the two-point function of the F-component of the hidden sector operators $O_{d,u}$ \cite{Komargodski:2008ax,DeSimone:2011va}.   
We have modified SoftSUSY 3.3.4 to accept these two extra UV parameters 
setting the new contributions
to the Higgs soft masses
at the messenger scale like in \eqref{NHmasses}. A priori, the mediation scale for these SUSY breaking 
effects would be unrelated with the one characterizing the GGM contributions but we assume them equal for simplicity.

If we want to preserve the calculability of the GGM framework we should impose a perturbativity bound on $\lambda_{u,d}$ which can be estimated in terms of our UV-parameters as
\begin{equation}
\vert\lambda_{u,d}\vert^2=\frac{k_{1}g_{1}^4(M_{\text{mess}})+3g_{2}^4(M_{\text{mess}})}{2(4\pi)^2}\left(\frac{\Lambda_S^2}{\Lambda_{H_{d,u}}^2}\right)\left(\frac{\delta m^2_{u,d}(M_{\text{mess}})}{m_{\tilde{E}_L}^2(M_{\text{mess}})}\right)\lesssim 1\ .
\end{equation} 
Assuming $\Lambda_{S}\simeq\Lambda_{H_{d,u}}$ and parametrizing with $c_{H}$ the $\mathcal{O}(1)$ numerical factor coming from the estimate of the numerator in the two-loop factor we get 
\begin{equation}
\delta m^2_{u,d}(M_{\text{mess}})\lesssim c_{H}(4\pi)^2{m_{\tilde{E}_L}^2(M_{\text{mess}})}\label{pertdelta}\ .
\end{equation}
We implement the perturbativity bound in our simulations by replacing $m_{\tilde{E}_L}^2(M_{\text{mess}})$ with $m_{\tilde{E}_L}^2(Q)$, where $Q\simeq M_{S}$ is the renormalization scale above $m_{Z}$ implemented in SoftSUSY 3.3.4 \cite{Allanach:2001kg}. Replacing $m_{\tilde{E}_L}^2(M_{\text{mess}})$ with $m_{\tilde{E}_L}^2(Q)$ can be a very bad estimate of the mass value of the left-handed sleptons, especially in the gaugino mediation region where the slepton masses receive large positive contributions from EW-gauginos.
Moreover, we are neglecting the RG evolution of the Yukawa couplings $\lambda_{u,d}$
which typically depends also on the details of the hidden sector dynamics.
%
 In order to take into account all these issues we will discuss the dependence of our results on the choice of the coefficient $c_{H}$ that we let vary from 4 to 1/100. To conclude this discussion we want to emphasize that the assumption of perturbativity for the extra-interactions \eqref{extraHiggs} is not strictly necessary from the model building perspective and several models which admit non-perturbative effects in the Higgs sector has already been proposed in the literature \cite{Csaki:2008sr, Nomura:2004zs}. We will see in the following how relaxing the hypothesis of perturbativity opens up new interesting regions of the GGM parameter space that might deserve further investigations.

The main focus of our study would be to understand which are the possible effects of the extra Higgs soft masses  on the low-energy spectrum and to show what are the possible viable spectra with an Higgs around 125 GeV. 

In doing that, we assume a complete GUT structure for the hidden sector, restricting our attention on regions of the parameter space in which $\delta m^2_{u}$ and/or $\delta m^2_{d}$ are dominant or of the same order compared to the standard gauge mediation contributions. Unlike the case of MPV $D_{Y}$-term contributions that will be discussed in section \ref{2.3}, here the GUT hypothesis is not crucial (even if theoretically appealing) but it allows us to show in a clear way the effects of having large Higgs soft masses which are leading to spectra with very distinctive features with respect to the standard CGGM ones.

In the presence of extra Higgs soft masses, the EWSB condition has to be reconsidered as suggested in \cite{Csaki:2008sr, DeSimone:2011va}. Since the Higgs mass constraint \eqref{Higgsmass} is imposing a lower bound on $\tan\beta$ around 5 and also requiring a large value of the stop mass, we can always expand \eqref{min1} for large $\tan\beta$ and explicitly write $\Sigma_u$ as the sum of the negative top-Yukawa contribution to the up Higgs mass and the positive gaugino-mediation contributions proportional to the chargino and the neutralino mass squared that we indicate as $K_{i}(M_{\tilde{\lambda}_{i}})$ with $i=1,2$ following the notation of \cite{Martin:1997ns}. Considering that $m^2_{Z}\ll\vert\Sigma_{u}\vert$ and neglecting terms of order $\mathcal{O}\left(\frac{\Sigma_u-\Sigma_d}{\tan^2\beta}\right)$ we get
\begin{equation}
\vert\mu\vert^2\simeq -(m^2_{H_{u}}+\sum_{i=1}^{2}K_{i}(M_{\tilde{\lambda}_{i}}))+\frac{3y_{t}^2}{4\pi^2}m^2_{\tilde{t}}\log\left(\frac{M_{\text{mess}}}{m_{\tilde{t}}}\right)-\delta m^2_{u}-\frac{\delta m^2_u- \delta m^2_d}{\tan^2\beta}\ . \label{NEWSB}
\end{equation}
 
In section \ref{Hcase1} we will see that a particularly interesting spectrum can be obtained when $\delta m^2_u$  is large and positive because accidental cancellations in \eqref{NEWSB} may lead to an exceptionally small value of $\mu$. It is clear from \eqref{NEWSB} that $\delta m^2_{d}$ would play a minor role in this mechanism, since its leading contribution is suppressed by a factor of $1/\tan^2\beta$ which is roughly of order $\mathcal{O}(1/100)$, because of the large value of $\tan\beta$.
Having a small $\mu$ would mostly modify the hierarchy in the gaugino sector allowing for regions of the parameter space with a light Higgsino NLSP first obtained in \cite{Csaki:2008sr, DeSimone:2011va}.
An upper bound on the magnitude of $\delta m^2_{u}>0$ can be derived from the EWSB condition \eqref{NEWSB} 
\begin{equation}
\delta m^2_{u}\lesssim  \frac{3y_{t}^2}{4\pi^2}m^2_{\tilde{t}}\log\left(\frac{M_{\text{mess}}}{m_{\tilde{t}}}\right)-(m^2_{H_{u}}+\sum_{i=1}^{2}K_{i}(M_{\tilde{\lambda}_{i}}))\ ,
\end{equation} 
since larger value of $\delta m^2_u$ would destabilize the EWSB vacuum in the MSSM. 

The two major effects of the extra Higgs soft masses in the sfermion sector are due to the violation of the $\Tr(Ym^2)\simeq0$ sum rule and to the enhancement of the Yukawa contributions to squark and slepton masses, as we will discuss in the following.
The most important phenomenological consequences of both these extra-terms would be visible in the uncolored sector, since all the squarks would be generically heavy in all the parameter space because of the Higgs mass constraint.

The Higgs extra-couplings violate the GGM sum rule already at 1-loop in the coupling constants $\lambda_{u,d}$ defined in \eqref{extraHiggs} and we have $\Tr(Ym^2)=S$ with $S=\delta m^2_{u}- \delta m^2_{d}$.  As a consequence, we get an additional $D_{Y}$ contribution to the RG flow equations of sfermion masses at 1-loop in the gauge couplings: 
\begin{equation}
\Delta\left(\frac{d}{dt} m_{\tilde{f}}^2\right)=2k_{1}\frac{Y_{\tilde{f}}g_{1}^2}{(4\pi)^2}(\delta m^2_u-\delta m^2_d)\ .
\label{Seq}\end{equation}
For $S<0$ the left-handed sleptons are driven lighter than the right-handed ones opening the possibility of having the sneutrino NLSP, while for $S>0$ we can get lighter right-handed sleptons with respect to the CGGM case.

Extra Higgs soft masses at the messenger scale would also generate a non standard shift in the Yukawa contributions of squark and sleptons that may revert the usual hierarchies between families. This contribution would be particularly important for the third generation sleptons which have sufficiently sizeable Yukawa couplings and are not constrained to be as heavy as the squarks. In particular 
\begin{equation}
\Delta X_{\tau}=2\vert y_{\tau}\vert^2 \delta m^2_{d}
\end{equation}
induces extra 1-loop contributions to the running of the third generation sleptons masses:
\begin{equation}
\Delta^{\prime}\left(\frac{d}{dt} m_{\tilde{l}_{L}^{3}}^2\right)=\frac{1}{(4\pi)^2}\Delta X_{\tau}\ ,\ \Delta^{\prime}\left(\frac{d}{dt} m_{\tilde{e}_{R}^{3}}^2\right)=\frac{2}{(4\pi)^2}\Delta X_{\tau}\ . \label{dYukawa}
\end{equation}
These contributions depend purely on $\delta m^2_d$ which, therefore, will play a major role in determining the physics of the sleptonic spectrum. For $\Delta X_{\tau}>0$ the third generation sleptons are driven lighter than the other ones, whereas for $\Delta X_{\tau}<0$ we can realize an inverted hierarchy in the sleptonic sector, where the third generation slepton masses are driven larger than the first two generations. 

Moreover, since the $\delta m^2_d$ term in the EWSB condition \eqref{NEWSB} is suppressed in the large $\tan\beta$ regime, it affects the EWSB condition less than $\delta m^2_u$ and for this reason we found that a favorable situation to maximize the effects on the slepton sector is to take $\vert\delta m^2_d\vert\gg\vert\delta m^2_u\vert$. In section \ref{Hcase2} we will consider $\delta m^2_{d}>0$ which implies $S<0$ and opens the possibility of having sneutrino NLSP with all the left-handed sleptons lighter than the right-handed ones. 

In section \ref{Hcase3} we will take $\delta m^2_d<0$ which implies $S>0$ and $\Delta X_{\tau}<0$. When the $\delta m^2_d$ contribution dominates over the standard gauge mediation ones, we are allowing for tachyonic masses for $H_{d}$ at the messenger scale. This choice would induce two striking phenomenological features in the IR, allowing for regions with selectron NLSP in gauge mediation and lowering the mass of the CP-odd and CP-even heavy scalars $A^0$, $H^0$, $H^{\pm}$. 
This second feature can be explained remembering that, in the large $\tan\beta$ regime, the equation \eqref{min2} 
implies
\begin{equation}
m^2_{A^{0}}\simeq 2(\vert\mu\vert^2+m^2_{H_{u}}+\sum_{i=1}^{2}K_{i}(M_{\tilde{\lambda}_{i}}))-\left(\frac{3y_{t}^2}{4\pi^2}m^2_{\tilde{t}}\log\left(\frac{M_{\text{mess}}}{m_{\tilde{t}}}\right)+\vert\delta m^2_{d}\vert\right)\gtrsim0\ , \label{lightA}
\end{equation}
where we have already taken into account that the dominant top-Yukawa contribution in $\Sigma_u$ and the value of $\delta m^2_d$ are negative. Clearly, a sizeable negative value for $\delta m^2_d$ would induce lighter masses for the heavy scalars and, consequently, an upper bound on $\vert\delta m^2_d\vert$ can be derived in this case by requiring non-tachyonic masses for $A^0$.

The case of tachyonic $\delta m^2_{u}$ has not shown any particularly interesting phenomenological features and it will not be discussed. Moreover, we have to mention that allowing for spectra with tachyonic Higgs soft masses and light sleptons would often imply the existence of unbounded from below (UFB) directions or charge-and-color-breaking (CCB) minima  
which may eventually destabilize the usual EWSB minimum \cite{Casas:1995pd, Evans:2008zx}. We leave a more detailed study of this issue in the context of gauge mediation for future investigations,  assuming that the $\delta m^2_{d}<0$ case is not affected by these problems and that the usual EWSB vacuum is at least metastable and long-lived compared to the age of the Universe \cite{Ellis:2008mc} and also that the presence of CCB vacua does not affect the cosmological history \cite{Carena:2008vj}.
  
As a final remark we notice that all the new features in the leptonic sector which are driven by the extra Higgs soft terms arise in the gaugino mediation region in which the gauge mediation contribution to the sfermion masses at the messenger scale are negligible. Similar effects on the MSSM spectrum were discussed in \cite{Buchmuller:2005ma,Evans:2006sj,Buchmuller:2006nx} motivated by extra-dimensional realizations of the gaugino-mediation mechanism. Here, we are showing that this kind of spectra are eventually realized in the gauge mediation framework in terms of a 2+1+3 dimensional parameter space. An interesting question would be to engineer calculable UV completions which can span over this enlarged parameter space.
Conversely, in the gaugino screening region, $\Lambda_S$ is restricted to be very large in order to achieve a sufficiently large stop mass to satisfy the Higgs mass constraint and, since we are assuming sfermion mass unification, this would imply that also the slepton masses will be always decoupled from the IR-physics very much like in the CGGM case.

\subsection{Large and positive $\delta m^2_u$: Higgsino NLSP}\label{Hcase1}
We consider the case in which $\delta m^2_u$ is large and positive in order to obtain accidental cancellations in 
\eqref{NEWSB} which would induce a very small value of $\mu$. Scanning the whole parameter space, we find a very narrow interval of values for $\delta m^2_u$ for which $\mu$ is below the TeV scale. This range of values satisfies the perturbativity bound \eqref{pertdelta} for a wide range of choices for $c_{H}$ which can go from 4 to $1/5$. 

In agreement with the fact that $\delta m^2_d$ contributions in \eqref{NEWSB} are suppressed for large $\tan\beta$, we do not find any correlation between the value of $\mu$ 
and  $\delta m^2_d$ which can be either negative or positive, varying from $-10^6 \text{ GeV}^2$ up to $10^6 \text{ GeV}^2$. 

In order to maximize the effect of accidental cancellations without caring about modification of the slepton sector we take $\delta m^2_u=\delta m^2_d=6.3\times 10^6 \text{ GeV}^2$ so that $S=0$ and we do not have extra $D_{Y}$ tadpoles \eqref{Seq} in the MSSM RG-flow equations. The only remaining effect would be the enhancement of the Yukawa contributions to the third generation sleptons \eqref{dYukawa} which make the stau lighter with respect to the CGGM case. 

\begin{figure}[ht]
\begin{center}
\subfigure{
\includegraphics[width=7.5cm]{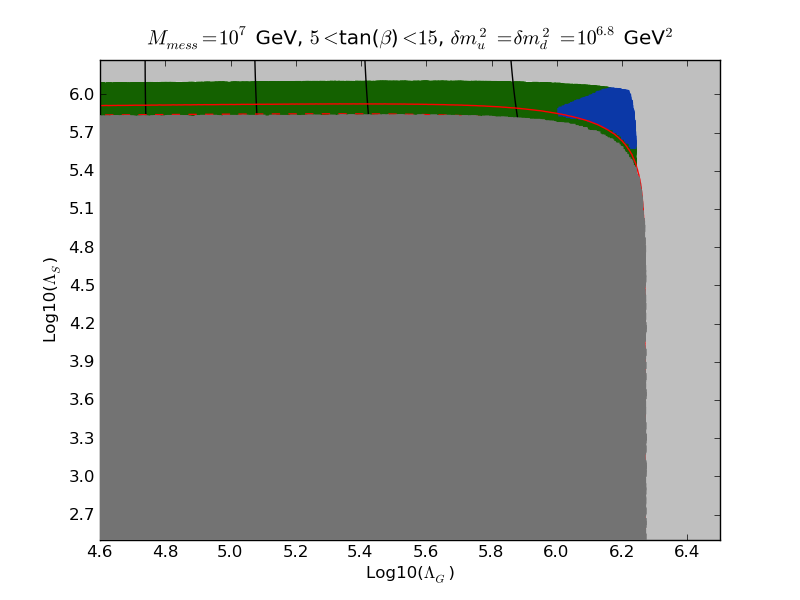}
}
\subfigure{
\includegraphics[width=7.5cm]{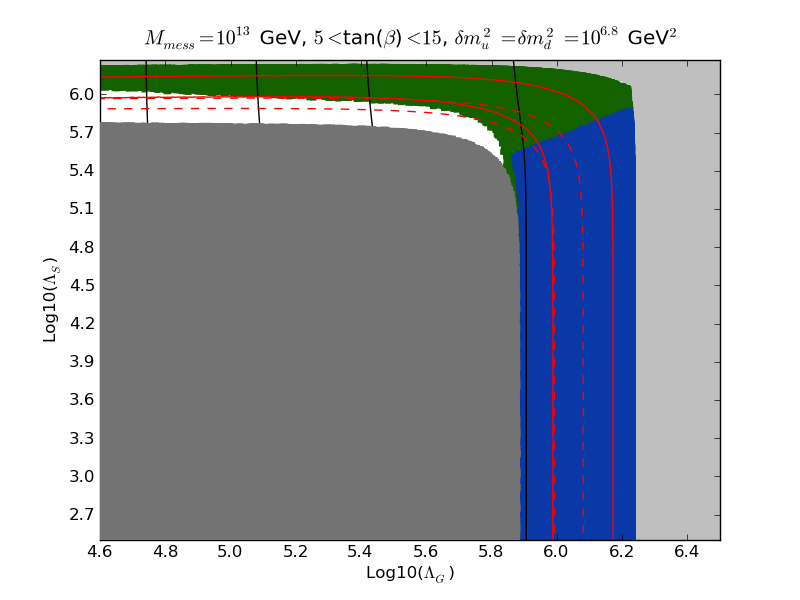}
}
\subfigure{
\includegraphics[width=7.5cm]{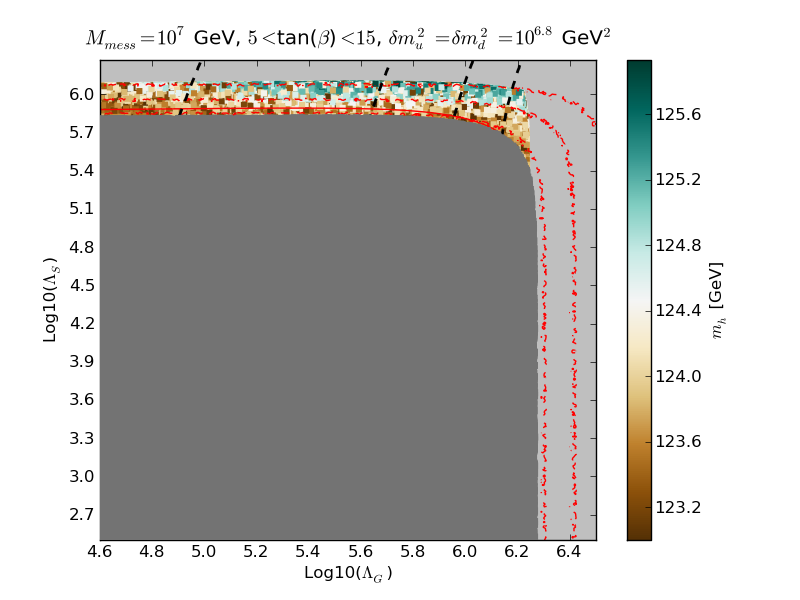}
}
\subfigure{
\includegraphics[width=7.5cm]{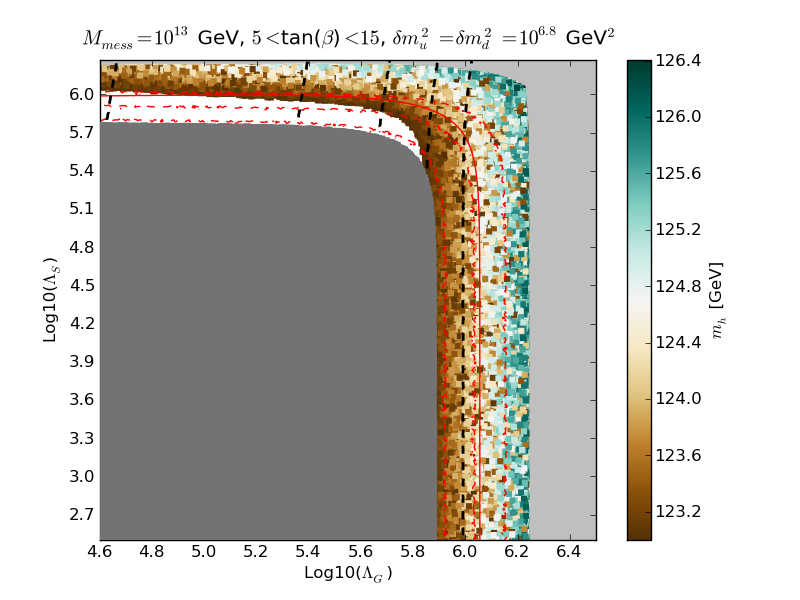}
}
\caption{
\label{udpos1}\footnotesize
\emph{First row}: Logarithmic plot in the $\Lambda_G, \Lambda_S$ plane. Explanations of the colors is 
in the text. The black, red, dashed-red contour plots identify the gluino, lightest stop, first generation masses
respectively. The scales of the contours for the gluino are $500$ GeV, $(1,2,5)$ TeV; for the stop are $6$ TeV on the left and also $4$ TeV on the right; for the first generation squarks are $6$ TeV and also $5$ TeV on the right.
\emph{Second row}: Logarithmic plot in the $\Lambda_G, \Lambda_S$ plane. The gradient indicates the Higgs mass on the allowed region. The red, dashed red and dashed black contours 
indicates $M_S ,\mu$ and $A_t$ respectively. The scales of the contours for $M_S$ are $6$ TeV on the left and $5$ TeV on the right; for $\mu$ are $(1,2,3)$ TeV; for $A_t$ are $(-0.2,-1,-2,-3)$ TeV on the left and also $-4$ TeV on the right.}
\end{center}
\end{figure}

In Figure \ref{udpos1} we restrict our analysis to the case of $\tan\beta= 10 \pm 5$, where the effects induced by Yukawa couplings are suppressed, and we show the plots for fixed value of $M_{\text{mess}} = 10^7, 10^{13}\text{ GeV}$. 
In the dark grey regions, where SoftSUSY did not converge, the equation \eqref{NEWSB} does not admit solution with positive $\vert\mu\vert^2$ because the $\delta m^2_u$ contribution becomes too large compared to the other ones, destabilizing the EWSB vacuum. In the short running case, this effect is completely cutting out the gaugino mediation region since the stop contribution is not sufficiently large to compensate both the effects of $\delta m^2_u$ and $\sum_{i=1}^{2}K_{i}(M_{\tilde{\lambda}_{i}})$.

From the plots in the second row of Figure \ref{udpos1} we see that the value of $\mu$ drastically decreases with respect to the CGGM case. The most interesting region in this scenario is the small band very close to the dark grey region, where $\mu\lesssim1\text{ TeV}$ and the minimal value of $\mu$ can approach 100 GeV in both the long running and the short running case. In the short running region, the band with very small $\mu$ has always a neutralino as NLSP (indicated in green) and it extends from the region where $\Lambda_G\simeq\Lambda_S$ on the right up to the gaugino screening region where $\Lambda_G\ll\Lambda_S$. Conversely, in the long running case we can have small $\mu$ only in the gaugino mediation region where the stau is typically the NLSP (indicated in blue).

Another relevant effect of having large and positive extra contributions to the Higgs masses both from $\delta m^2_u$ and $\delta m^2_d$ is that the mass of the pseudoscalar $A^0$ increases considerably with respect to CGGM as was noticed in \cite{DeSimone:2011va} being always more than 3 TeV in our case.

\begin{figure}[ht]
\begin{center}
\subfigure{
\includegraphics[width=7.5cm]{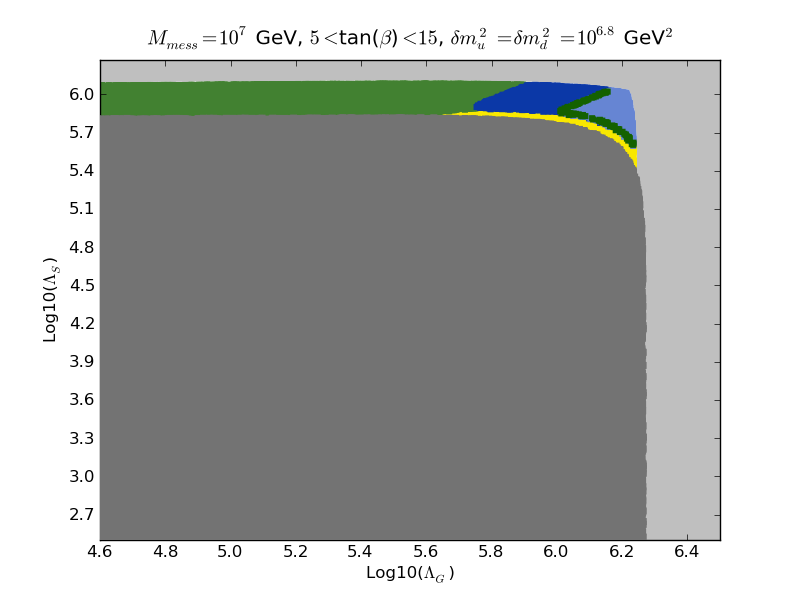}
}
\subfigure{
\includegraphics[width=7.5cm]{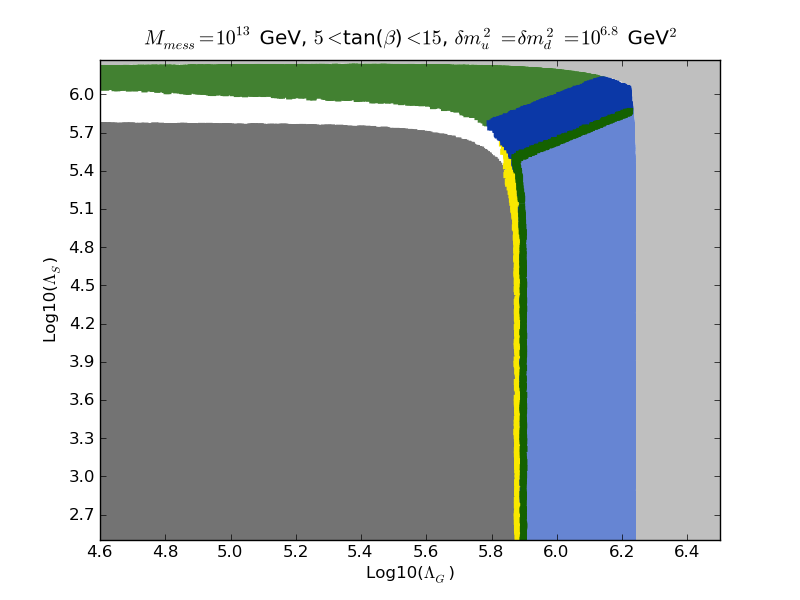}
}
\caption{\footnotesize Logarithmic plot for the NNLSP in the $\Lambda_G, \Lambda_S$ plane. 
NNLSP colors are light green for the second lightest neutralino, blue for the stau, green for the lightest neutralino pale blue for the smuon and yellow for the lightest chargino.
}
\label{udpos2}\footnotesize
\end{center}
\end{figure}
In Figure \ref{udpos2} we see that the NNLSP species and the features of the gaugino spectrum are very much dependent on how much $\mu$ can be small compared to the electroweak gaugino masses $M_{\tilde{\lambda}_{1,2}}$, which are substantially determined by $\Lambda_G$ in gauge mediation \eqref{gmass}. 
Using the formulas for the neutralino and chargino mass eigenvalues derived in \cite{Martin:1993ft}, we see that for $\mu\ll M_{\tilde{\lambda}_{1,2}}$ we get three almost degenerate neutralino/charginos with masses proportional to $\mu$ at the bottom of the spectrum. In particular we have $\vert m_{\tilde{N}_{1}}\vert\simeq\vert m_{\tilde{N}_{2}}\vert\simeq \vert m_{\tilde{C}_{1}}\vert\simeq\mu$ in the limit in which we neglect all the effects coming from EWSB. 
The EWSB corrections can always be treated as a perturbation in most of the allowed region, since $M_{\tilde{\lambda}_{1,2}}$ are sufficiently large. Hence, taking into account EWSB will not change much the gaugino spectrum for $\mu\ll M_{\tilde{\lambda}_{1,2}}$ which is characterized by an neutral Higgsino NLSP with the other neutral and charged Higgsino almost degenerate NNLSP and the Bino and the Winos decoupled from the IR physics. 

In the long running case, this kind of spectrum is a generic feature of the small $\mu$ region since accidental cancellations occur for large values of $\Lambda_G$. The typical spectrum in the small $\mu$ region has stau NLSP around 600 GeV with Higgsino NNLSP almost degenerate which corresponds to the green band in the gaugino mediation region of Figure \ref{udpos2}. Lowering $\Lambda_G$, we reach a region at the boundary of SoftSUSY convergence where the Higgsino becomes the NLSP as can be seen from the tiny green band in the first row of Figure \ref{udpos1} which corresponds to a yellow band in Figure \ref{udpos2} signaling a charged Higgsino NNLSP. In this region also the stau is very light. However, obtaining a good solution to the EWSB condition so close to the boundary of instability is very sensitive to the precise value of $\tan\beta$. 

In the short running case, we can have a region with small $\mu$ and small $\Lambda_G$ which has the standard gaugino screening spectrum obtained in the CGGM case: the Bino is NLSP and we have an almost degenerate neutral Wino NNLSP, a very light charged Wino and a quite light gluino. 
Moreover the small $\mu$ in this region pushes down the mass of 
both charged and neutral Higgsinos.
The presence of light Higgsinos in the spectrum could in principle modify the collider signatures of this scenario 
and make it distinguishable from the CGGM case.

Increasing $\Lambda_G$ up to $10^{5.8} \text{ GeV}$, we get $\mu\ll M_{\tilde{\lambda}_{1,2}}$ for both short and long running so that the Bino and the Winos are decoupled and the masses of neutral and charged Higgsinos can be arbitrarily light, varying quite rapidly from less than 100 GeV to 900 GeV in the interesting region. 
In this scenario all the colored sparticle are heavy and the gluino is heavier than 3 TeV (5 TeV for short running) as a consequence of the GUT-complete structure of the hidden sector. Consequently, the colored production channel is very much suppressed and our scenario cannot be constrained with the existing ATLAS and CMS searches for Z+jets+MET final states \cite{ATLAS-CONF-2012-152, Chatrchyan:2012qka}.

Because of the lightness of the Higgsinos, an interesting production channel for both long and short running might be the Drell-Yan production of pairs mediated by an electroweak gauge boson which was studied in \cite{DeSimone:2011va}. In the long running case also a light stau can eventually be produced in Drell-Yan processes enhancing the detectability of this scenario with respect to the short running case where the stau are always hevier than 700 GeV.     
 
In the end, even if the room for discovery at LHC is reduced and this scenario might appear very fine-tuned, this example remains the simplest situation which satisfies the Higgs mass constraint having a very small value of $\mu$. Having worked out the basic consequences of accidental cancellations in the EWSB condition will be very useful in the following, where these features will indeed appear quite generically in different contexts.

\subsection{Large and positive $\delta m^2_d$: Sneutrino co-NLSP}\label{Hcase2}
We consider the case in which $m^2_{u}=0$ and $m^2_{d}$ is large and positive. This would imply a sizeable negative contribution in the equation \eqref{Seq} which can make the left-handed sleptons lighter than the right-handed ones. 

Since the $\delta m^2_{d}$ contributions to the EWSB condition are suppressed by $\mathcal{O}(1/\tan^2\beta)$ the values of $\mu$ would remain unchanged with respect to the CGGM case being larger or equal to 2 TeV. 

\begin{figure}[ht]
\begin{center}
\subfigure{
\includegraphics[width=7.5cm]{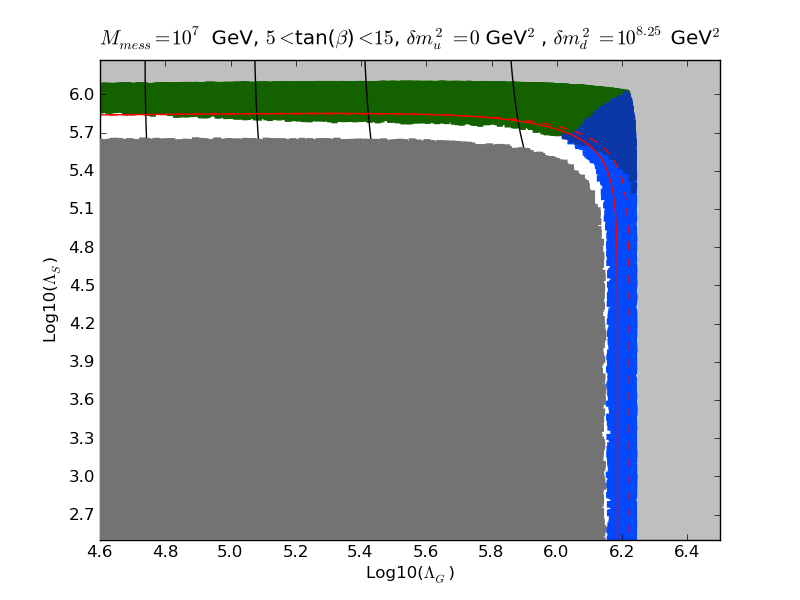}
}
\subfigure{
\includegraphics[width=7.5cm]{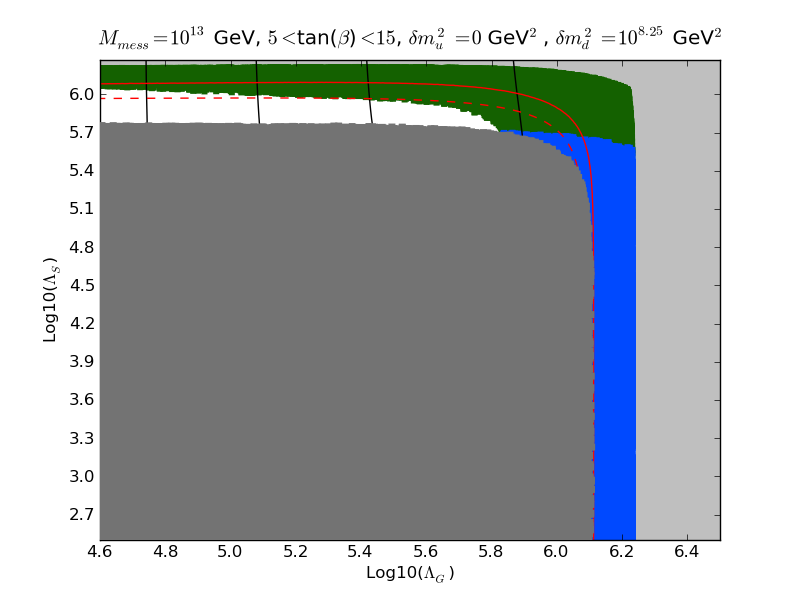}
}
\caption{
\label{dpos1}\footnotesize
Logarithmic plot in the $\Lambda_G, \Lambda_S$ plane. Explanations of the colors is 
in the text. 
The black, red, dashed-red contour plots identify the gluino, lightest stop, first generation masses
respectively.
The scales of the contours are $500$ GeV, $(1,2,5)$ TeV for the gluino, $5$ TeV for the stop and $6$ TeV for the first generation squarks.}
\end{center}
\end{figure}

In order to maximize the new effects on the spectrum and present our results in the usual $(\Lambda_{G}, \Lambda_{S})$ plots, we fix $\delta m^2_{d}\simeq1.8\times10^8 \text{ GeV}^2$. In Figure \ref{dpos1} we show the plots for fixed value of $M_{\text{mess}}=10^7,10^{13} \text{ GeV}$ and $\tan\beta=10\pm5$, restricting our analysis to the case of moderate values for $\tan\beta$ in which the mixing effects in the stau mass matrix are not the dominant ones.

First of all, it is important to mention that our choice of $\delta m^2_{d}$ is really at the border of the allowed parameter space and it may imply the presence of strongly coupled effects in the extra-interactions \eqref{extraHiggs}. In fact, we checked that most of the sneutrino NLSP points would disappear requiring the perturbativity bound \eqref{pertdelta} to be satisfied with $c_{H}\lesssim1$.

The dark grey regions where SoftSUSY did not converge get enlarged with respect to the CGGM case. In this region the sneutrino becomes tachyonic at the EW scale.
In the gaugino screening regime where $\Lambda_{G}\lesssim\Lambda_{S}$ we get the usual green region of neutralino NLSP. In this region all the scalars will be decoupled and the physics substantially unchanged with respect to the CGGM case. For this reason we will not discuss this region any further. 

Conversely, in the gaugino mediation regime the light blue region of sneutrino NLSP is replacing the typical region of stau NLSP of the CGGM case in almost all the parameter space. A small blue area with stau NLSP appears in the short running case in the MGM area in which $\Lambda_S$ gets larger and the usual gauge mediation contributions are the dominant ones.

The appearance of sneutrino NLSP is a generic consequence of having the left-handed sleptons lighter than the right-handed ones at low energy, since the mass splitting between the left-handed sleptons is a model independent feature of the MSSM \cite{Martin:1997ns} essentially proportional to the D-term contributions produced by the EWSB  
\begin{equation}
m^2_{\tilde{e}_{L}}-m^2_{\tilde{\nu}_{L}}\simeq\left(1-\frac{2}{\tan^2\beta}\right)m^2_{W}\ . \label{sneutrinoNLSP}
\end{equation} 

The possibility of realizing this scenario using extra Higgs soft masses in gauge mediation was previously investigated in \cite{DeSimone:2011va}. Here we are seeing that the Higgs mass constraint is pushing the region of sneutrino NLSP close to the non-perturbative regime for the extra Higgs coupling which is out from the hypothesis of \cite{DeSimone:2011va}. 
\begin{figure}[ht]
\begin{center}
\subfigure{
\includegraphics[width=7.5cm]{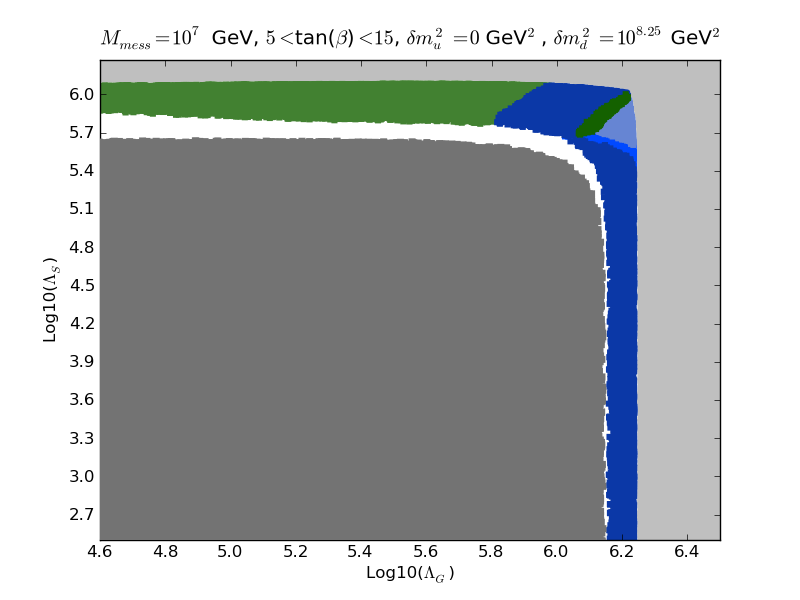}
}
\subfigure{
\includegraphics[width=7.5cm]{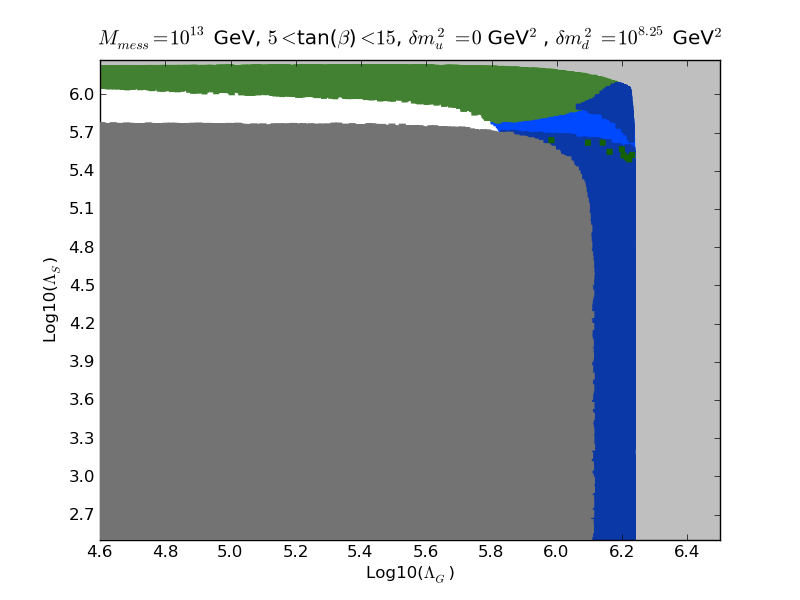}
}
\subfigure{
\includegraphics[width=7.5cm]{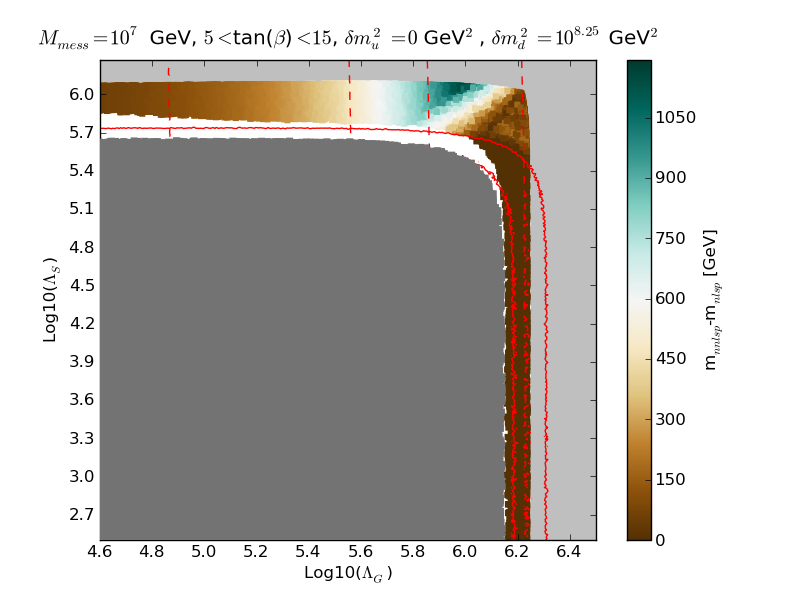}
}
\subfigure{
\includegraphics[width=7.5cm]{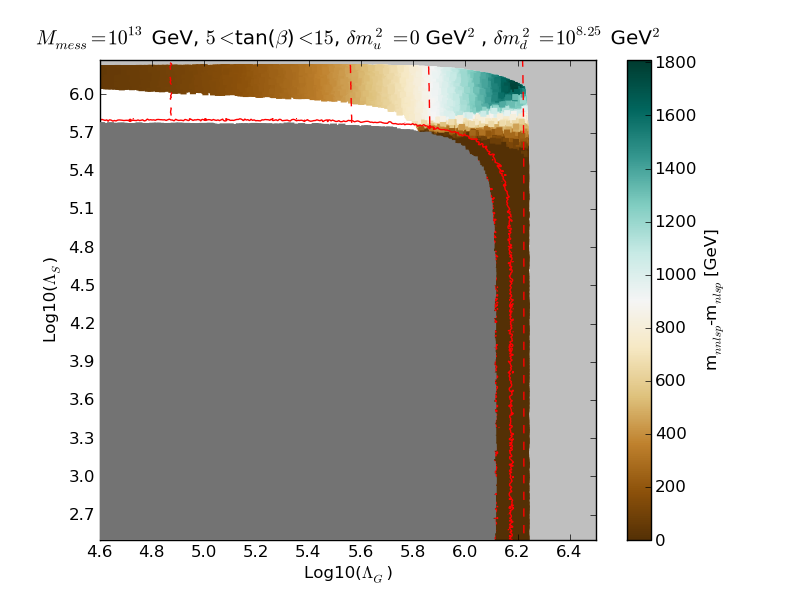}
}
\caption{
\label{dpos2}\footnotesize
\emph{First row}: Logarithmic plot for the NNLSP in the $\Lambda_G, \Lambda_S$ plane. 
NNLSP colors are light green for the second lightest neutralino, blue for the stau, green for the lightest neutralino, pale blue for the smuon and light blue for the sneutrino.
\emph{Second row}: Logarithmic plot in the $\Lambda_G, \Lambda_S$ plane. 
The gradient represents the mass difference $m_{\text{NNLSP}}-m_{\text{NLSP}}$.
The solid and dashed contours identify the NLSP masses, sneutrino and neutralino
respectively.
The scales of the contours are $(100,500)$ GeV and $(1,2)$ TeV
for the neutralino and for the sneutrino $900$ GeV and 1.5 TeV.
}
\end{center}
\end{figure}    

In the first row of Figure \ref{dpos2} we show the NNLSP species and in the second row we give an idea of the structure of the spectrum plotting the mass difference between NLSP and NNLSP and the contours for the two most common NLSP species which are the neutralino and the sneutrino. 

A very peculiar feature of this scenario is the large splitting between the third generation and the other two which is induced by large and positive extra-Yukawa contribution in the RG-flow of the sleptons \eqref{dYukawa}. As a consequence, the sneutrino NSLP is always the tau-sneutrino $\tilde{\nu}_{\tau}$ and the lightest of the left-handed sleptons is the stau $\tilde{\tau}_{1}$ which is lighter than the other left-handed slepton $\tilde{l}_{L}$ of around $200$ GeV in the short running case and between $150$ and $500$ GeV in the long running case.

Actually, the Yukawa contributions are enhanced so much that $\tilde{\tau}_{1}$ is driven lighter than the sneutrinos of the first two generations $\tilde{\nu}_{l}$. In the short running case $m_{\tilde{\nu}_{\tau}}-m_{\tilde{\tau}_{1}}\simeq -3.4 \text{ GeV}$ in all the gaugino mediation region, while the splitting $m_{\tilde{\nu}_{\tau}}-m_{\tilde{\nu}_{l}}$ decreases increasing $\Lambda_{G}$ and goes from $-200$ GeV to $-50$ GeV. In the long running case the splitting between the tau-sneutrino and the stau increases to $m_{\tilde{\nu}_{\tau}}-m_{\tilde{\tau}_{1}}\simeq -5.7 \text{ GeV}$, whereas the splitting $m_{\tilde{\nu}_{\tau}}-m_{\tilde{\nu}_{l}}$ goes from $-150$ to $-15$ increasing the value of $\Lambda_{G}$. 

The right-handed sleptons of the first two generation $m_{\tilde{l}_{R}}$ are split from $\tilde{\tau}_{1}$ by more than 600 GeV for short running and more than 1 TeV for long running, being completely decoupled from the IR physics.
The splitting can be slightly alleviated for the heavier stau mass eigenvalues $m_{\tilde{\tau}_{2}}$ which however is always decoupled as the other right-handed sleptons in the region of smaller $\Lambda_G$.

These spectra would be very interesting from the point of view of collider signature because the large splitting between left-handed and right-handed sleptons and the presence of a tau-sneutrino co-NLSP with the stau would favor the leptophilic signals at LHC described in \cite{Katz:2009qx} where the right-handed sleptons are assumed not to participate in the decay chain. 

The feature of having the third generation slepton doublet much lighter that the other two can be enhanced taking larger values for $\tan\beta$ and it is a generic consequence of having large and positive $\delta m^2_d$ which  enhances the Yukawa contributions \eqref{dYukawa}. This kind of spectra would favor multi-$\tau$ final states at colliders which have been studied extensively in the literature \cite{Covi:2007xj,Ellis:2008as,Medina:2009ey,Santoso:2009qa,Katz:2010xg}. 

Unluckily, gaugino and sfermion mass unification at $M_{\text{mess}}$ would imply relatively heavy sneutrino NLSP, typically around $600$ GeV in the short running case and $500$ GeV in the long running case. Moreover, the gaugino masses will be at the multi-TeV scale in the gaugino mediation region. Both these effects suppress the colored and the electroweak production of sleptons NLSP at LHC. However, we expect that these difficulties would be easy to overcome without changing the main physical features of this scenario in the full EGGM parameter space.  

\subsection{Large and negative $\delta m^2_d$: Selectron NLSP}\label{Hcase3}

We take into account the possibility of having $\delta m^2_{u}=0$ and $\delta m^2_{d}$ large and negative. As discussed at the beginning of this section, this choice would lower the right-handed slepton masses through $S>0$ contributions and, at the same time, it would allow for an inverted hierarchy among the sleptons by reversing the usual sign of the Yukawa contributions \eqref{dYukawa} to the third generation sleptons. 

The lightest slepton in the spectrum will be the one with the smallest Yukawa coupling and we obtain a region with selectron and smuon co-NLSP in the gaugino mediation region, the two being sufficiently split from the lightest stau eigenvalue which is driven larger by sizeable Yukawa contributions \eqref{dYukawa} controlled by $\delta m^2_{d}$. This effect is an unusual one in gauge mediation and it is enhanced in the large $\tan\beta$ regime.

\begin{figure}[ht]
\begin{center}
\subfigure{
\includegraphics[width=7.5cm]{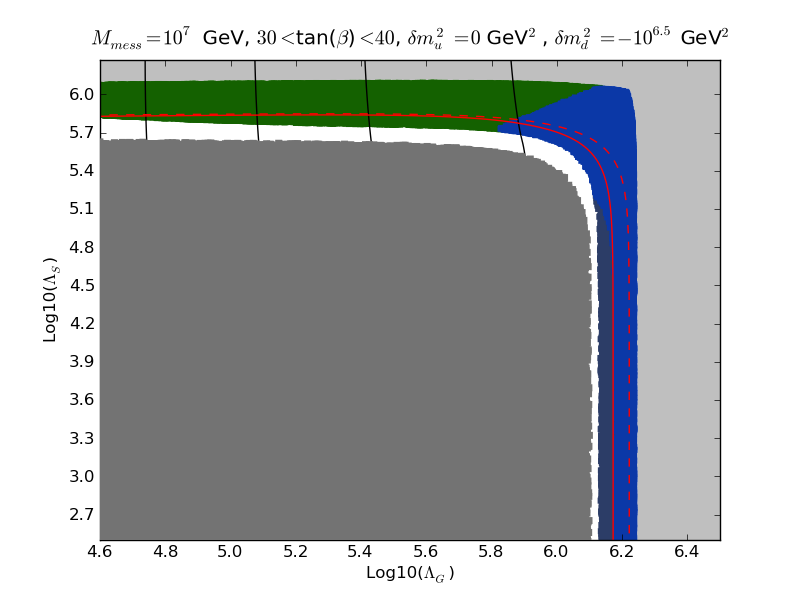}
}
\subfigure{
\includegraphics[width=7.5cm]{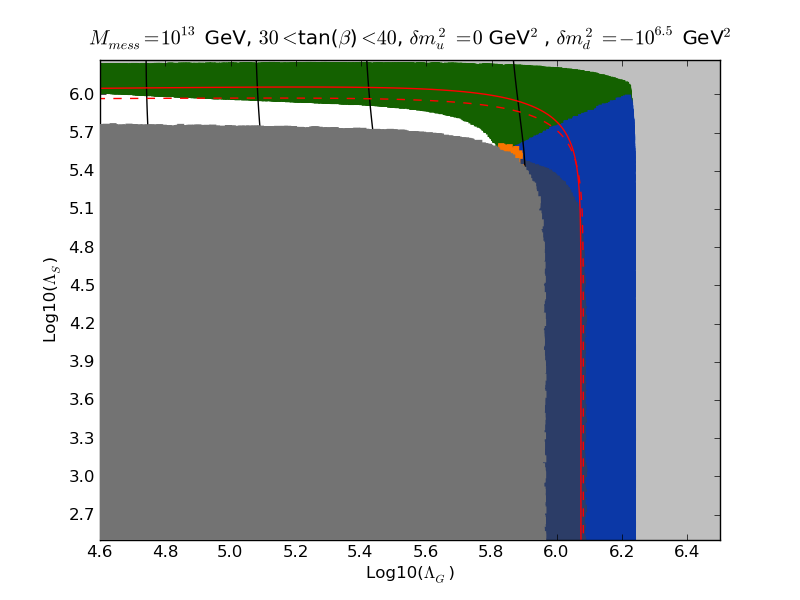}
}
\caption{
\label{dneg1}\footnotesize
Logarithmic plot in the $\Lambda_G, \Lambda_S$ plane. Explanations of the colors is 
in the text. 
The black, red, dashed-red contour plots identify the gluino, lightest stop, first generation masses
respectively.
The scales of the contours are $500$ GeV, $(1,2,5)$ TeV for the gluino, $5$ TeV for the stop and $6$ TeV for the first generation squarks.
}
\end{center}
\end{figure}

For this reason in Figure \ref{dneg1} we focus on large $\tan\beta$ scenarios fixing $\tan\beta=35\pm5$ for both $M_{\text{mess}}=10^7, 10^{13}$. We fix $\delta m^2_{d}= -3.2\times10^6\text{ GeV}^2$ in order to present the usual $(\Lambda_G, \Lambda_S)$ plots. 

Our choice of $\delta m^2_{d}$ satisfies the perturbativity bound \eqref{pertdelta} and, in fact, we checked that we can have regions of the parameter space with selectron NLSP in a wide range of choices for $c_{H}$ which can go from 4 to $1/5$. However, we are not aware of any UV-complete setup realizing this mechanism in the context of gauge mediation and it would be interesting to address this question in terms of weakly coupled realization of the hidden sector.     

The dark grey regions where SoftSUSY do not converge are due to the appearance of tachyonic masses for $A^{0}$, $H^0$ and $H^{\pm}$. The heavy scalars are generically driven lighter because of the tachyonic extra contribution to the down Higgs as can be inferred from \eqref{lightA}. This effect can have dramatic consequences on the MSSM, leading to spectra which do not satisfy the decoupling limit condition \cite{Gunion:2002zf} and, eventually, to the instability of the EWSB vacuum. Avoiding tachyonic masses for $A^0$ and $H^0$ give us an upper bound for $\vert\delta m^2_{d}\vert$ which results more constraining than the perturbativity bound  \eqref{pertdelta}. Our choice of $\vert\delta m^2_{d}\vert=3.2\times 10^6\text{ GeV}^2$ is at the boundary of the allowed region and it has been chosen in order to maximize the effects on the low energy slepton spectrum. 

Like in section \ref{Hcase2} the value of $\mu$ is substantially unchanged with respect to the CGGM case, confirming the observation that the effects of $\delta m^2_{d}$ on the EWSB condition \eqref{NEWSB} are always suppressed in the large $\tan\beta$ regime.  The squark spectrum is also unchanged with respect to the CGGM case and hence decoupled from the low energy spectrum in the whole parameter space. 

In the gaugino screening region we have the usual green region with neutralino mostly Bino NLSP. No new features arise in this region which would be phenomenologically indistinguishable from the corresponding CGGM region.
Conversely, in the gaugino mediation region we get a new dark blue region with selectron NLSP when the $\delta m^2_{d}$ is dominating over the usual gaugino mediation contributions. The heavy scalars are also quite light in this region being, however, always heavier than $800$ GeV in the short running case and $700$ GeV in the long running case and, hence, ensuring the validity of the decoupling limit condition \cite{Gunion:2002zf}. Increasing $\Lambda_S$, $A^0$ and $H^0$ can even become lighter than the NLSP at the border of the region of SoftSUSY convergence. In the long running case this effect becomes more clear and we have indicated this effect
with an orangish color. 
\begin{figure}[ht]
\begin{center}
\subfigure{
\includegraphics[width=7.5cm]{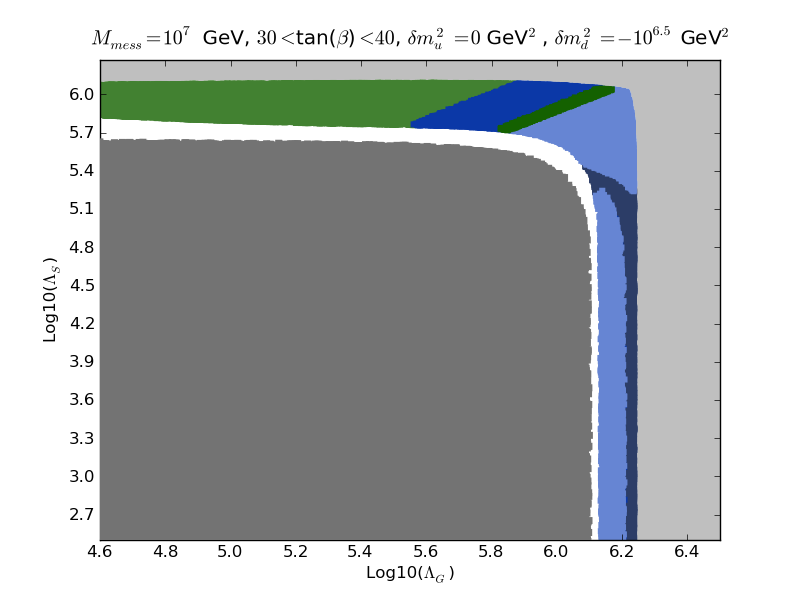}
}
\subfigure{
\includegraphics[width=7.5cm]{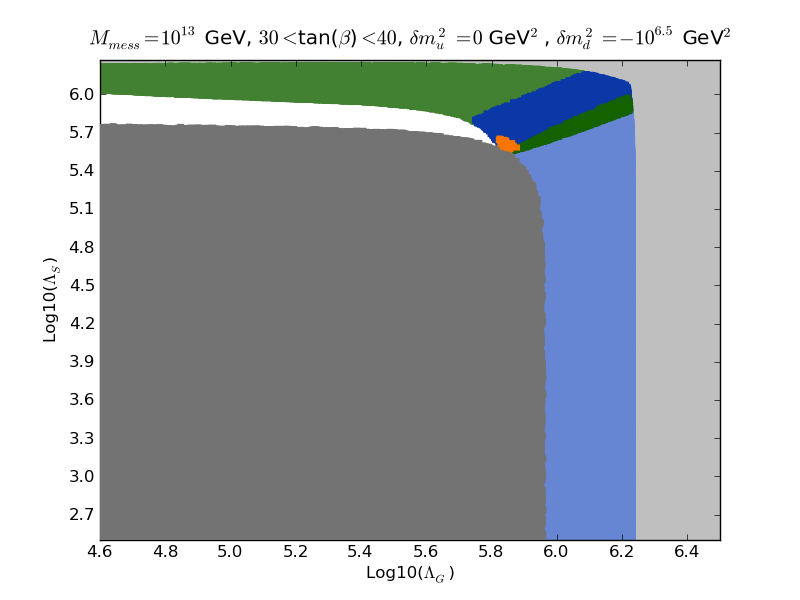}
}
\subfigure{
\includegraphics[width=7.5cm]{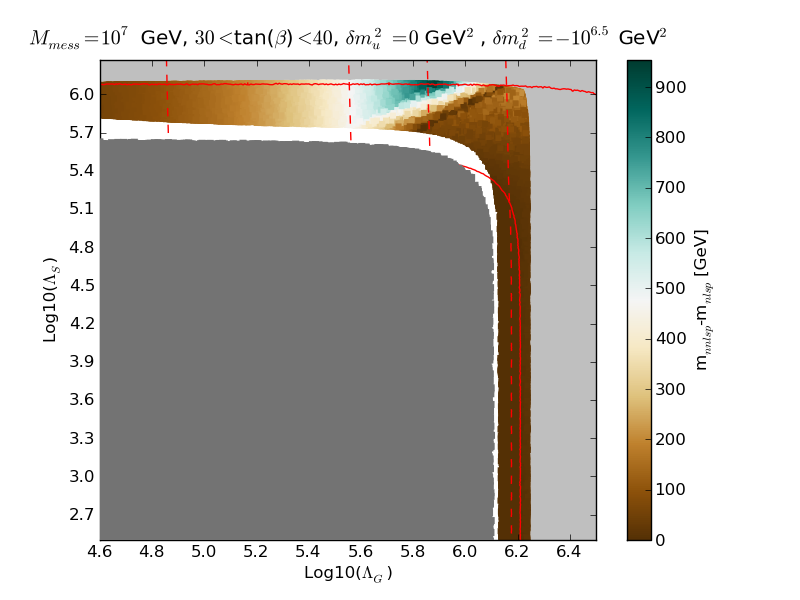}
}
\subfigure{
\includegraphics[width=7.5cm]{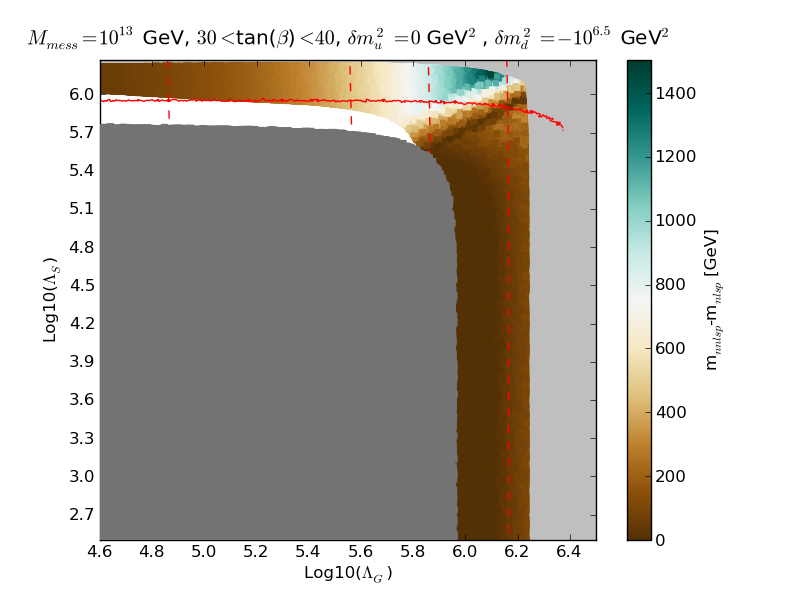}
}
\caption{
\label{dneg2}\footnotesize
\emph{First row}: Logarithmic plot for the NNLSP in the $\Lambda_G, \Lambda_S$ plane. 
NNLSP colors are light green for the second lightest neutralino, blue for the stau, green for the lightest neutralino, pale blue for the smuon, dark blue for the selectron, and orange for $H^0$.
\emph{Second row}: Logarithmic plot in the $\Lambda_G, \Lambda_S$ plane. 
The gradient represents the mass difference $m_{\text{NNLSP}}-m_{\text{NLSP}}$.
The solid and dashed contours identify the NLSP masses, stau and neutralino
respectively.
The scales of the contours are $(100,500)$ GeV and $(1,2)$ TeV
for the neutralino and for the stau $600$ GeV and $2$ TeV.}
\end{center}
\end{figure} 

In the first row of Figure \ref{dneg2} we display the NNLSP species and in the second row the mass difference between the NNLSP and the NLSP with the contours for the values of the two most frequent NLSP, which are the neutralino and the stau. The contours for the stau mass value give an indication of the average scale in the leptonic sector. 

In the short running case the right-handed selectron and the smuon can be co-NLSP, being split from the stau by $m_{\tilde{e},\tilde{\mu}}-m_{\tilde{\tau}}\simeq -17 \text{ GeV}$. The splitting is substantially reduced lowering the value of $\tan\beta$ and the stau will be again the lightest of the sleptons for $\tan\beta=10\pm5$. In the long running case the splitting between generations is enhanced and we can get regions where $m_{\tilde{e},\tilde{\mu}}-m_{\tilde{\tau}}\simeq -140 \text{ GeV}$.
 
\begin{figure}[ht]
\begin{center}
\subfigure{
\includegraphics[width=7.5cm]{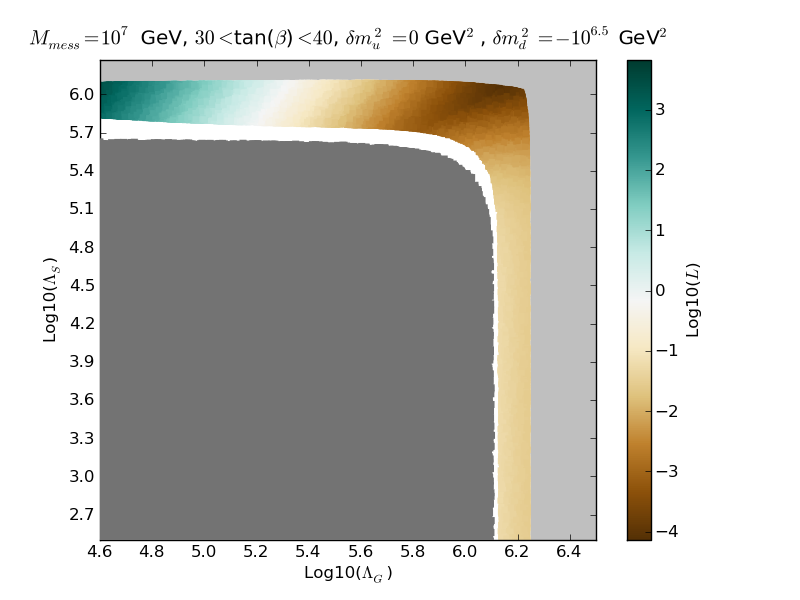}
}
\subfigure{
\includegraphics[width=7.5cm]{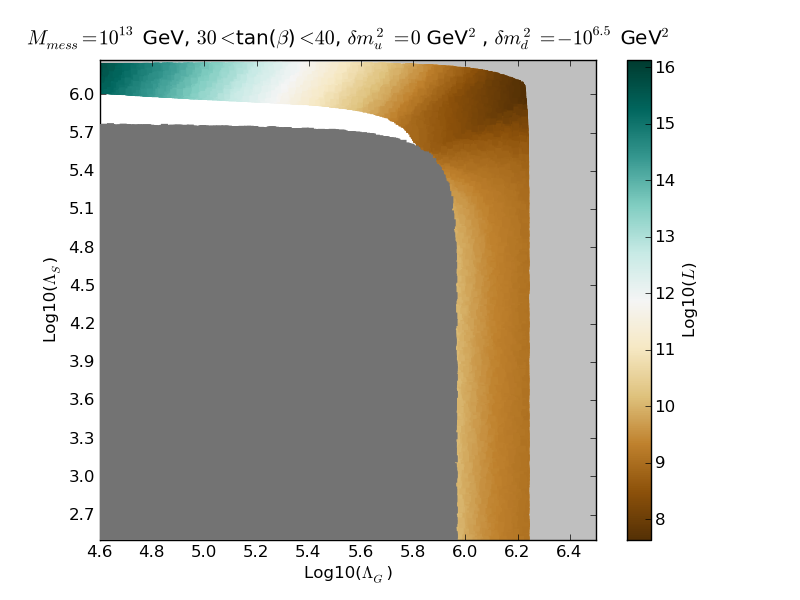}
}
\caption{
\label{dneg3}\footnotesize
Logarithmic plot in the $\Lambda_G, \Lambda_S$ plane. 
The gradient indicates the decay length of the two body NLSP decay into 
gravitino plus NLSP partner.
Different regions have different NLSP type, and the decay length is computed accordingly.}
\end{center}
\end{figure}
In Figure \ref{dneg3} we give an estimate of the decay length for the NLSP two-body decay into its superpartner and the gravitino using the formula \eqref{decaylength}. We see that in the short running case a prompt decay of the selectron and smuon co-NLSP is allowed while in the long running case the NLSP are always very long lived. 

In the short running case would be interesting to study in more detail the collider phenomenology of the region with selectron and smuon NLSP which may lead to very clear multi-leptons signals at LHC which are already under considerations in both ATLAS and CMS experiments \cite{ATLAS-CONF-2012-153,CMS-PAS-SUS-12-026,ATLAS-CONF-2012-154}.

In particular, since all the colored sparticle including the gluino are decoupled in the gaugino mediation region, 
the most relevant 
processes would be the EW production of gauginos, subsequently decaying into sleptons, or
the direct production of sleptons through neutral and charged electroweak currents which were studied in \cite{Bozzi:2004qq}. However, the assumption of sfermion and gaugino mass unification is making both the electroweak gauginos and the sleptonic spectrum quite heavy. In particular, in the short running case the lightest NLSP mass is around $490$ GeV with both the neutralino and the chargino around $1.8$ TeV. As a consequence, both the cross sections for the production through electroweak gauginos and for the direct slepton production would be very suppressed in these scenarios.

The situation can be improved by relaxing the GUT assumption, and a careful study of the possible collider signatures at LHC is needed to understand which part of the parameter space would be sensible to LHC direct searches \cite{NewProject}.

 \section{Messenger-Parity Violation}\label{2.3}

A complete parametrization of all the possible contributions arising from a generic model of gauge mediation should take into account the possibility that a non-zero $D$-tadpole for the $U(1)_{Y}$ is generated at the messenger scale \cite{Meade:2008wd}. This term gives extra contributions to the sfermion masses, proportional to their hyper-charge, that can be parametrized with an extra real parameter $\Lambda_{D}$ as in \eqref{D-tad}.

These effects are possibly dangerous because they typically arise at one loop and they can make some sleptons tachyonic at the electroweak scale. For this reason, in the standard definition of the GGM parameter space, one assumes a parity invariance of the hidden sector, the so-called messenger parity, which set $\Lambda_{D}$ to zero at all orders in perturbation theory.

However, it has been noticed in \cite{Dimopoulos:1996ig} that if the hidden sector realizes a complete representation of a GUT gauge group, then the 1-loop contributions to the $D_{Y}$ term are zero because they are proportional to $\Tr Y$ in the hidden sector which is zero for the GUT hypothesis. 

This observation opens up the possibility of having messenger parity violating (MPV) GUT-complete models which allow for less dangerous two-loop $D_{Y}$-tadpole contributions. These new contributions are proportional to $\sum_{i}\Tr Y_a q^2_{a_{i}}$ which is non-vanishing\footnote{Here $a$ is an index which runs over all the components of the messenger fields, $i=1,2,3$ is an index associated to the gauge group under which the component $a$ transforms and $q^{2}_{a_{i}}$ is the quadratic Casimir for the representation to which the components $a$ belong.} and they have been first computed in a weakly coupled model of messengers in \cite{Dimopoulos:1996ig} and included in the GGM framework in \cite{Argurio:2012qt}. Working with messengers in the $\bf{5}$ or in the $\bf{10}$ of $SU(5)$ we can determine explicitly the group theoretical factor obtaining
\begin{equation}
\delta m^2_{\tilde{f}}(M_{\text{mess}})=k_{1}\frac{g_{1}^2(M_{\text{mess}})}{(4\pi)^2}Y_{\tilde{f}}\left(\frac{5k_{1}}{36}\frac{g^2_{1}(M_{\text{mess}})}{(4\pi)^2}+\frac{3}{4}\frac{g^2_{2}(M_{\text{mess}})}{(4\pi)^2}-\frac{3}{4}\frac{g_{3}^2(M_{\text{mess}})}{(4\pi)^2}\right)\Lambda_D^2\ .\label{extraD}
\end{equation}
 
In the model presented in \cite{Dimopoulos:1996ig} the $D_{Y}$-tadpole contributions to the sfermion masses \eqref{extraD} are always subleading with respect to the usual gauge mediation ones \eqref{sfmass}. However, in \cite{Argurio:2012qt} a weakly coupled model in which these contributions can be dominant with respect to the gauge mediation ones has been constructed and, precisely in this context, it will be interesting to study the phenomenological consequences of messenger parity violation (MPV) on the soft spectrum at the electroweak scale. We have modified SoftSUSY 3.3.4 to accept \eqref{extraD} as extra contribution to the soft spectrum.
 
From the study of the CGGM case we know that, having assumed GUT completeness, the parameter space is reduced by the Higgs mass constraint to two physically different regions: the gaugino screening region $(\Lambda_{S}>\Lambda_{G})$ in which $\Lambda_S\simeq10^6 \text{ GeV}$ independently on any other parameter in order to obtain a very heavy stop mass, and the gaugino mediation region $(\Lambda_{G}>\Lambda_{S})$ in which $7\times10^5\lesssim\Lambda_{G}\lesssim1.5\times10^6 \text{ GeV}$ depending on the value of $M_{\text{mess}}$, where the large stop mass is obtained via gluino 1-loop contributions. 

The MPV contributions will not change this general picture because their effect on the squark masses, and in particular on the up-type squarks, is suppressed with respect to the one on the uncolored sparticles by a factor of order $\mathcal{O}(1/100)$. 

However we will see how the MPV contributions can lead to very different spectra in the leptonic sector with respect to the CGGM ones. In particular, we can have sizeable effects on the slepton masses when
$\Lambda_D\gtrsim\Lambda_S$ and $\Lambda_D\gtrsim\Lambda_G$.

From the general expression \eqref{extraD} we see that the $D_{Y}$-tadpole contributes with opposite sign to the right-handed and left-handed sleptons: with $\Lambda_D^2>0$ we get a negative contribution to the right-handed sleptons soft masses and a positive one to the left-handed sleptons and viceversa for $\Lambda_D^2<0$. Since the sign of the MPV contributions can be either positive or negative on general grounds, we considered both the cases in our scan.

The case of positive $\Lambda_{D}^2$ opens potentially interesting regions that have a very light stau in the gaugino mediations region. In particular, taking $\Lambda_D^2\geq10^{10}\text{ GeV}^2$ we can lower the stau mass to 250 GeV for short running and to 190 GeV for long running, leaving the rest of the spectrum substantially unmodified with respect to the CGGM case. Hence, adding $D_{Y}$-tadpole contributions in the hidden sector can be a useful way of lowering the mass of right-handed sleptons in the gaugino mediation region without taking very large values of $\tan\beta$. However, these scenarios will be typically difficult to probe at LHC because the gauginos will be as heavy as in the CGGM case, suppressing the stau production. 

For this reason, we will not display the results for $\Lambda_D^2>0$, focusing our attention to the case of $\Lambda_D^2<0$ where the left-handed sleptons are driven lighter and a new region with sneutrino NLSP shows up in the low energy spectrum as already found in section $4.2$.
 
The $D_{Y}$ tadpoles contribute also to the Higgs soft masses generating extra terms for the up and down Higgs with opposite sign that we can roughly estimate keeping only the contribution proportional to $g_{3}^2$ in \eqref{extraD}
\begin{equation}
\delta m^2_{H_{u}}=-\delta m^2_{H_{d}}\simeq-\frac{9}{40}\frac{g_{1}^2g_{3}^2}{(4\pi)^4}\Lambda_D^2\ .
\end{equation}
 
These two contributions cancel out in the EWSB condition \eqref{min2} which depend only on $m^2_{{H}_{u}}+m^2_{{H}_d}$ and, consequently, the masses of the heavy scalars $A^{0},H^0, H^{\pm}$ would be essentially unchanged with respect to the CGGM case.
  
Expanding the EWSB condition \eqref{min1} for large $\tan\beta$ we find
\begin{equation}
\vert\mu\vert^2\simeq -m^2_{H_{u}}+\vert\Sigma_{u}\vert-\left(\frac{2+\tan^2\beta}{\tan^2\beta}\right)\delta m^2_{H_{u}}\geq0\ . \label{min1D}
\end{equation}
For $\Lambda_{D}^2>0$ $\delta m^2_{H_{u}}$ is  negative and it will be positive for $\Lambda_{D}^2<0$. 
In the first case the 
$D_{Y}$-tadpole contribution would sum up to the radiative contribution in $\vert\Sigma_u\vert$ coming from the top-Yukawa giving us a larger $\mu$ term with respect to the CGGM case.
In the second case, the $D_{Y}$-tadpole bring down the $\mu$ term with respect to CGGM without, however, giving rise to accidental cancellations in the whole region which satisfies the Higgs mass constraint.

As a final remark, we notice that the MPV contributions to the Higgs soft masses induce also new effects on the RG equations of the sfermion masses which are very much similar to the effects discussed in the previous section. 
On the one hand, the breaking of the GGM sum-rule $\Tr ( Y m^2)\simeq0$ at the two-loop level generates an extra $D_{Y}$ contribution to the running of soft masses \eqref{Seq} proportional to $S=2\delta m^2_{H_{u}}$. On the other hand, the presence of extra Higgs soft masses induces extra-Yukawa contributions \eqref{dYukawa} proportional to $\Delta X_{\tau}=-\delta m^2_{H_{u}}$. 
For $\Lambda_D^2<0$ we get $S>0$ which gives a subleading contribution to the slepton masses which counteracts the leading contribution \eqref{extraD} making the right-handed sleptons lighter than the left-handed ones. This effect arises at 3-loop and thus is always negligible, being further suppressed by an extra loop factor $\frac{g_{1}^2}{(4\pi)^2}$ .
The extra Yukawa contribution has $\Delta X_{\tau}<0$ for $\Lambda_D^2<0$ and it is suppressed by $\frac{y_{\tau}^2}{(4\pi)^2}$. This effect drives the sleptons of the first two generations slightly lighter than the third generation ones, mitigating the splitting between different generations.    

An interesting extension of this setup, which we leave for further investigations, will be to consider the effect of 1-loop $D_{Y}$-tadpole arising in scenarios in which we do not assume gaugino mass unification.

\subsection{Large and negative $\Lambda_{D}^2$: Sneutrino co-NLSP}\label{Dnegative}

We consider the case of negative $\Lambda_D^2$ in which the D-tadpole contributions lead to left-handed sleptons lighter than the right-handed ones. From \eqref{sneutrinoNLSP} we know that a generic feature of this scenario is to have sneutrinos co-NLSP with the left-handed sleptons.
\begin{figure}[ht]
\begin{center}
\subfigure{
\includegraphics[width=7.5cm]{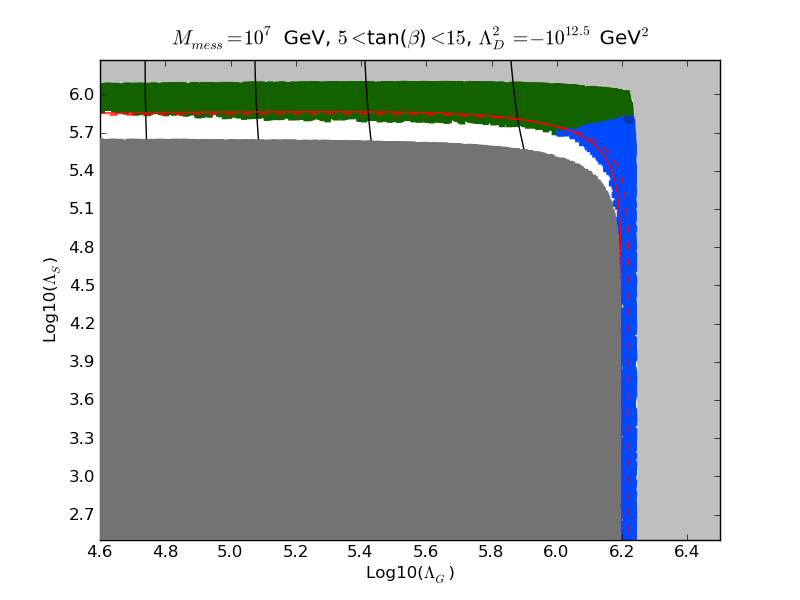}
}
\subfigure{
\includegraphics[width=7.5cm]{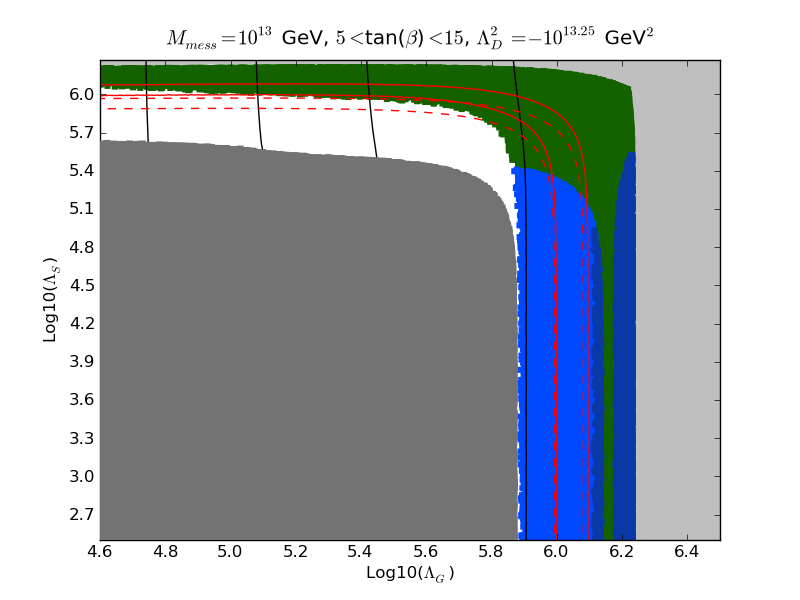}
}
\caption{
\label{D1}\footnotesize
Logarithmic plot in the $\Lambda_G, \Lambda_S$ plane. Explanations of the colors is 
in the text. 
The black, red, dashed-red contour plots identify the gluino, lightest stop, first generation masses
respectively.
The scales of the contours are $500$ GeV, $(1,2,5)$ TeV for the gluino, $5$ TeV for the stop and also $4$ TeV on the right and $6$ TeV for the first generation squarks and also 5 TeV on the right.
}
\end{center}
\end{figure} 
In the Figure \ref{D1} we show the plots for fixed values of $M_{\text{mess}}=10^7,10^{13}$ and $\tan\beta=10\pm5$. We fix $\tan\beta$ to moderate values because we want to minimize the mixing effects in the stau mass matrix which would drive lighter one of the stau mass eigenvalues, making it less obvious to disentangle the D-term effects on the slepton spectrum. We fix $\Lambda_D^2\simeq 3.16\times10^{12}\text{ GeV}^2$ in the short running case and $\Lambda_D^2\simeq 1.78\times10^{13}\text{ GeV}^2$ in the long running case, putting ourselves at the boundary of the allowed parameter space in order to maximize the D-tadpole effects on the spectrum. 

In the dark grey region where SoftSUSY did not converge the sneutrinos become tachyonic destabilizing the EWSB vacuum. In the allowed region the value of $\mu$ is always larger than $1.4$ TeV in the short running case and $1.8$ TeV in the long running case so that we get values of $\mu$ slightly lighter than in the CGGM case without getting any accidental cancellation in the EWSB condition \eqref{min1D}. The stop and the first generation squark contours in Figure \eqref{D1} are essentially identical to the ones of the corresponding cases in CGGM, confirming the observation that the MPV contributions are very suppressed for the squarks. 

The spectrum in the green region of neutralino NLSP is very similar to the one of CGGM  in the regime of gaugino screening, with the Bino NLSP, the Wino NNLSP and the gluino mass going from 350 GeV to up to 5 TeV as can be seen from the gluino mass contours. All the rest of the spectrum is decoupled and this case will be undistinguishable form the CGGM one so that we do not discuss it any further.

Conversely, in the gaugino mediation region we get the expected new features in the leptonic spectrum. The appearence of the sneutrino NLSP (indicated in light blue) is signaling the fact of having left-handed sleptons lighter than the right-handed ones. 

In the short running case the sneutrino NLSP region is the only viable one, whereas for long running we can have regions of very large $\Lambda_G$ where the usual gauge mediation contributions dominate over the D-term contributions and the NLSP is again
a mostly right handed stau. In the transition between the two regimes we have a region where all the leptonic spectrum is very degenerate and the neutralino is NLSP. However, since all the spectrum becomes very heavy for large $\Lambda_G$ (all the sparticles are above 1.5 TeV) the corresponding region would be unaccessible at colliders.  For this reason we are going to focus our discussion on the regions where the sneutrino is NLSP for both short and long running. 

In these regions the spectrum is very similar to the one already discussed in section 4.2 with all the right-handed sleptons splitted always more than 1 TeV. The only difference compared to the section 4.2 is that we get a flavor democratic spectrum characterized by left-handed sleptons and sneutrinos almost degenerate with masses within a range of 10 GeV. The precise hierarchy of this very compressed spectrum is very much dependent on the precise value of $\tan\beta$.  

The flavor democratic case of sneutrino co-NLSP was studied from the perspective of collider signature in \cite{Katz:2009qx} and the possibility of distinguishing it from a flavor-biased scenarios like 4.2 was the subject of further investigations \cite{Katz:2010xg}. It is remarkable that within the same range of values for $\tan\beta$ two kind of UV completions which can motivate the appearence of a sneutrino co-NLSP generically lead to two different kind of spectra that are, in principle, distinguishable. 

The minimal value of the left-handed stau mass is $m_{\tilde{\tau}_{1}}\simeq 530 \text{ GeV}$ with $m_{\tilde{\tau}_{1}}-m_{\tilde{\nu}_{\tau}}\simeq 5.5 \text{ GeV}$ in the short running case and $m_{\tilde{\tau}_{1}}\simeq 430 \text{ GeV}$ with $m_{\tilde{\tau}_{1}}-m_{\tilde{\nu}_{\tau}}\simeq 6.5 \text{ GeV}$ in the long running case. All the gauginos are decoupled in the gaugino mediation region so that the only possible production channel would be the Drell-Yan production of left-handed sleptons reducing a lot the detectability of this scenario.
In principle, the situation can be improved relaxing the assumption of GUT completeness as we will see in the following sections. However, allowing for hidden sector which are not GUT-complete would give rise to MPV effects already at 1-loop. A careful study of the phenomenological consequences of these terms could be interesting and it is left for future studies.

\section{Splitting the Colored Sector}
In the previous sections we discussed scenarios 
where the hidden sector is characterized by a complete GUT structure.
These scenarios can realize gauge and mass unification at the GUT scale in the easiest way,
and can be easily constructed in term of weakly coupled models of gauge mediation
with pairs of vector-like messengers belonging to complete representations of the GUT gauge group.
The gaugino and scalar masses are determined in these cases by only two independent scales
$\Lambda_G$ and $\Lambda_S$.

In this section we relax this assumption and we allow for
different supersymmetry breaking scales associated to the different
gauge group factors of the MSSM.
This possibility can be realized in explicit models with several
messengers in different representations 
of the GUT gauge group \cite{Carpenter:2008w,Carpenter:2008he}.
Note that in this context the gauge coupling unification is a delicate matter,
since the gauge couplings could typically become non-perturbative before the unification scales.
In the following we simply assume that a scenario with six independent 
parameters ($\Lambda_{G_i}$, $\Lambda_{S_i}$) 
is realizable and analyze the phenomenological 
consequences on the weak scale sparticle spectrum,
without addressing the issue of its explicit UV realization
in terms of weakly coupled models, which we postpone for future studies.

In order to keep the numerical scan and the phenomenological study 
feasible, we restrict to representative 
subcases of the complete parameter space.
In the previous sections we observed that the Higgs mass constraint 
is satisfied only for large values of the SUSY-breaking scales, 
implying an heavy sparticle spectrum, especially in the colored scalar sector.
Indeed, we have observed that the large Higgs mass is eventually obtained
via a large stop mass.
Since we are interested in the consequences of the Higgs mass bound
on the GGM sparticle spectrum, the supersymmetry breaking scales
of the $SU(3)$ gauge group will then play a dominant role.
In this section we study the possibility of disentangling the soft term parameters
associated to the $SU(3)$ gauge factor of the Standard Model, i.e. $\Lambda_{G_3}$ and
$\Lambda_{S_3}$, from the other $\Lambda$'s.
The total dimension of the parameter space is six: four supersymmetry breaking scales
($\Lambda_{G_{1,2}} \equiv \Lambda_{G_1}=\Lambda_{G_2}, \Lambda_{G_3},\Lambda_{S_{1,2}}\equiv \Lambda_{S_1}=\Lambda_{S_2},\Lambda_{S_3}$)
plus $\tan \beta$ and $M_{\text{mess}}$.
For simplicity we consider the case of a single mediation
scale for the different gauge groups.

In such a scenario
we expect to be able to satisfy the Higgs
mass bound by increasing $\Lambda_{G_3}$ and/or $\Lambda_{S_3}$,
enlarging the masses of the colored sparticles up to the upper limit of $10$ TeV.
The rest of the parameter space will then be essentially unconstrained, 
at the edge of the collider bounds.

We analyze two simple subcases.
In the first one we
disentangle the gluino mass scale $\Lambda_{G_3}$ from the
other gaugino scales, keeping $\Lambda_S$ unified. 
In region with sufficiently large $\Lambda_S$, 
the gluino mass can be arbitrary light and we can realize scenario with gluino NLSP.
On the other hand, by setting $\Lambda_{G_3}$ large, 
we will get a spectrum with all the colored sparticles very heavy but the
un-colored ones very light.

In the second case we disentangle
$\Lambda_{S_3}$ from the other two scalar scales, keeping the $\Lambda_{G_i}$ unified.
Once again, by fixing the squarks mass scale large, the Higgs mass bound can
be easily satisfied, 
and the other supersymmetry breaking parameters result unconstrained,
generating spectra with a light gluino and light un-colored sparticles.
We also explore the possibility of having tachyonic
UV boundary conditions for the squarks, 
with the purpose of reaching scenarios with large stop mixing.

We expect that
our investigation already highlights most of the interesting features that can be obtained by 
exploring the complete six dimensional parameter space.

\subsection{Splitting Gaugino Mass Scales}
In this subsection we analyze the case with two independent 
gaugino supersymmetry breaking
scales $\Lambda_{G_{1,2}},\Lambda_{G_3}$
and one scalar mass scale  $\Lambda_S$.
Here the gluino mass is set by $\Lambda_{G_3}$ and is not related to the other
gaugino masses which are determined by $\Lambda_{G_{1,2}}$.

Since the Higgs mass is mainly influenced by the stop-top corrections,
we now discuss the main contributions to the stop masses 
as we move along this three dimensional parameter space.
The stop mass matrix at the electroweak scale is
\be
\label{stopmassma}
m_{\tilde t}^2=\left(
\begin{array}{cc}
m_{Q_3}^2+ m_t^2+\Delta_{u_L} & v (y_t A_t \sin \beta- \mu y_t \cos \beta) \\
v (y_t A_t \sin \beta- \mu y_t \cos \beta) & m_{u_3}^2+ m_t^2+\Delta_{u_R}
\end{array}
\right)\ .
\ee
The diagonal entries $m_{Q_3}^2$ and $m_{u_3}^2$ are set by $\Lambda_S$ at
the messenger scale and receive gaugino mediation contributions
along the RG flow. $m_{t}=173.5 \text{ GeV}$ and the extra contributions $\Delta_{u_{L,R}}\simeq\mathcal{O}(M_{Z}^2)$ generated by D-term interactions after the EWSB can be neglected since we are always in the limit where $m_{t},M_{Z}\ll m_{Q_3}, m_{u_3}$.
The off-diagonal entries are determined by the $A_t$ term, the $\mu$ term and $\tan \beta$.
Since $\tan \beta$ is always large in our cases to satisfy the Higgs mass bound,
the term proportional to $\mu$ is suppressed and will not play any role in the stop mass matrix.
The $A_t$ term is vanishing at the messenger scale and is induced at one loop along the RG flow
by the gaugino masses, via the equation
\be
\label{Aterm}
 \frac{d}{dt} A_t=y_t^2 \left ( \frac{2 g_3^2}{3 \pi^2} M_{\tilde  \lambda_3}+ \frac{3 g_2^2}{8 \pi^2} M_{\tilde  \lambda_2}+\frac{13 g_1^2}{120 \pi^2} M_{\tilde  \lambda_1} \right)
+\mathcal{O}(A_t,A_b)\ ,
\ee
where here we omitted terms proportional to the $A$ term of the stop and of the sbottom.
For $\Lambda_{G_i}$ of the same order,
the hierarchy between the gauge coupling of the MSSM implies that
the gaugino mass contributions to the diagonal entries and to the
$A_t$ term are set predominantly by $\Lambda_{G_3}$ and then by $\Lambda_{G_{1,2}}$.

This discussion anticipates the main features we expect for the stop masses
moving along the three dimensional parameter space.

By fixing $\Lambda_{G_{1,2}}$ and varying $\Lambda_{G_3}$ and $\Lambda_S$,
we expect the same shape in the stop mass contours that we found in the CGGM case.

By fixing $\Lambda_{G_3}$ and varying $\Lambda_{G_{1,2}}$ and $\Lambda_S$, the 
features can be very different.
Indeed, the gluino mediation contribution is now fixed, and can be
subleading with respect to the effects of the other gauginos 
if there is a large hierarchy
between  $\Lambda_{G_{1,2}}$ and $\Lambda_{G_3}$.
For values of $\Lambda_{G_{1,2}}$ larger than $\Lambda_S$ and $\Lambda_{G_3}$,
the gaugino mediation contribution from the electroweak sector, especially from the 
$SU(2)$ part, becomes dominant.
The $A_t$ term gets larger with increasing $\Lambda_{G_{1,2}}$, 
induced primarily by the $SU(2)$ gaugino mass $M_2$ in (\ref{Aterm}),
and so the off diagonal component of the stop mass matrix increases. 
On the other hand, the right-handed diagonal entry, 
$m_{u_3}^2$, is sensitive only to the Bino mass $M_1$.
Having set $\Lambda_{G_1}=\Lambda_{G_2}$ we are always in the regime in which
$M_2 \gtrsim M_1$. Hence the two effects only partially compensate each other,
and the resulting lightest stop mass gets effectively reduced for very large values
of $\Lambda_{G_{1,2}}$.
This effect is enhanced for larger $\tan \beta$, which raises further 
the off diagonal entry of the mass matrix proportional to $A_t$, and for longer RG flow.
These features will show up in the following numerical analysis.

Another interesting quantity which characterizes the sparticle spectrum is the 
$\mu$ term, by setting the Higgsino mass and by entering into the neutralino and chargino
mass matrices.
The $\mu$ term is determined from the EWSB condition (\ref{min1}), and accidental cancellation
can make it very small in some corner of the parameter space, a situation we
have already encountered in section 4.1.
Recall that in the large $\tan \beta$ regime, assuming $M_{Z}$ small compared to all the other contributions, the EWSB condition can be written as
\begin{equation}
\label{minimumbis}
\vert\mu\vert^2\simeq -(m^2_{H_{u}}+\sum_{i=1}^{2}K_{i}(M_{\tilde{\lambda}_{i}}))+\frac{3y_{t}^2}{4\pi^2}m^2_{\tilde{t}}\log\left(\frac{M_{\text{mess}}}{m_{\tilde{t}}}\right)\ .
\end{equation}
When $\Lambda_{G_{1,2}}$ is smaller or of the same order than 
$\Lambda_{G_3}$, we expect the same behaviour than in the CGGM case:
the UV contribution $m^2_{H_{u}}$ is compensated by the large 
stop correction, determining the $\mu$ term,
and the gaugino mediation contributions 
$K_{i}(M_{\tilde{\lambda}_{i}})$
are subleading.
Instead, when $\Lambda_{G_{1,2}}$ is very large, 
the gaugino mediation effects
$K_{i}(M_{\tilde{\lambda}_{i}})$
become relevant,
generating large contributions 
proportional to $M_1$ and $M_2$.
As a consequence, there is a further cancellation 
among 
the terms 
in equation (\ref{minimumbis}), and eventually $|\mu|^2$ is small.
For $\Lambda_{G_{1,2}} \gg \Lambda_S$ this effect is so large that 
 the condition (\ref{minimumbis}) cannot be satisfied and the EWSB vacuum
is destabilized.

This discussion about the EWSB condition 
will be particularly relevant for the regime of gaugino mediation, where the
$\Lambda_{G_i}$ effects dominate over $\Lambda_S$.
Instead, in the regime of gaugino screening, where $\Lambda_S$ is larger
than the $\Lambda_{G_{i}}$, 
the EWSB mechanism is realized like in the CGGM case and the value of $\mu$ is fixed determined by the partial cancellation between $m^2_{H_{u}}$ and the stop contribution in \eqref{minimumbis}.

The interesting aspect of the gaugino screening region is that
the gluino mass can be now arbitrarily light, being independent of the other
gaugino masses.
The expressions for the gluino and the Bino masses of GGM (\ref{gmass})
imply that we have gluino NLSP whenever the two 
supersymmetry breaking gaugino scales satisfy the relations
$\Lambda_{G_3}/\Lambda_{G_{1,2}}\leq g_1^2/g_3^2$ \cite{Raby:1997bpa, Arvanitaki:2012ps}.
For values of $\Lambda_{G_3}/\Lambda_{G_{1,2}}$ slightly larger than this inequality,
the gluino will be the 
NNLSP, and the Bino the NLSP.
In this case the decay of the gluino to the NLSP neutralino is a three body decay 
through a virtual squark to 
quark, antiquark and neutralino, whose decay rate is 
approximately
\cite{ArkaniHamed:2004fb,ArkaniHamed:2004yi}
\be
\Gamma(\tilde g \to q \bar q \tilde{N}_{1}) = \frac{\mathcal{N} \alpha \alpha_s M_{\tilde \lambda_3}^5}{192 \pi \sin \theta_W^2 
m_{\tilde q}^4}
\ee
and is suppressed for large squark masses.
The numerical prefactor $\mathcal{N}$ takes into account the possible different decay channels into squarks.

This has to be confronted with the decay rate \eqref{decayrateNLSP}
of the two body decay into gluon and gravitino.
The squarks are very heavy in the allowed portion of the parameter space, to satisfy the Higgs mass constraint,
but we have imposed an upper bound of $10$ TeV on all the sparticle masses.
Thus, even considering the extremal limit with $10$ TeV squarks and very low messenger mass $M_{\text{mess}} =10^6$ GeV,
the decay into gluon plus gravitino results always negligible compared to the 
three body decay, if the latter is kinematically allowed.
So we conclude that the gluino pair production at collider typically leads 
to decay cascades into MSSM sparticles and at least four jets on all the parameter space, 
unless the gluino is the NLSP.

In the following, we scan the parameter space
($\Lambda_{G_{1,2}},\Lambda_{G_3},\Lambda_S,\tan\beta,M_{\text{mess}}$)
in the ranges explained in the introduction.
Once again we choose to fix three of these parameters and present the results in
two dimensional plots with contours.
We have selected the region of parameter space where the
spectra differs significantly from the ones we have found in the CGGM case.

\subsubsection{Large $\Lambda_{G_{1,2}}$: Gluino NLSP}
The first interesting case is realized by fixing $\Lambda_{G_{1,2}}$ to the large value
$7.94 \times 10^{5}$ GeV, corresponding to a Bino mass of $1$ TeV,
and scan over $\Lambda_{G_3}$ and $\Lambda_{S}$. 
In this situation a large stop mass, necessary to raise the Higgs mass,
 can be obtained either through a large $\Lambda_S$ scale,
or through a large gaugino mediated contribution, dominated mainly by the gluino mass.
Indeed the value of $\Lambda_{G_{1,2}}$ is not large enough to make the stop
sufficiently heavy through gaugino mediation, since those contributions
are nevertheless suppressed by weak couplings and we are not
pushing $\Lambda_{G_{1,2}}$ to extreme high values.

The shape of the allowed region is then analogous to the CGGM case. 
For small $\Lambda_{G_3}$ the allowed region is obtained for high $\Lambda_S$,
whereas if $\Lambda_{G_3}$ is sufficiently large to induce large stop masses, the UV boundary 
condition for the scalars $\Lambda_S$ can be small.
In the portion of the parameter space characterized by gaugino mediation, 
the spectrum and the phenomenology
is very similar to CGGM. Instead, in the gaugino screening region
 new interesting possibilities can be realized.
In particular,
given the fixed large value for $\Lambda_{G_{1,2}}$, in the region where $\Lambda_{G_3}\leq g_{1}^2/g_{3}^3 \Lambda_{G_{1,2}}$
the NLSP is the gluino.

\begin{figure}[ht]
\begin{center}
\subfigure{
\includegraphics[width=7.5cm]{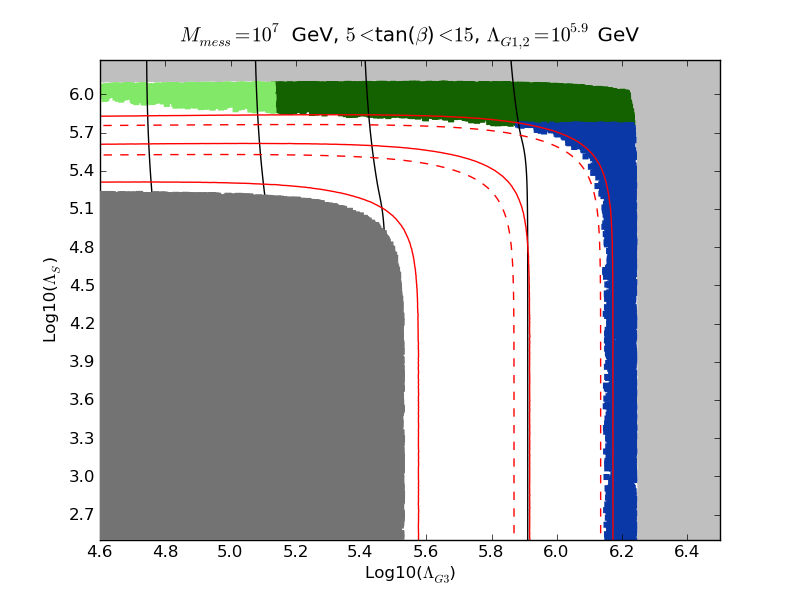}
}
\subfigure{
\includegraphics[width=7.5cm]{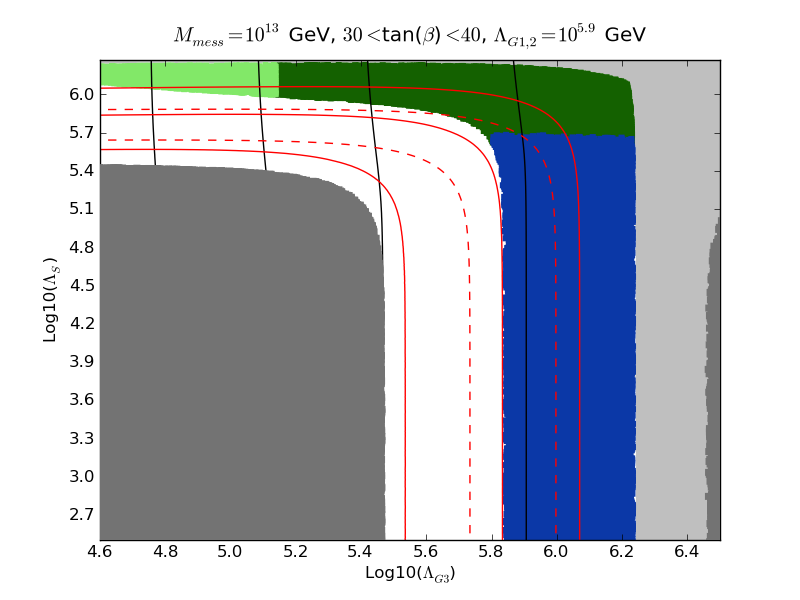}
}
\caption{
\label{GGMG3a}\footnotesize
Logarithmic plot for the NLSP in the $\Lambda_{G_3}, \Lambda_S$ plane. 
Very light green is the gluino, green is neutralino,
blue is stau.
The black, red, dashed-red contour plots identify the gluino, 
the lightest stop and first generation masses
respectively.
The scales of the contours are $(0.5,1,2,5)$ TeV for the gluino,
$(1.5,3,5)$ TeV for the lightest stop and $(3,5)$ TeV 
for the first generation squarks in the left plot, while 
$(0.5,1,2,5)$ TeV for the gluino,
$(1.5,3,5)$ TeV for the lightest stop and $(3,5)$ TeV 
for the first generation squarks
in the right one.
}
\end{center}
\end{figure}

These features are shown in Figure \ref{GGMG3a}, where we present only two 
exhaustive cases for $M_{\text{mess}}$ and $\tan\beta$.
In the light green region the gluino is the NLSP, in the green region the lightest neutralino, and
in the blue region the stau.
The black contours identify the gluino mass at $500$ GeV, $1,2,5$ TeV, so the gluino NLSP
region extends to values of the gluino mass up to $O(1)$ TeV.
The other contours refer to the stop and to the first generation squarks and they 
strongly resemble the ones in the CGGM case, being colored sparticles 
and hence influenced mainly by the gluino mediation contribution.
Note that the dark grey region of non convergence of SoftSUSY is larger than in the CGGM case.
In that dark grey region $\Lambda_{G_3}$ and $\Lambda_{S}$ are too small compared to $\Lambda_{G_{1,2}}$
and EWSB cannot occurs.

Except for this aspect,
the shape of the contours for the Higgs mass
bound and the $A_t$, $\mu$ and $M_S$ quantities
are very similar
to the CGGM case, since the squark sector and 
the gluino mass give the most relevant contributions to the Higgs mass corrections,
so we do not present them here.

\begin{figure}[ht]
\begin{center}
\subfigure{
\includegraphics[width=7.5cm]{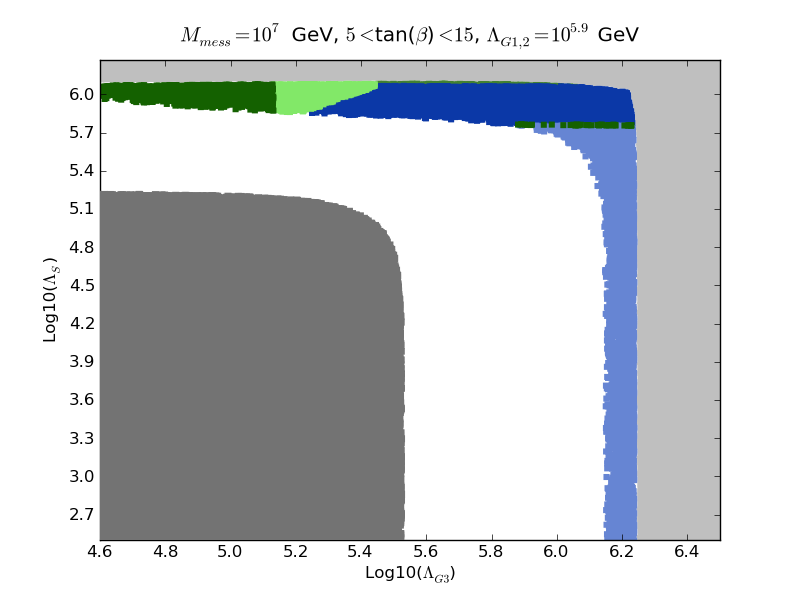}
}
\subfigure{
\includegraphics[width=7.5cm]{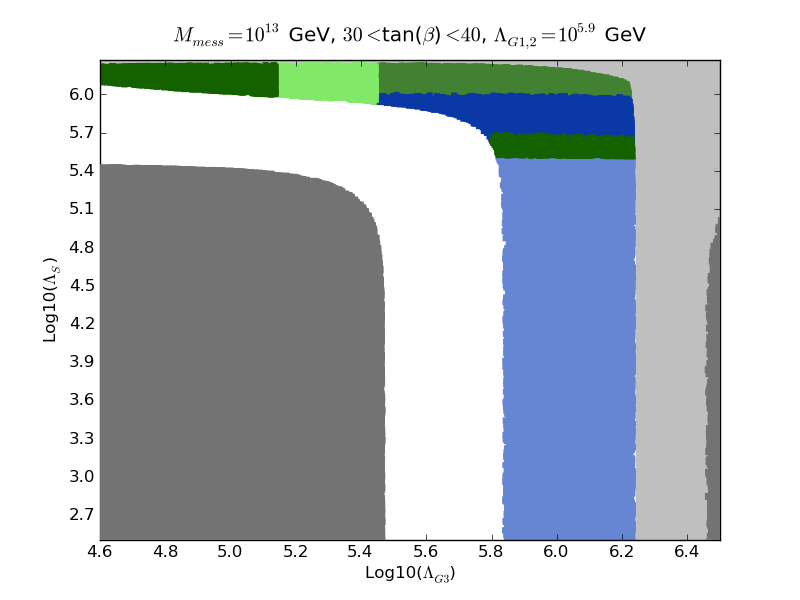}
}
\caption{
\label{GGMG3c}\footnotesize
Plot for the NNLSP in the $\Lambda_{G_{3}}, \Lambda_S$ plane. 
Very light green is the gluino, light green is the second lightest neutralino, 
blue is the lightest stau, green 
is the lightest neutralino, pale blue is the smuon.
dark grey is for regions where SoftSUSY did not converge.
}
\end{center}
\end{figure}

\begin{figure}[ht]
\begin{center}
\subfigure{
\includegraphics[width=7.5cm]{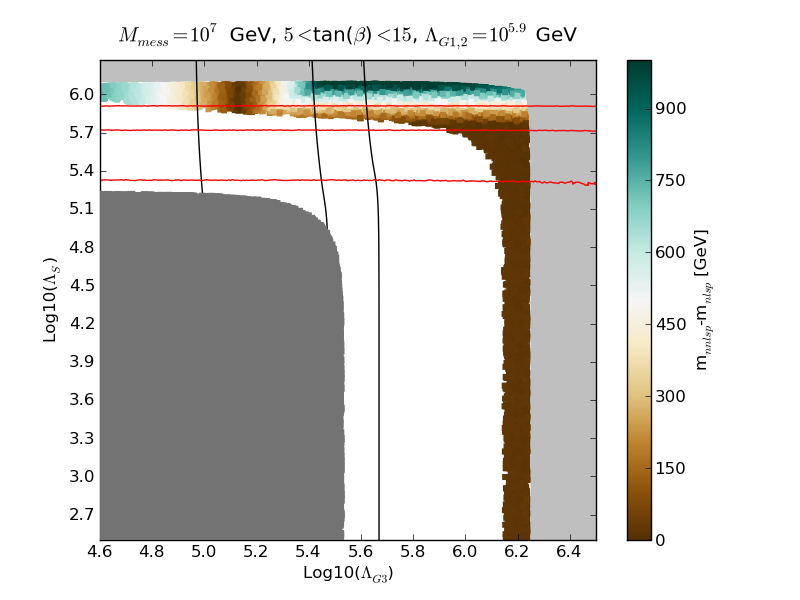}
}
\subfigure{
\includegraphics[width=7.5cm]{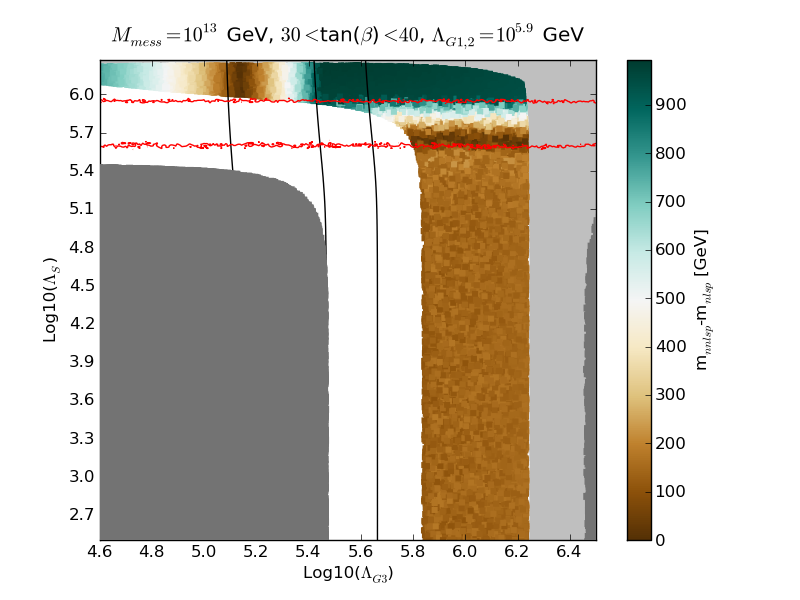}
}
\caption{
\label{GGMG3d}\footnotesize
Gradient plot for the mass difference
$m_{NNLSP}-m_{NLSP}$ in the $\Lambda_{G_{3}}, \Lambda_S$ plane.
The black and red contour plots identify the gluino and
stau mass respectively.
The contours
are $(0.8,2,3)$ TeV for the gluino and
$(0.2,0.5,1,1.5)$ TeV for the stau in the left plot,
while 
$(1,2,3)$ TeV for the gluino and
$(1,2)$ TeV for the stau
in the right plot.
}
\end{center}
\end{figure}

In Figure \ref{GGMG3c} we show the NNLSP type, and in Figure
\ref{GGMG3d} the
difference in mass between the NNLSP and the NLSP, with contours for the NLSP masses.
The contours plot refer to the gluino and stau masses, black and red respectively.
The lightest neutralino, which is mostly Bino in the entire allowed parameter space,
has constant mass around 1 TeV.
In case with $M_{\text{mess}}=10^7$ GeV
the gluino mass varies between $190$ GeV and $10$ TeV, while the stau mass
varies between $315$ GeV and $2.2$ TeV.
In case of long RG flow
the gluino mass varies between $152$ GeV and 10 TeV, while the stau mass
varies between $470$ GeV and $3.8$ TeV.
In the case of gluino NLSP, the NNLSP is the Bino.
In the case of neutralino NLSP, the NNLSP can be either the gluino, or the neutral Wino, or the stau.
In the region of stau NLSP, the NNLSP is the smuon with almost degenerate mass
when $\tan \beta$ is small.

\begin{figure}[ht]
\begin{center}
\subfigure{
\includegraphics[width=7.5cm]{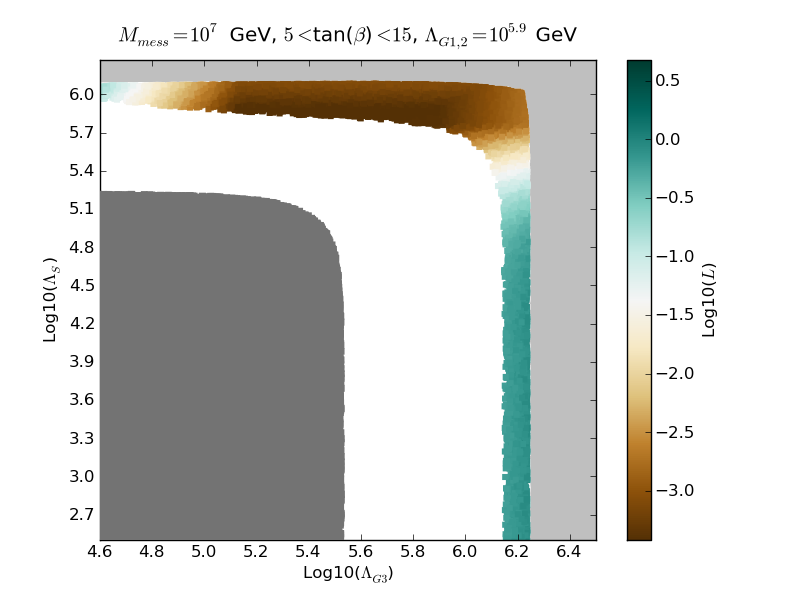}
}
\subfigure{
\includegraphics[width=7.5cm]{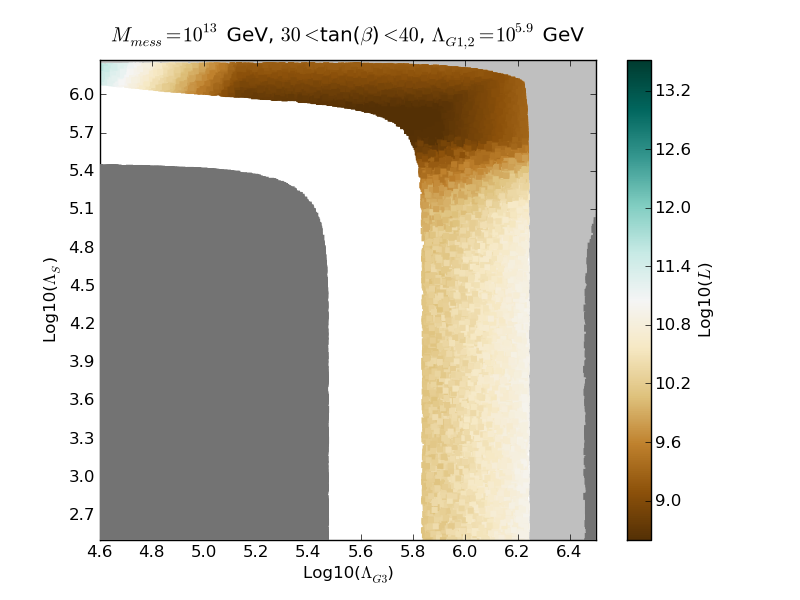}
}
\caption{
\label{GGMG3e}\footnotesize
Gradient plot for the decay length of the NLSP in the
$\Lambda_{G_{3}}, \Lambda_S$ plane.
The NLSP decay is prompt for $M_{\text{mess}}=10^7$ GeV and long for 
$M_{\text{mess}}=10^{13}$ GeV, independently on the nature of the NLSP.
}
\end{center}
\end{figure}

To complete the phenomenological characterization of this scenario we show in
Figure \ref{GGMG3e} the decay length of the NLSP in meters.
In the long running case, the decay is always outside the detector.
In the short running case the decay can be displaced or prompt.
Note that we could also consider a scenario with smaller messenger mass, 
where the same pattern
for the soft masses is realized, but with always promptly decaying NLSP.
In the case of the gluino NLSP the decay is to gluons plus grativinos.
Also in the large $M_{\text{mess}}$ case, 
the lifetime of the gluino is nevertheless short enough to not have cosmological
consequences 
\cite{ArkaniHamed:2004fb}.

The most interesting region is clearly the one with gluino NLSP.
In this region the NNLSP neutralino can be significantly heavier than the gluino,
realizing a scenario where all the sparticles except the gluino are effectively
decoupled for collider physics.
In the case of gluino NLSP the
LHC direct searches in the two possibilities of long-lived or prompt decaying 
differ significantly.

For long-lived gluino the current strongest constraints come from ATLAS and CMS searches on
R-hadrons \cite{Aad:2012pra, Chatrchyan:2012dxa} and from CMS searches of $\text{high-p}_{T}$ isolated tracks with large ionization energy loss $dE/dx$ \cite{Chatrchyan:2012sp}. For promptly decaying gluino the bound is set by limits on jets plus missing energy signatures in both CMS and ATLAS experiments \cite{ATLAS:2012ona, :2012mfa}. In both cases a conservative lower bound around $800-850$ GeV on the gluino mass was obtained in \cite{Kats:2011qh} considering $1/\text{fb}$ of data from LHC.  Including the most recent current searches, the gluino mass bound is around 1 TeV which implies that almost all the gluino NLSP region in our plots can be already excluded by LHC direct searches besides a small region which lies beyond the 1 TeV contour for the gluino in Figure 14. Regions with gluino NLSP with heavier mass can be easily obtained fixing $\Lambda_{G_{1,2}}$ to an higher value which would imply heavier masses for the Bino and the Winos.

Another interesting characteristic of this scenario is the 
region with 
neutralino NLSP and gluino NNLSP,
which differs from CGGM case
since now the gluino NNLSP can be almost degenerate in mass with the neutralino.
Both sparticle are however quite heavy, around $1$ TeV.
The EW production is suppressed, since the Winos are heavy, 
with mass which is generically twice the Bino mass, 
because we are sticking to the relation
$\Lambda_{G_1}=\Lambda_{G_2}$.
The neutralino production at LHC 
is then mainly 
through pair production of the NNLSP gluinos decaying via virtual squarks to
quark, antiquark and neutralino.
The spectrum is the one of a simplified model for neutralino mostly Bino NLSP,
promptly decaying in the case of small $M_{\text{mess}}$, produced via
an NNLSP gluino with an arbitrary small mass.
If the gluino mass is very close the the Bino mass, the 
signal of jets plus missing energy
(plus eventually photons in the case of promptly decaying Bino)
 can be softened, 
and the scenario more difficult to exclude at colliders.

\subsubsection{Large $\Lambda_{G_{3}}$: Light Un-colored Spectrum}
\label{largeG3}
An alternative possibility for this scenario
is to fix $\Lambda_{G_3}$ and show the results in the 
$\Lambda_{G_{1,2}},\Lambda_S$ plane.
We set $\Lambda_{G_3}=1.7 \times 10^6$ GeV, resulting in a gluino mass around $10$ TeV.
This gluino mass induces large squark masses and $A_t$, facilitating the
accomplishment of a heavy Higgs.
We expect that in this case we can reach regions with
small values of $\Lambda_{G_{1,2}}$, allowing then for a 
light un-colored spectrum.
Once again we fix $M_{\text{mess}}$ and $\tan \beta$ 
to two representative values.

\begin{figure}[ht]
\begin{center}
\subfigure{
\includegraphics[width=7.5cm]{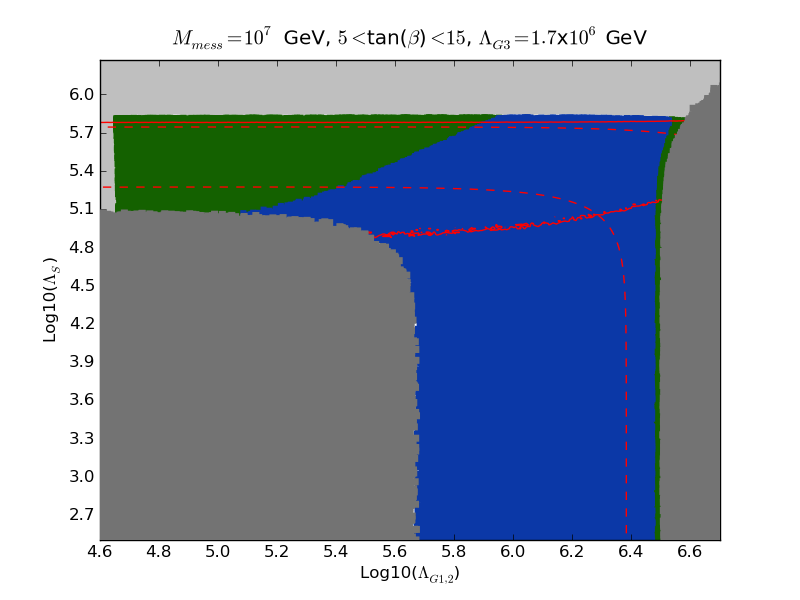}
}
\subfigure{
\includegraphics[width=7.5cm]{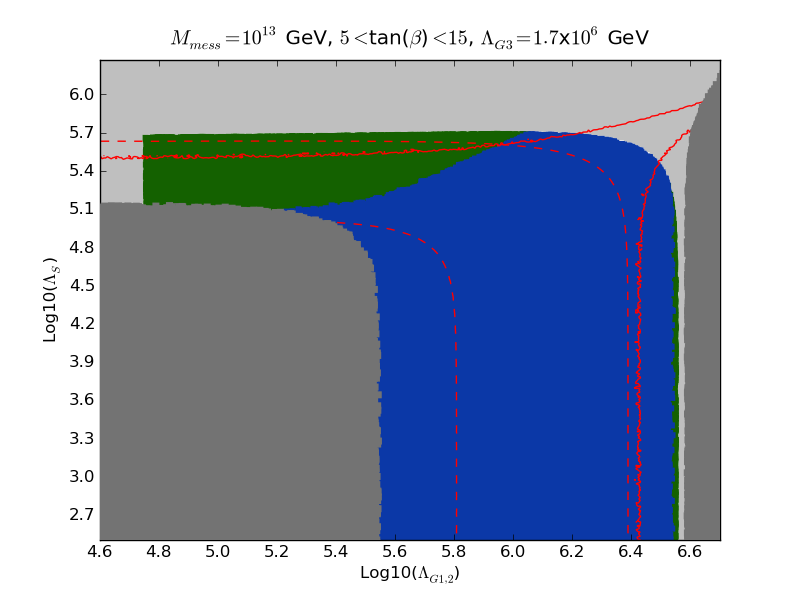}
}
\caption{
\label{GGMG12a}\footnotesize
Logarithmic plot in the $\Lambda_{G_{1,2}}, \Lambda_S$ plane. 
Blue is stau and green is neutralino.
The red, dashed-red contour plots identify the lightest stop and first generation masses
respectively.
The scales of the contours are $(5.6,7)$ TeV for the stop and $6.2,7.5$ TeV 
for the first generation squarks in the left plot, while 
$(6.7,7.2)$ TeV for the stop and $(8,8.4)$ TeV 
for the first generation squarks in the right plot.
}
\end{center}
\end{figure}

In Figure \ref{GGMG12a} we show the plots with the
NLSP types, and the contours for lightest stop and first generation squarks mass.
When  $\Lambda_{G_{1,2}} < \Lambda_S$
the squark masses depend only on $\Lambda_S$.
For values of $\Lambda_{G_{1,2}}$ larger than $\Lambda_S$ 
the gaugino mediation contribution from the electroweak sector
becomes important.
This explains the shape of the first generation squark mass contour.
The lightest stop is instead essentially independent of $\Lambda_{G_{1,2}}$,
getting even slightly lighter for larger $\Lambda_{G_{1,2}}$.
The explanation of this phenomenon was given below equation (\ref{Aterm}),
and is related to the 
 off diagonal entry of the stop mass matrix 
 which is indeed 
enhanced 
for longer RG flow.

The NLSP is either the stau or the neutralino depending on the ratio
of $\Lambda_{G_{1,2}}/\Lambda_S$ similarly to the CGGM case. 
However, note that now $\Lambda_{G_{1,2}}$ is allowed to take smaller values
compared to $\Lambda_G$ in the CGGM case, so the stau mass can be much lower.
For values of $\Lambda_{G_{1,2}}$
much larger than $\Lambda_S$ (and also with
$\Lambda_{G_{1,2}} > \Lambda_{G_3}$) there is a new tiny region with neutralino
NLSP. 
In this region we are in the regime explained in the introduction after equation 
(\ref{minimumbis}). The accidental cancellation in the EWSB condition leads to 
a very small $\mu$ and the NLSP is the neutral Higgsino.
For values of $\Lambda_{G_{1,2}}$ even larger,
the EWSB cannot be realized, and
we obtain a dark grey region signaling that  
 SoftSUSY failed to produce a spectrum.

\begin{figure}[ht]
\begin{center}
\subfigure{
\includegraphics[width=7.5cm]{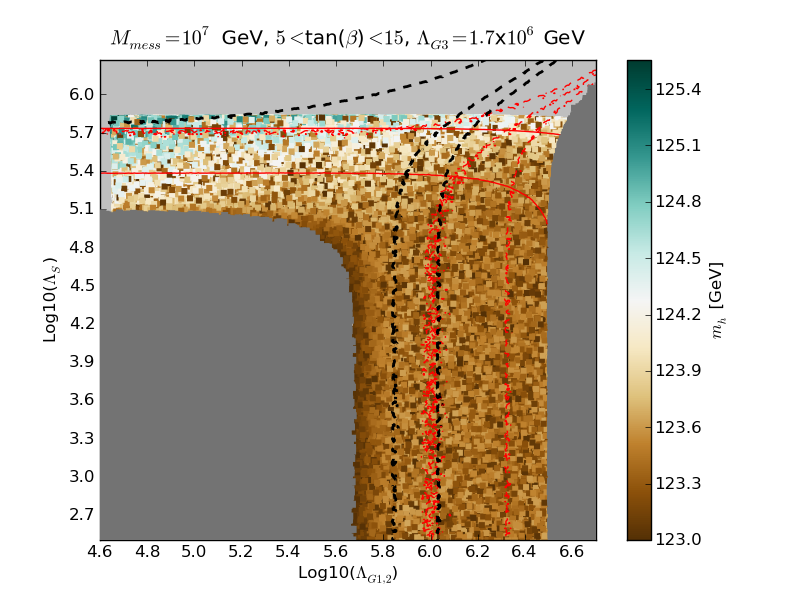}
}
\subfigure{
\includegraphics[width=7.5cm]{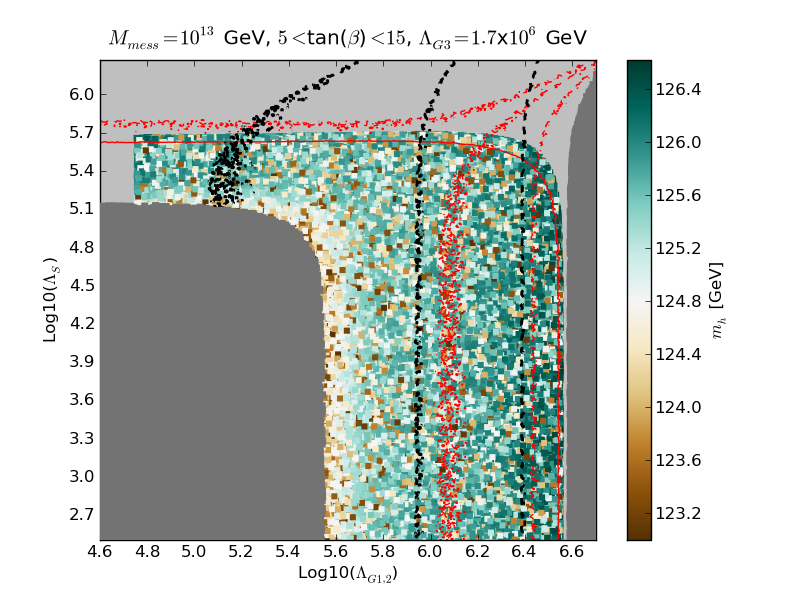}
}
\caption{
\label{GGMG12b}\footnotesize
Gradient plot in the $\Lambda_{G_{1,2}}, \Lambda_S$ plane for the Higgs mass. 
The dashed black, red,
dashed-red contour plots identify $A_t$, $M_S$ 
and $\mu$
respectively.
The scales of the contours are $(-3.1,-3.4,-3.5)$ TeV for $A_t$,
$(6,7)$ TeV 
for $M_S$ 
and
$(2,2.5,3)$ TeV for $\mu$ in the left plot
 while
 $(-5.5,-6,-7)$ TeV for $A_t$, 
 $(7.3,7.6)$ TeV 
for $M_S$ 
and
$(3.5,4.5,5)$ TeV for $\mu$  in the right one.
}
\end{center}
\end{figure}

These observations are confirmed in Figure 
\ref{GGMG12b} where the plots about the Higgs mass are showed.
We analyze the shape of the contours for the 
$A_t$ term (dashed black), for $M_S$ (red) 
and for the $\mu$ term (dashed red),
starting with 
the case of $M_{\text{mess}}=10^7$ GeV.
The $A_t$ is essentially constant,
the contours being at around $3$ TeV. 
It increases slightly with increasing $\Lambda_{G_{1,2}}$,
because of the contribution induced by the gauginos.
The average stop mass is almost constant and also slightly increase for larger 
$\Lambda_{G_{1,2}}$ because of the gaugino mediation contribution.
The $\mu$ term becomes smaller if $\Lambda_{G_{1,2}}$
increases because of the arguments given below equation (\ref{minimumbis}),
and reaches its minimal value along the edge with the dark grey region.
These features are present also in the case of $M_{\text{mess}}=10^{13}$ GeV,
where the longer RG flow only implies that $A_t$, $M_S$ 
and 
$\mu$ vary more significantly along the parameter space, and are more sensitive
to the exact value of $\tan \beta$, resulting in fuzzy contours.

\begin{figure}[ht]
\begin{center}
\subfigure{
\includegraphics[width=7.5cm]{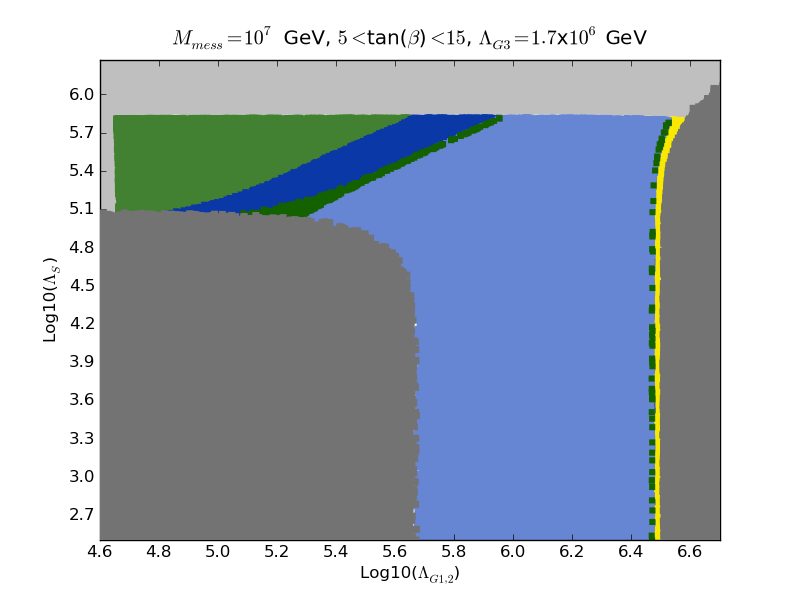}
}
\subfigure{
\includegraphics[width=7.5cm]{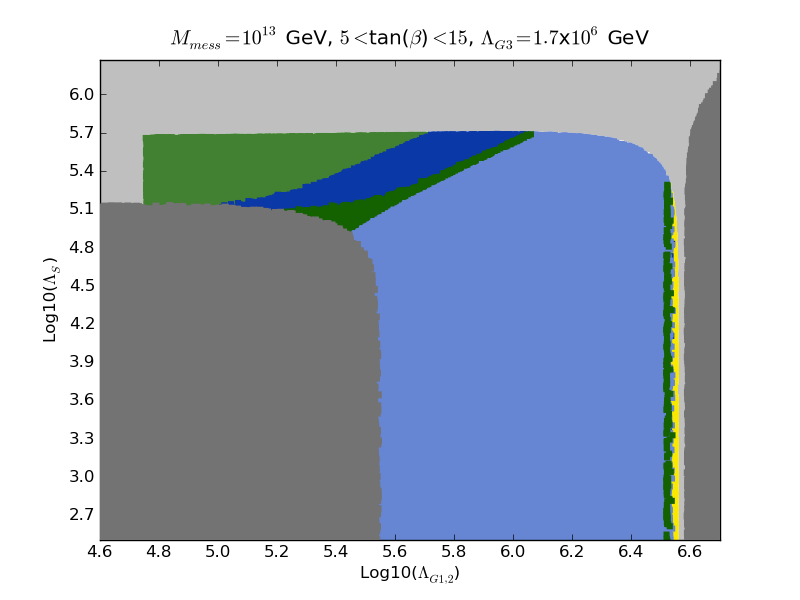}
}
\caption{
\label{GGMG12c}\footnotesize
Plot for the NNLSP in the $\Lambda_{G_{1,2}}, \Lambda_S$ plane. 
Light green is the second lightest neutralino, blue is the lightest stau, green 
is the lightest neutralino, pale blue is the smuon, and yellow is the lightest chargino.
Dark grey is for regions where SoftSUSY did not converge.
}
\end{center}
\end{figure}

\begin{figure}[ht]
\begin{center}
\subfigure{
\includegraphics[width=7.5cm]{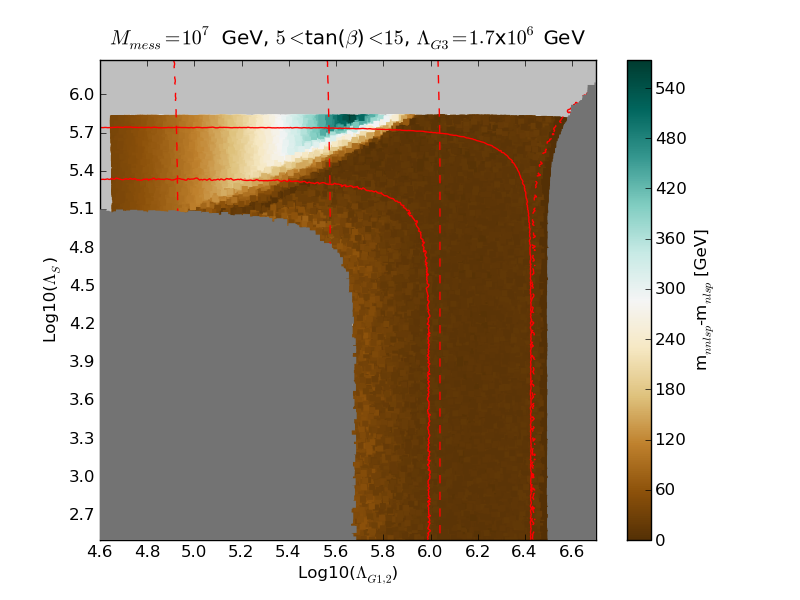}
}
\subfigure{
\includegraphics[width=7.5cm]{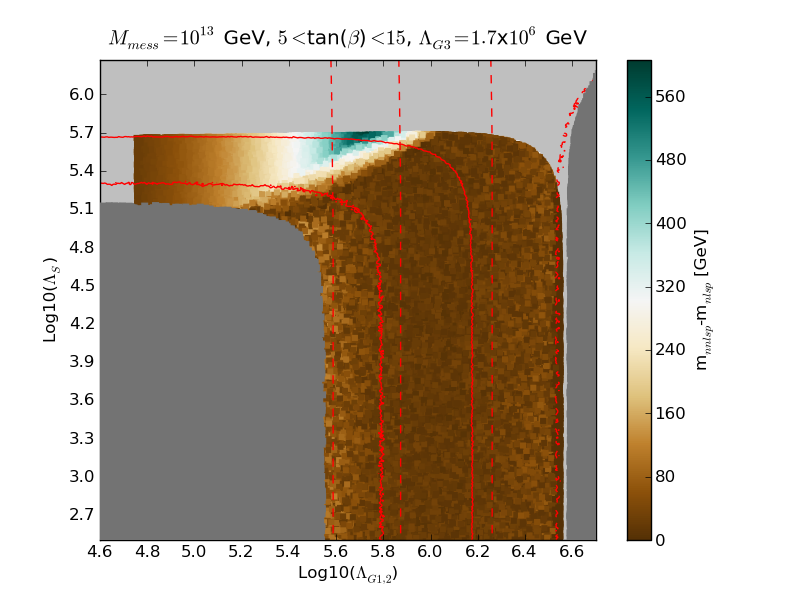}
}
\caption{
\label{GGMG12d}\footnotesize
Gradient plot for the mass difference
in $m_{NNLSP}-m_{NLSP}$ in the $\Lambda_{G_{1,2}}, \Lambda_S$ plane.
The dashed red and red contour plots identify the neutralino and
stau mass respectively.
The contours
are $(0.1,0.5,1)$ TeV for the neutralino and
$(0.4,1)$ TeV for the stau in the left plot,
while 
$(0.5,1,2.5)$ TeV for the neutralino and
$(0.5,1,2)$ TeV for the stau in the right plot.
The rightmost neutralino contours is still the
contours associated to the heaviest mass.
They signal the fact that the neutralino becomes light again 
for larger values of $\Lambda_{G_{1,2}}$, for the reasons explained in the text.
}
\end{center}
\end{figure}

In Figure \ref{GGMG12c} we show the NNLSP types, and in Figure
\ref{GGMG12d} the gradient of the mass difference between the NNLSP and the NLSP,
with contours for the neutralino and for the stau masses.

In the $\Lambda_{G_{1,2}} <\Lambda_S$ region the Bino is the NLSP and
the NNLSP is the neutral Wino. 
When $\Lambda_{G_{1,2}}  \simeq \Lambda_S$ there is the transition between neutralino NLSP
and stau NLSP, with the other one being NNLSP.
For most of the region with $\Lambda_{G_{1,2}} > \Lambda_S$, the NLSP is the stau 
with smuon and selectron co-NNLSP.
Note that, contrary to the CGGM case, here the stau can be very light; the minimum
value for the stau mass in the allowed region is $136$ GeV for 
$M_{\text{mess}}=10^7$ GeV
and $174$ GeV
for the $M_{\text{mess}}=10^{13}$ GeV case.
This is due to the fact that having disentangled $\Lambda_{G_{1,2}}$ from 
$\Lambda_{G_{3}}$, now $\Lambda_{G_{1,2}}$ 
can be small even satisfying the Higgs mass bound,
and hence
the gaugino mediation contribution to the sleptons is less relevant.

Finally, in the right band the neutralino is again the NLSP 
with almost degenerate chargino NNLSP.
In this region the $\mu$ term is small and 
in particular we have $\mu\ll M_{\tilde{\lambda}_{1,2}}$ so that the lightest neutralino
is the Higgsino with the other neutral Higgsino and the charged Higgsino nearly degenerate NNLSP. The other neutralinos (Bino and Winos) and the charged Wino
are very heavy, since their mass is set by $\Lambda_{G_{1,2}}$.

We already encounter a similar situation in section 4.1, where accidental cancellations
were due to extra contributions to the Higgs soft masses.
Here the mechanism is different, and relies on very large electroweak gaugino
mass scales $\Lambda_{G_{1,2}}$ which maximize the $K_{1,2}$ contributions in the EWSB condition \eqref{minimumbis}.
As a consequence the Bino, the Winos and also the sleptons are very heavy, 
essentially decoupled from collider physics,
realizing a simplified model with only Higgsinos accessible at collider which is very similar to the one obtained in section 4.1 for short running case and $\Lambda_G$ sufficiently large.

As already discussed in section 4.1, this kind of scenario has colored production very much suppressed by the heaviness of the gluino and the squarks and this is even more true in the present case where we fixed the gluino mass at around 10 TeV.
Colored production is thus negligible and the CMS and ATLAS constraints on the gluino-Higgsino simplified model based on Z+jets+MET searches \cite{Chatrchyan:2012qka, ATLAS-CONF-2012-152} are not constraining in our case.    
We believe that this region deserves further studies for collider signature, particularly in the case where the NLSP is promptly decaying (i.e. with short RG flow $M_{\text{mess}}=10^7$ GeV).

\begin{figure}[ht]
\begin{center}
\subfigure{
\includegraphics[width=7.5cm]{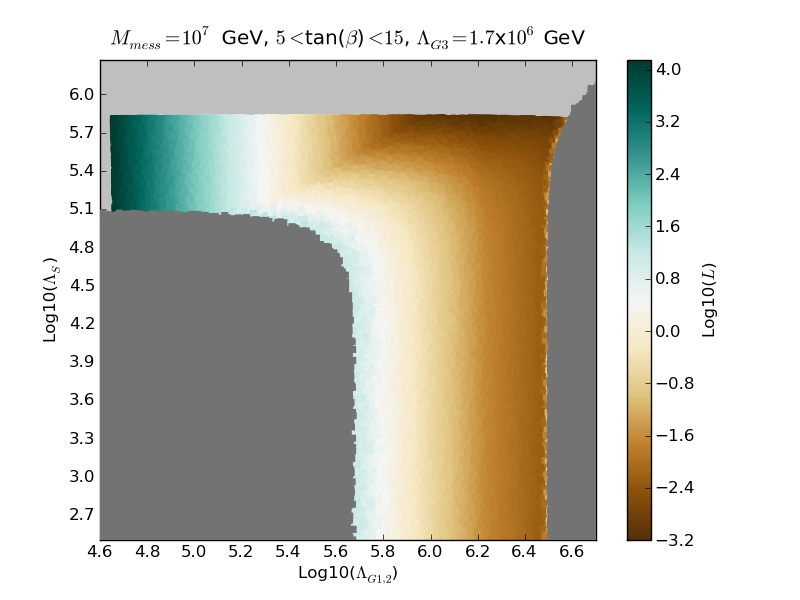}
}
\subfigure{
\includegraphics[width=7.5cm]{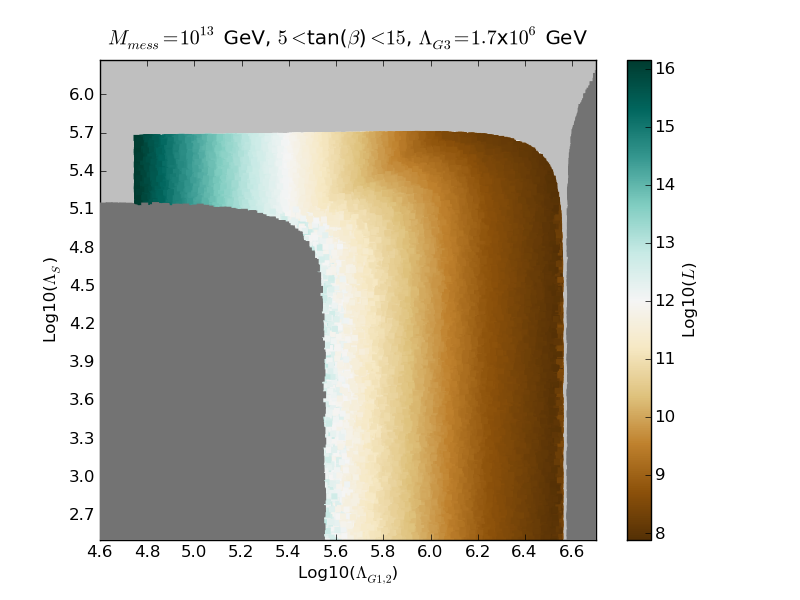}
}
\caption{
\label{GGMG12e}\footnotesize
Gradient plot for the decay length of the NLSP in the
$\Lambda_{G_{1,2}}, \Lambda_S$ plane. 
}
\end{center}
\end{figure}

The decay length of the NLSP is shown in figure \ref{GGMG12e}.
The NLSP escapes the detector for $M_{\text{mess}}=10^{13}$ GeV whereas
the decay can happen either inside or outside the detector, depending
on the parameters, for the $M_{\text{mess}}=10^{7}$ GeV case.

Except for the Higgsino NLSP region,
most of the parameter space in
short (or long) running is essentially 
the typical gauge mediated one of promptly decaying (or long-lived) Bino NLSP 
or stau NLSP.
The NLSP masses can be quite small, at the edge of the collider
bounds.
Since we have fixed $\Lambda_{G_3}$ to be large and the squarks are 
heavy to satisfy the Higgs mass constraint, the whole colored spectrum 
can be considered decoupled for collider physics and 
the LHC production would be mainly electroweak.
Nevertheless, differently than in the CGGM case, here the Winos 
can be very light, since their
masses are not anymore tied to the gluino mass.
The EW production can be significant, and we expect that LHC searches
can already reduce considerably the allowed portion of the parameter space.

The entire regions with $\Lambda_{G_{1,2}}< 10^{5.9}$ GeV in the short running case
and with  $\Lambda_{G_{1,2}}< 10^{5.7}$ GeV in the long running case
can be considered generators of simplified models
with Bino or stau NLSP, with pure EW production, which we now discuss.

The $\mu$ term is always very large, so the second lightest neutralino and the 
lightest chargino are Winos, with mass given by $M_{\tilde \lambda_2}$
evaluated at the EW scale, 
which is roughly twice the Bino mass $M_{\tilde \lambda_1}$.
The most interesting case for LHC searches is the one of short running
where the NLSP is promptly decaying or decay with displaced vertex in a large portion of the
parameter space. Also in this scenario we can always obtain a promptly-decaying NLSP by reducing further $M_{\text{mess}}$ up to its lower bound which is around $10^6 \text{ GeV}$.

In the Bino NLSP region, which decays promptly for short running,
an analysis similar to the one in \cite{Kats:2011qh} would 
give an LHC lower bound on the neutralino mass around $200$ GeV.
In this region the stau mass can vary from the Bino mass up to
$1.1$ TeV, where it can be considered as decoupled from collider
perspectives. The other sleptons are degenerate with 
the stau when $\Lambda_{S}$ is very large, since the diagonal components in the mass
matrix dominates. When $\Lambda_{S} < 10^{5.4}$ GeV the mass splitting can reach 
$100$ GeV.
When the sleptons are light they can be produced and decay to neutralino,
providing an interesting channel of production for the NLSP, with extra leptons in the
final state.

In the short running case, 
when $10^{5.7}\;\text{GeV}<\Lambda_{G_{1,2}}< 10^{5.9}\;\text{GeV}$ 
the NLSP is the stau, whose mass varies between $136$ and $300$ GeV. 
The Wino mass lies between $1.2$ and $2$ TeV,
and the neutralino mass between $600$ GeV to $1$ TeV.
The relevant production process is EW direct pair production
of sleptons, which is accessible due to the low mass of the NLSP.

In the long running case, 
when $10^{5.5}\;\text{GeV}<\Lambda_{G_{1,2}}< 10^{5.7}\;\text{GeV}$,
the typical stau NLSP mass is between $174$ and $500$ GeV.
The Wino mass varies between $700$ and $1.2$ TeV, and the neutralino
mass between $450$ and $600$ GeV.
The slepton mass splitting between the stau and
the co-NNLSP selectron and smuon can be large, around $100$ GeV.
In this case the decay of the selectron/smuon to the goldstino plus
electron/muon is suppressed by the large mediation scale $M_{\text{mess}}$,
see eqs (\ref{decayrateNLSP}).
The selectron/smuon pair production can then lead to interesting
three body decays $\tilde l_R \to \tilde \tau_1 \tau l $ through virtual neutralinos,
resulting in multi-lepton final states \cite{Ambrosanio:1997bq}.

Finally, when $\Lambda_S \simeq \Lambda_{G_{1,2}}$, there is the
interesting region of stau and neutralino co-NLSP, with stau slightly heavier,
which in this case is accessible to collider physics, since the typical NLSP 
mass can be as low as $140$ GeV
for the short running case and $180$ GeV for the long running one. 
The production will be mainly EW, with Wino mass at least around 
$300$ GeV or $500$ GeV, in the short and in the long RG flow cases
respectively. 
Consider the case of long running, $M_{\text{mess}}=10^{13}$ GeV, where both
neutralino and stau decay 
to the gravitino, i.e. $\tilde \tau \to \tau \tilde G$ , $\tilde{N}_{1} \to \gamma \tilde G$
are extremely suppressed.
The lightest stau is a mixture of left and right gauge eigenstates, and the mixing
is large,
being it proportional to $\mu \tan \beta$.
%
If the mass difference between the stau
and the neutralino is smaller than the tau mass, 
$\Delta_{\tau}\equiv m_{\tilde \tau}-m_{\tilde{N}_{1}}< m_{\tau}$, the
two body decay $\tilde \tau \to \tilde{N}_{1} \tau$ is not kinematically allowed.
So the stau has to decay with a three body process, through a virtual
tau decaying into neutrino and W boson, plus the long-lived neutralino,
or it can 
also decay via four body decay to $\tilde \tau_1 \to \nu_{\tau} \nu_{e} e \tilde{N}_{1}$ and
$\tilde \tau_1 \to \nu_{\tau} \nu_{\mu} \mu \tilde{N}_{1}$.
The branching ratios among these processes are controlled by $\Delta_{\tau}$, 
which also determines if the decay
will happen
outside or inside the detectors, leading in the second case to 
peculiar final states.
This scenario has been investigated recently in \cite{Jittoh:2005pq, Citron:2012fg} in the context
of gravity mediation, where the neutralino is the true LSP.
In our case the neutralino is not the LSP and it is not a viable dark matter
candidate, so the corresponding cosmological bound should not be applied.
However, for what concerns the collider signature, 
our realization is analogous to the one presented in those papers,
since the neutralino is longlived.
Along our parameter space, we can smoothly modify the 
value of $\Delta_{\tau}$, so we expect to be able to realize
various scenarios with different branching ratios and 
decay length.
It would be very interesting to perform a dedicated analysis 
for the collider signatures on this particular portion of the GGM parameter sapce.

\subsection{Splitting Scalar Mass Scales}
In this subsection we keep unified gaugino mass scales 
$\Lambda_{G_1}=\Lambda_{G_2}=\Lambda_{G_3}\equiv \Lambda_G$
and we disentangle the $SU(3)$ scalar mass scale $\Lambda_{S_3}$ 
from the other two 
$\Lambda_{S_1}=\Lambda_{S_2}\equiv \Lambda_{S_{1,2}}$.
Generically the UV boundary condition for the stop mass is set 
by the three scalar mass scales $\Lambda_{S_i}$. 
In the present scenario we can modify the stop mass (and also the other squark masses)
by varying $\Lambda_{S_3}$ keeping fixed the other $\Lambda_{S_{1,2}}$. In this way
we modify the squark masses without affecting the slepton sector.
This suggest two possible alternative directions.

One possibility is to try to maximize the $A_t$ term keeping the stop light, 
aiming at naturalness.
In order to raise the Higgs mass we would need to obtain a scenario with
maximal stop mixing, where $\frac{X_t}{M_S} \simeq 2$.
However, the $A_t$ term is negligible at the messenger scale in gauge mediation,
and both $A_t$ term and stop masses are induced along the RG flow by gluino masses,
making difficult to maximize their ratio.
A two-loop analysis of the MSSM RG equation is necessary to study analytically
this issue. In \cite{Brummer:2012ns} it has been shown that with positive mass squared for
the squarks and vanishing $A_t$ term at the messenger scale it is not possible
to reach a scenario of maximal stop mixing. 
Hence to obtain relative small stop mass at the EW scale we have to
impose UV boundary condition which are tachyonic for the squarks,
corresponding to negative $\Lambda_{S_3}^2$, 
a scenario which has been first suggested in 
\cite{Dermisek:2006ey}.
Like in the previous case of tachyonic UV masses for the Higgses,
we are going to ignore possible issues with CCB minima supposing 
that the usual EWSB vacuum is at least metastable and 
long-lived compared to the age of the Universe \cite{Ellis:2008mc,Carena:2008vj}.
We explore this possibility numerically in the following subsection,
but we find only a small improvement in the smallest value of the 
stop mass compatible with a $125$ GeV Higgs, resulting in $m_{\tilde t_1} \simeq 2$ TeV which implies an unavoidable fine-tuning in the EWSB condition.

The alternative possibility, which is more in the spirit of the present work, 
is to not pursue naturalness and admit that the large Higgs
mass is obtained entirely by heavy stop contribution. 
This effect can be obtained by setting
$\Lambda_{S_3}$ large. As a result the other supersymmetry breaking parameters $\Lambda_G$
and $\Lambda_{S_{1,2}}$ are essentially unaffected by the Higgs mass constraint.
Note that in this last scenario the squarks are necessarily very heavy, 
beyond the LHC reach, and the only possible light colored sparticle is the gluino.

In the following
we scan over the parameter space 
($\Lambda_G,\Lambda_{S_{1,2}},\Lambda_{S_3},,\tan\beta,M_{\text{mess}}$),
in the ranges explained in the introduction, and we fix three of these parameters to
present bi-dimensional plots with contours.

\subsubsection{Large positive $\Lambda_{S_3}^2$: Enlarged CGGM-like scenario}
If we do not stick to argument related to naturalness
a simple possibility is to
explore regions of the parameter space with large $\Lambda_{S_3}$.
In this case the squarks and in particular the stop are very heavy,
resulting in a large Higgs mass. 
The rest of the sparticle spectrum is characterized 
by the other two scales $\Lambda_{S_{1,2}}$ $\Lambda_G$, which
are essentially not constrained by the Higgs mass bound.

\begin{figure}[ht]
\begin{center}
\subfigure{
\includegraphics[width=7.5cm]{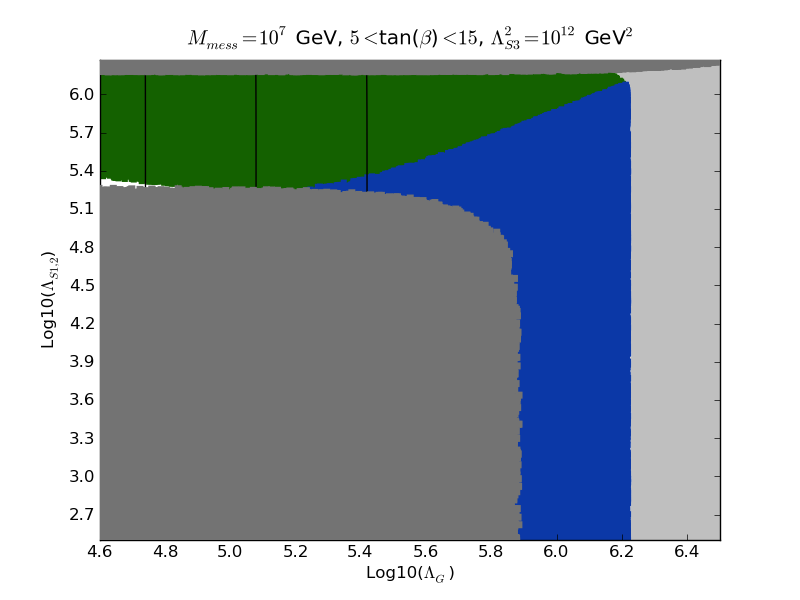}
}
\subfigure{
\includegraphics[width=7.5cm]{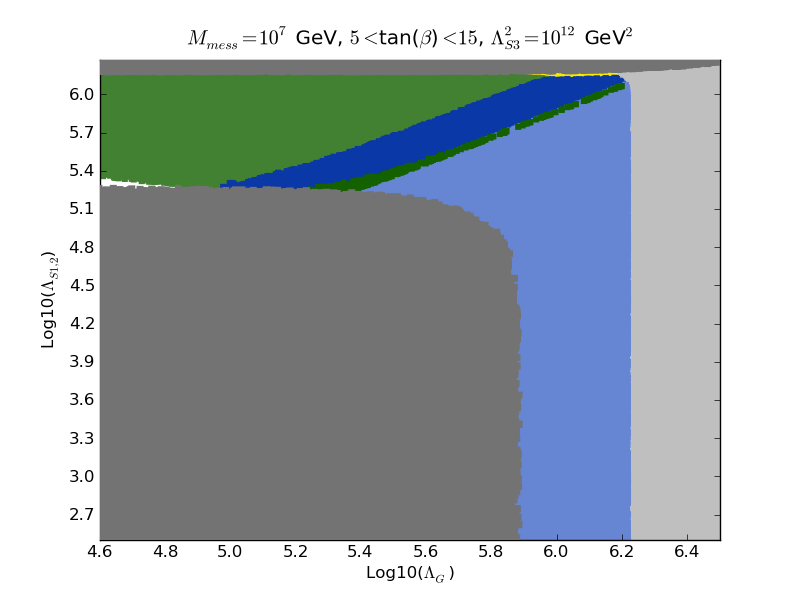}
}
\subfigure{
\includegraphics[width=7.5cm]{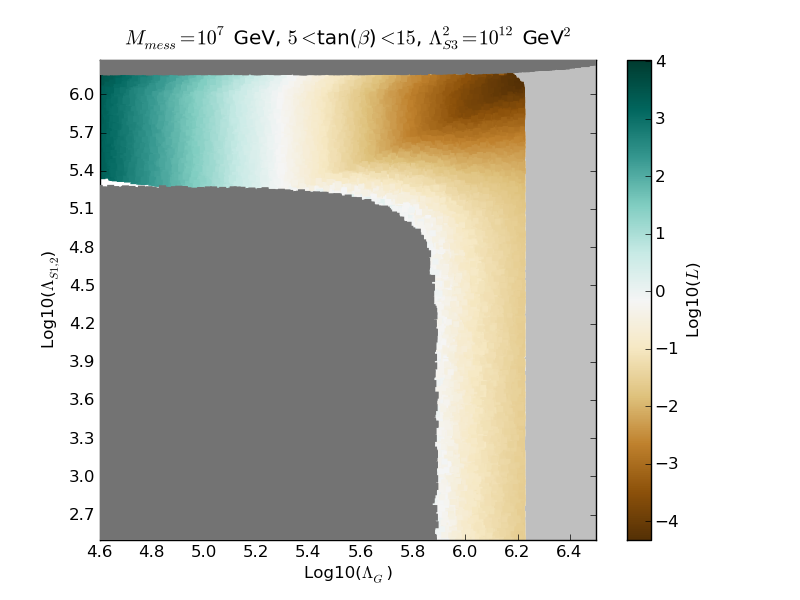}
}
\subfigure{
\includegraphics[width=7.5cm]{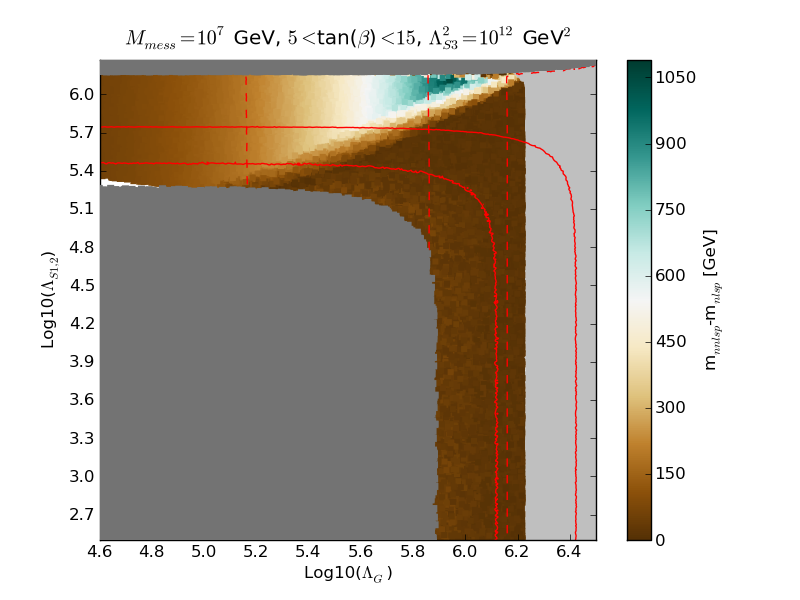}
}
\caption{
\label{largeLS3}\footnotesize
Logarithmic plot in the $\Lambda_G, \Lambda_{S_{1,2}}$ plane. 
In the first plot we show the NLSP type and black contours for the gluino mass at $0.5,1,2$ TeV.
The second plot show the NNLSP types, the third plot the NLSP decay length. In the fourth plot we show the 
value for $m_{NNLSP}-m_{NLSP}$. The red and dashed red contours identify the stau
mass and the neutralino mass, at $(0.5,1)$ TeV and $(0.2,1,2)$ TeV respectively.
}
\end{center}
\end{figure}

In the following we plot the results of our scan in the 
$\Lambda_{S_{1,2}}$, $\Lambda_G$ plane, by fixing $\Lambda_{S_3}=10^6$ GeV,
and we comment on the similarities and differencies with the CGGM 
case discussed in section \ref{cGGM}.
For simplicity we consider only one case for fixed $\tan \beta$ and $M_{\text{mess}}$,
and we show all the results in figure \ref{largeLS3}.

In the first plot of Figure \ref{largeLS3} we display the NLSP type with contours for the gluino mass. The
squark masses are almost constant along the parameter space, of the order of $7$ to $10$ TeV,
increasing for larger values of $\Lambda_G$ because of the gaugino mediation
contribution.
The NLSP can be the stau or the neutralino, as in the CGGM case. The
parameter space in the $\Lambda_G$, $\Lambda_{S_{1,2}}$ plane has opened up compared
to the CGGM with unified $\Lambda_{S_i}$, for the reasons explained before. 
Indeed the Higgs mass constraint does not play a relevant role, since there is no white region.
The allowed portion of the parameter space extends to the border with dark grey region, where SoftSUSY failed
to converge.

In the third figure \ref{largeLS3} we show the estimated decay length of the NLSP into 
standard model partner plus gravitino. 
The decay length is typically small in the stau NLSP region wheras it can be
short or long in the Bino NLSP region. In the stau neutralino 
co-NLSP region we can have either prompt decay or displaced vertex.
Generically, we can realize the same low energy spectra distribution 
with smaller
$M_{\text{mess}}$ if we would like to have promptly decaying NLSP
on all the parameter space.

An interesting difference compared to the CGGM case is that 
the sleptons can be significantly 
light in the allowed region,
making this scenario
more accessible in
terms of collider physics.
In CGGM the requirement of heavy squarks 
forced the entire scalar spectrum to be very heavy,
and the lightest stau was essentially always heavier than $500$ GeV. 
Here instead the stau can be as light as $220$ GeV, 
both in the gaugino mediated region, 
where it is the NLSP, but also 
in the gaugino screening region, where the NLSP
is the Bino.
This feature is shown in the fourth plot in Figure \ref{largeLS3} where we show the stau and neutralino
mass contours together with their mass difference.
The NNLSP type is shown in the second plot of Figure \ref{largeLS3}. 
In the stau NLSP region the co-NNLSP are the selectron and smuon, 
with splitting from the stau that can vary from 
few to $100$ GeV. 
In the Bino NLSP region, the NNLSP is predominantly 
neutral Wino.
The collider signatures of most of the allowed region
are the traditional ones of gauge mediated scenarios with moderately light
spectrum, 
except for having quite heavy squarks.
Differently than in section \ref{largeG3}, the gluino mass is not constrained to be large
to obtain heavy squarks. 

For instance, an interesting portion of the parameter space is the one with 
stau NLSP and $\Lambda_{G} \simeq \Lambda_{S_{1,2}}$,
with smallest possible stau mass around $220$ GeV.
The stau decay is inside the detector or displaced.
The gluino, Wino and Bino masses are around $2.1$ TeV, $700$ GeV and $350$ GeV respectively.
The selectron/smuon co-NNLSP are heavier than the stau by around $80$ GeV,
so the spectrum cannot be considered flavour democratic.
The stau pair production can be purely EW or through colored production via gluinos,
and both channels can be relevant for discovery at LHC.

In the upper right corner of the allowed region 
the NNLSP is the lightest chargino.
Indeed, with increasing $\Lambda_{S_{1,2}}$ 
the positive contributions to the Higgs soft mass
get larger. These contributions partially cancel the negative corrections
induced by the stop, and the 
 resulting $\mu$ term is very small.
 We have already encountered the same mechanism of 
accidental cancellation in the previous section,
 in the case of large $\Lambda_{G_3}$ and $\Lambda_{G_{1,2}}$.
 The $\mu$ term can be as small as $200$ GeV in the upper extrema
 of the allowed region.
There the lightest neutralino is mostly Higgsino, its
 mass is essentially set by $\mu$, and the other neutral and charged Higgsino are
 almost degenerate. 
 The chargino NNLSP is manifest only in the region of large $\Lambda_{G}$,
 where the gaugino mediation contribution to the Higgs soft terms enhance the
 effect even more, but the region with small $\mu$ 
 and consequent Higgsino NLSP extends horizontally for all values of $\Lambda_{G}$.
 So we can find benchmark points for simplified models with 
 relatively light gluino and with Higgsino NLSP, leading to Z+jets+MET signals for
 the Z-rich case or even to Higgs production in the Higgs-rich case \cite{Kats:2011qh}.

In summary, 
this GGM scenario with large 
$\Lambda_{S_3}$
generates
low energy spectra
 that satisfy the Higgs mass bound and that present
colored sparticle production accessible at LHC, being the gluino not 
too heavy.
Moreover, in this subsection the gluino mass was tied to the Bino and Wino 
mass by the GUT hypothesis for the gauginos.
One can instead envisage scenario with both $\Lambda_{G_3}$ and $\Lambda_{S_3}$ decoupled,
where a large  $\Lambda_{S_3}$ sets the squark mass very large, and the other parameters
results essentially unconstrained by the Higgs mass bound.
We could then realize spectra with neutralino or stau NLSP with arbitrary low gluino mass.

\subsubsection{Negative $\Lambda_{S_3}^2$}
Trying to minimize the stop mass, we restrict to the region where $\Lambda_{S_3}^2$
is negative. 
We fix the value of $\Lambda_{S_{1,2}}$ and we show the plots in the 
$\Lambda_{S_3}^2$, $\Lambda_G$ plane.
We select the case of long running, necessary to induce the $A_t$ term, so we 
set $M_{\text{mess}}=10^{13}$ GeV.

\begin{figure}[ht]
\begin{center}
\subfigure{
\includegraphics[width=7.5cm]{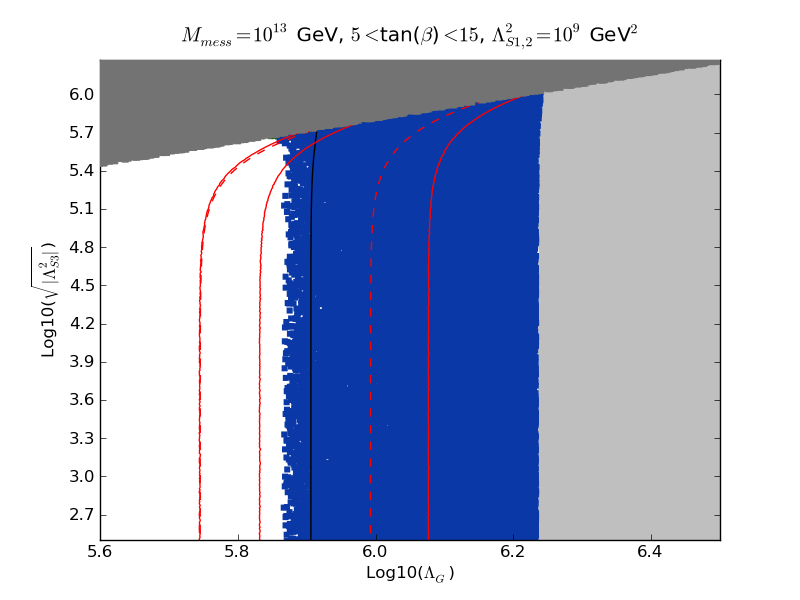}
}
\subfigure{
\includegraphics[width=7.5cm]{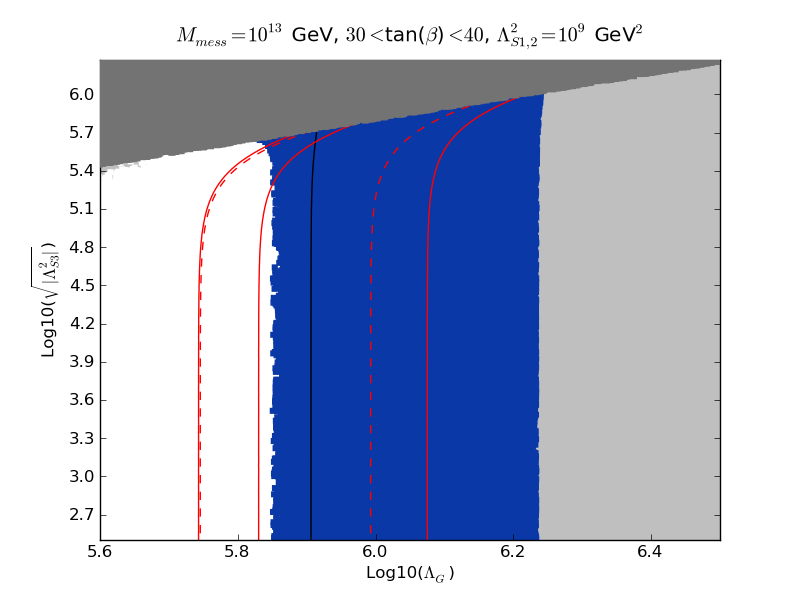}
}
\caption{
\label{S3negNLSP}\footnotesize
Logarhitmic plot in the $\Lambda_{S_3}$, $\Lambda_G$ plane fixing
$\Lambda_{S_{1,2}}^2=10^9$ GeV$^2$. In the dark grey region SoftSUSY did not converge,
in the light grey region the soft spectrum is too heavy to be interesting, while in the white region
the Higgs mass is smaller than $123$ GeV. In the blue region the Higgs mass is in the range
$123$-$127$ GeV and the NLSP is the stau.
The black contour follow the gluino mass at $5$ TeV. The red contours are the lightest stop mass
at $(2.5,3,5)$ TeV, while the red dashed contours are the first generation squark masses
at $3$ and $5$ TeV. 
}
\end{center}
\end{figure}

In Figure \ref{S3negNLSP} we show the NLSP and the contours for the
lightest stop, gluino mass, and first generation squarks (see the caption for details).
Note that $\Lambda_{S_3}^2$ is always negative, getting larger in modulus in the 
top region of the plot. 
The blue colored region,
with stau NLSP, is the only one with Higgs mass within $123$ and $127$ GeV.

The leftmost contour for the stop is at $2.5$ TeV, so we obtain an allowed region with
stop mass smaller than $2.5$ TeV.
The contours for the squark masses 
are independent from $\Lambda_{S_3}$ for small values of $\Lambda_{S_3}$,
since in that regime those masses are given mainly by the gluino mediation contribution,
and by $\Lambda_{S_{1,2}}$, which is constant.
The gluino mass is essentially determined only by $\Lambda_G$.
For larger values of $\tan \beta$, the region satisfying the Higgs mass constraint
gets slightly larger, as in previous cases.
In the allowed region the NLSP is always a long lived stau, 
and the smuon is the NNLSP.

\begin{figure}[ht]
\begin{center}
\subfigure{
\includegraphics[width=7.5cm]{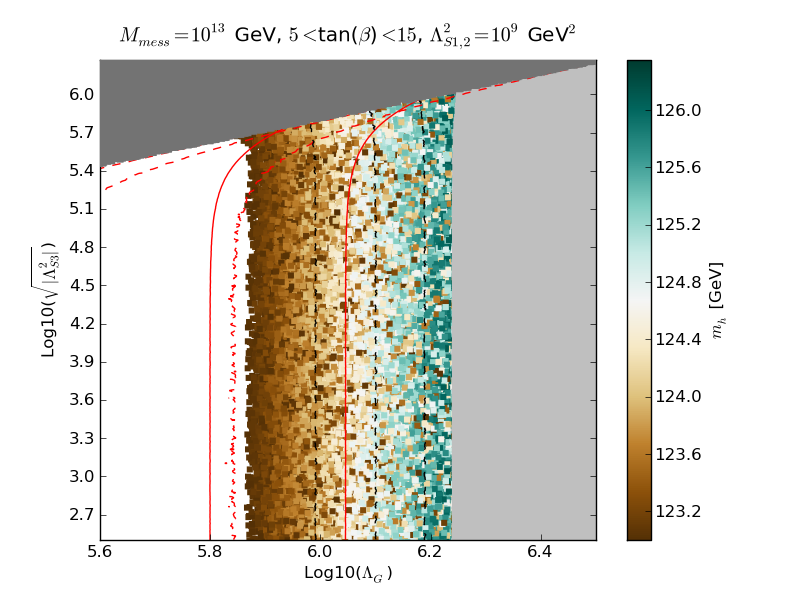}
}
\subfigure{
\includegraphics[width=7.5cm]{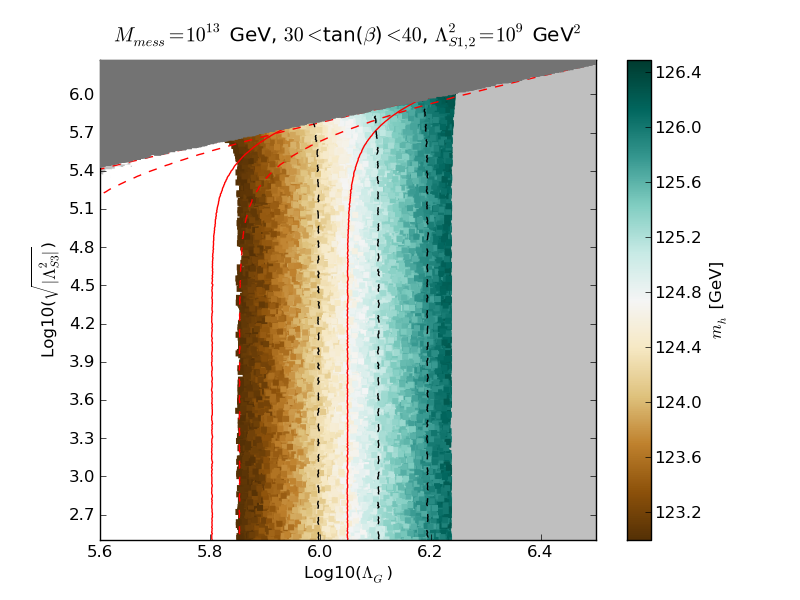}
}
\caption{
\label{S3neghiggs}\footnotesize
Gradient plot for the Higgs mass in the $\Lambda_{S_3}$, $\Lambda_G$ plane fixing
$\Lambda_{S_{1,2}}^2=10^9$ GeV$^2$.
The red contours for $M_S$ are at $3$ and $5$ TeV.
The dashed red contours identify $\mu$ at $500$ GeV, $1$ and $2$
TeV. The dashed black contours are for $A_t$ at $(-4,-5,-6)$ TeV.
}
\end{center}
\end{figure}

In Figure \ref{S3neghiggs} we show the gradient plot for the Higgs mass, with contours
for $\mu$, $A_t$ and $M_S$. The average stop mass contours have the expected shape, 
dominated by the gluino mass contribution for $|\Lambda_{S_3}|<\Lambda_G$.
The $A_t$ term depends essentially only on the gluino mass, being induced at one loop
during the RG flow (see eqs (\ref{Aterm})). Finally, the $\mu$ term decreases for decreasing stop mass.
This effect can be understood with reasonings similar to the ones around 
formula (\ref{minimumbis}).
The negative contribution to $\Sigma_u$ induced by the stop are less important for 
light stop, and are partially cancelled by the positive UV boundary value for $m_{H_u}^2$.
This partial cancellation leads to a small $\mu$ term in the region of light stop mass.\\

Given the previous results, we now fix $\Lambda_{S_3}$ to an optimized
value to obtain the smallest possible stop mass. We
set $\Lambda_{S_3}^2=-2.19 \times 10^{11}$ GeV$^2$ and we scan over $\Lambda_{G}$
and $\Lambda_{S_{1,2}}$.
This choice can appear to be fine tuned, but our purpose here is just to explore
remote corners of the parameter space of GGM in order to obtain 
the lightest possible squarks.

\begin{figure}[ht]
\begin{center}
\subfigure{
\includegraphics[width=7.5cm]{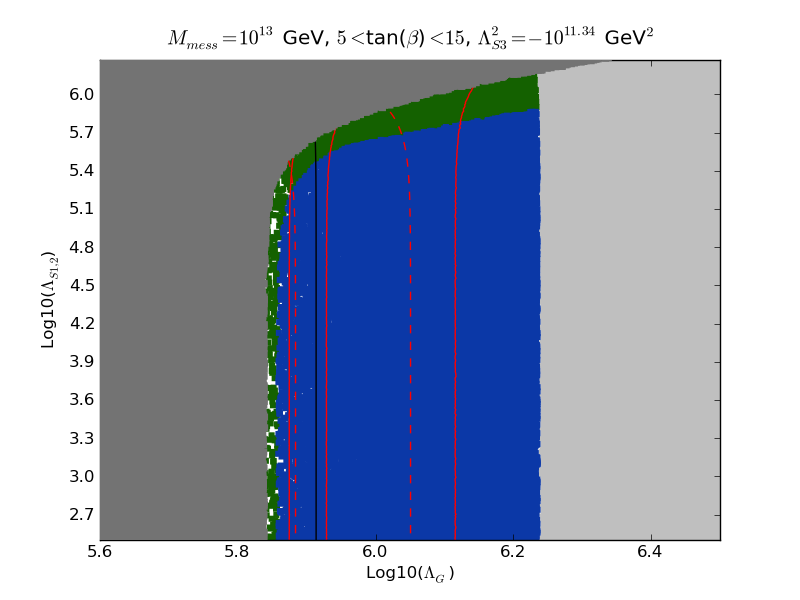}
}
\subfigure{
\includegraphics[width=7.5cm]{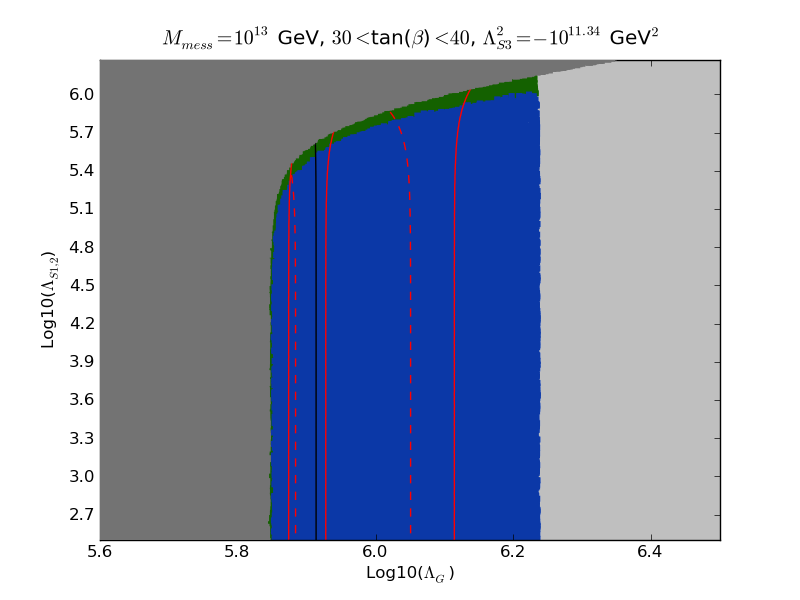}
}
\caption{
\label{S3negNLSPbis}\footnotesize
Logarithmic plot in the $\Lambda_{S_{1,2}}$, $\Lambda_G$ plane fixing
$\Lambda_{S_{3}}^2$. In the dark grey region SoftSUSY did not converge,
in the light grey region the soft spectrum is too heavy to be interesting. 
In the blue and green region the Higgs mass is in the range
$123$-$127$ GeV and the NLSP is the stau and the neutralino respectively.
The black contour follow the gluino mass at $5$ TeV. The red contours are the lightest stop mass
at $(2.5,3,5)$ TeV, while the red dashed contours are the first generation squark masses
at $3$ and $5$ TeV. 
}
\end{center}
\end{figure}

In Figure \ref{S3negNLSPbis} we show the usual NLSP plot with contours for
stop, gluino, and first generation squark masses. 
In the regime we are studying the squark masses are 
essentially independent on $\Lambda_{S_{1,2}}$, being dominated by the large value
of  $\Lambda_{S_3}$ and by the gluino mediation contribution.
The smallest value for the
stop mass is around $2.2$ TeV. 
The leftmost contour for the stop mass is at $2.5$ TeV. 
In the interesting region at the left of this contour the possible NLSP are both neutralino and
stau.
The stau NLSP region gets dominant for larger
value of $\tan \beta$, because of the mixing in the stau mass matrix.

\begin{figure}[ht]
\begin{center}
\subfigure{
\includegraphics[width=7.5cm]{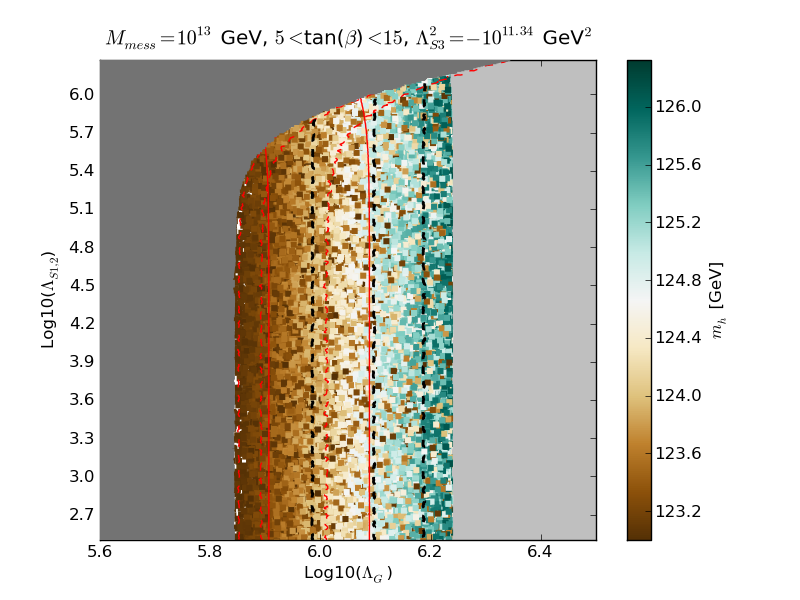}
}
\subfigure{
\includegraphics[width=7.5cm]{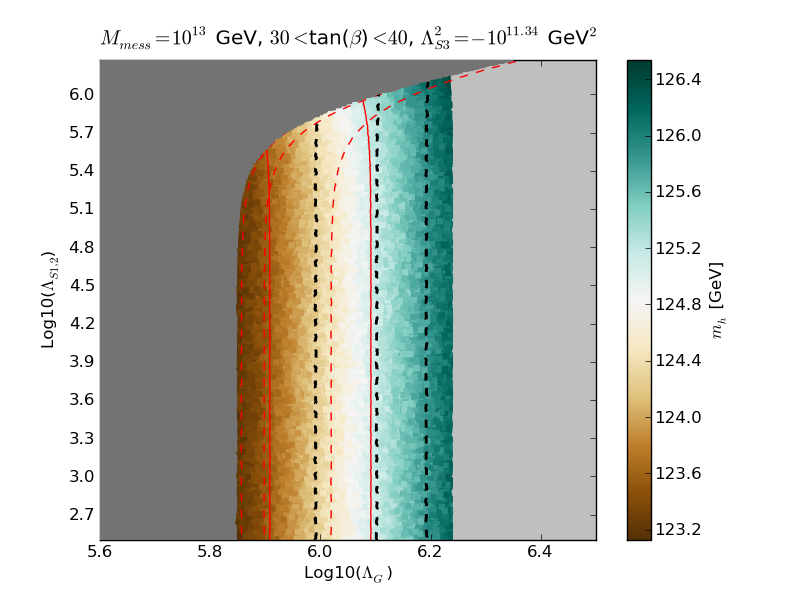}
}
\caption{
\label{S3negHiggsbis}\footnotesize
Gradient plot for the Higgs mass in the $\Lambda_{S_{1,2}}$, $\Lambda_G$ plane fixing
$\Lambda_{S_{3}}^2$
The red contours for $M_S$ are at $3$ and $5$ TeV.
The dashed red contours identify $\mu$ at $500$ GeV, $1$ and $2$
TeV. The dashed black contours are for $A_t$ at $(-4,-5,-6)$ TeV.
}
\end{center}
\end{figure}

It is then interesting to analyze in Figure \ref{S3negHiggsbis} the Higgs mass gradient and the
contours for the $\mu$ term, together with formula (\ref{minimumbis}),
in order to understand the rest of the dark grey region. 
Indeed, the dark grey region follows the dashed red contour of a small $\mu$ term
at $500$ GeV. For large value of $\Lambda_{S_{1,2}}$ the 
negative contribution to $\Sigma_u$ from the stop 
is too small compared to the positive contribution to $m_{H_u}^2$ proportional
to $\Lambda_{S_{1,2}}^2$, and EWSB cannot occur.
The situation is ameliorated by increasing $\Lambda_G$, since this increases the stop mass,
and then increases the negative contribution to $\Sigma_u$.
The $M_S$ and $A_t$ contours are determined by $\Lambda_G$ only as expected.

This discussion also indicates that 
in correspondence of the region of small $\mu$ 
the NLSP 
of Figure \ref{S3negNLSPbis}
is an Higgsino,
which is 
the usual feature of the low-energy spectrum 
in the presence of accidental cancellation in the EWSB condition.

\begin{figure}[ht]
\begin{center}
\subfigure{
\includegraphics[width=7.5cm]{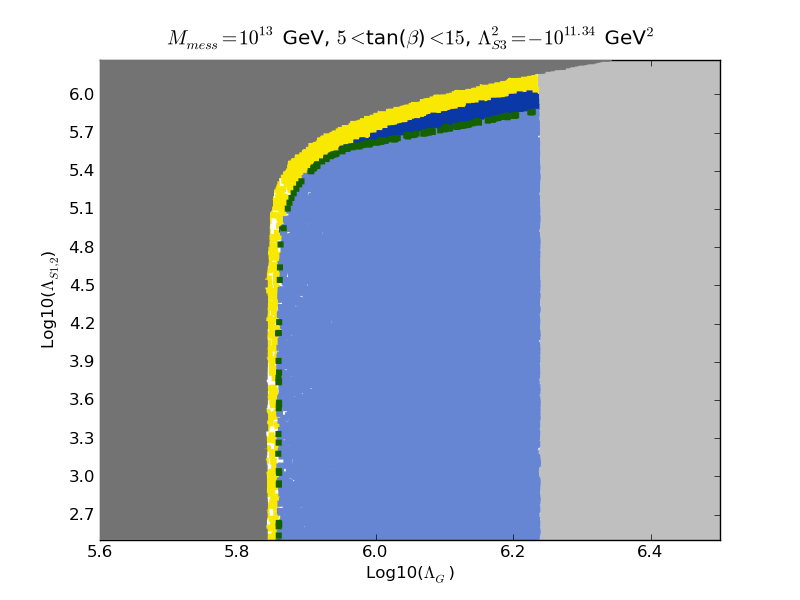}
}
\subfigure{
\includegraphics[width=7.5cm]{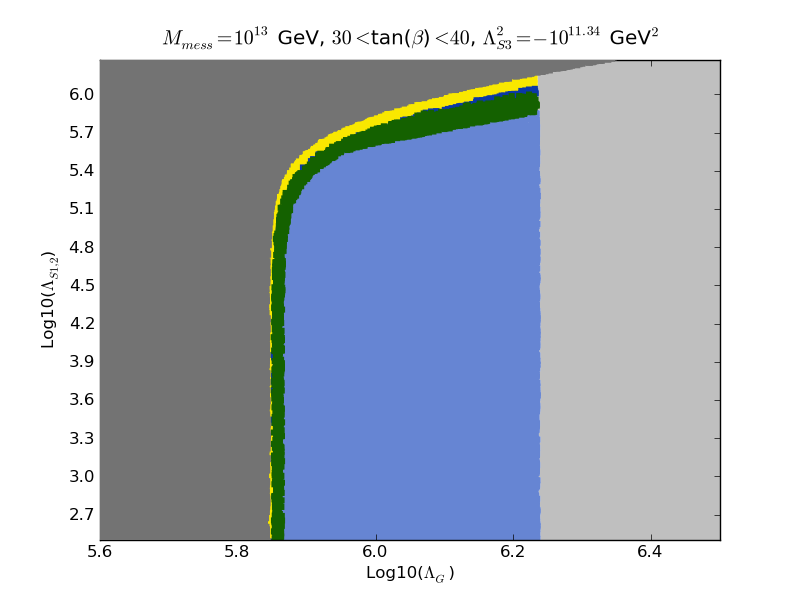}
}
\caption{
\label{S3negNNLSPbis}\footnotesize
NNLSP plot in the $\Lambda_{S_{1,2}}$, $\Lambda_G$ plane fixing
$\Lambda_{S_{3}}^2$.
The pale blue corresponds to smuon, the green to the lightest neutralino, the blue
to the stau, and the yellow to the lightest chargino.
}
\end{center}
\end{figure}

\begin{figure}[ht]
\begin{center}
\subfigure{
\includegraphics[width=7.5cm]{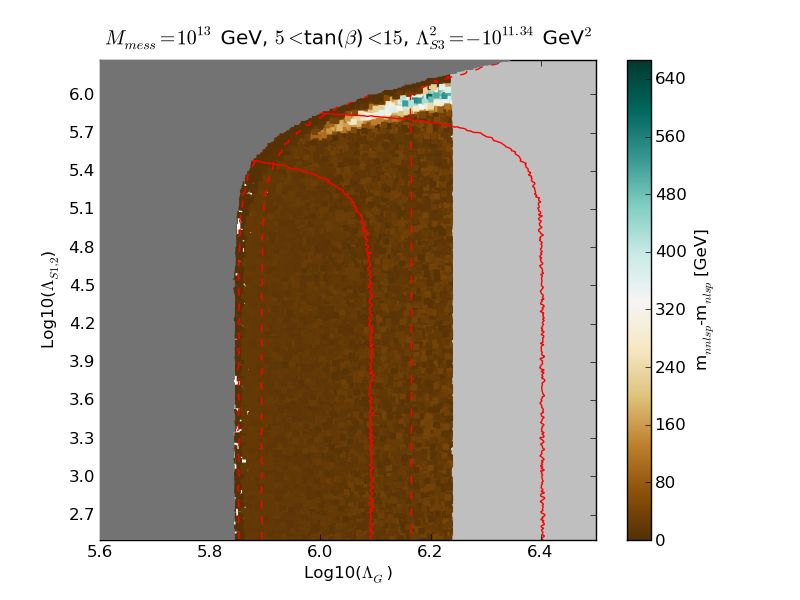}
}
\subfigure{
\includegraphics[width=7.5cm]{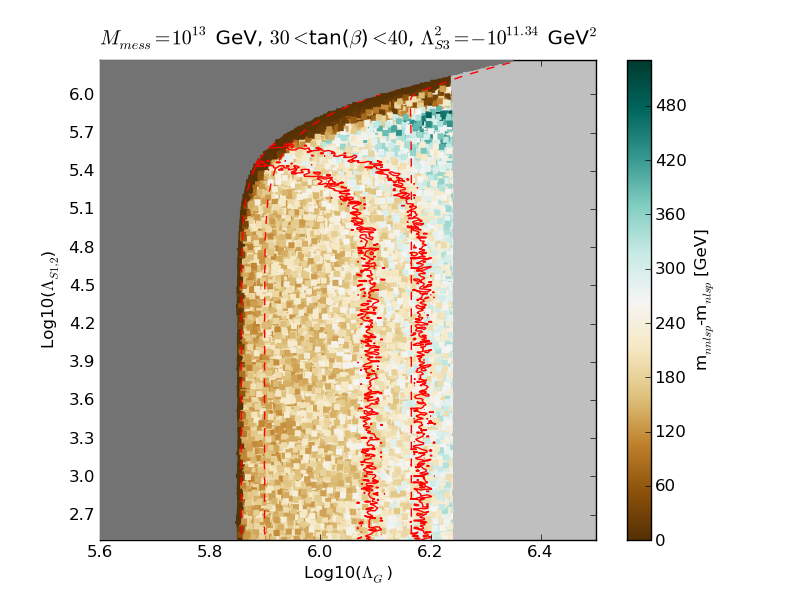}
}
\caption{
\label{S3negdiffbis}\footnotesize
Gradient plot 
for the mass difference between the NNLSP and the NLSP,
in the $\Lambda_{S_{1,2}}$, $\Lambda_G$ plane fixing
$\Lambda_{S_{3}}^2$.
In the left plot, the dashed red contours identify the neutralino mass at 
$500$ GeV, $1$ and $2$ TeV, while the red contours is the stau mass at 
$1$ and $2$ TeV. In the right plot,
the dashed red contours identify the neutralino mass at 
$500$ GeV, $1$ and $2$ TeV, while the red contours is the stau mass at 
$800$ GeV and $1$ TeV.
}
\end{center}
\end{figure}

In Figure \ref{S3negNNLSPbis} we show the NNLSP species. 
In the region of light
stop the $\mu$ term is very small and 
the NNLSP is the lightest neutral Higgsino in the blue region of stau NLSP, 
or the charged Higgsino in the green region of Higgsino NLSP.
The NLSP is always long lived and stable for collider physics, since we are
in the large $M_{\text{mess}}$ case.
Finally in Figure \ref{S3negdiffbis} we show the
mass difference between the NNLSP and the NLSP mass, with contours
for the stau and the lightest neutralino mass eigenvalue.
The stau mass contours are determined mainly by the gaugino mediated contribution,
and are sensitive to the scalar mass $\Lambda_{S_{1,2}}$ only for large value of 
$\Lambda_{S_{1,2}}$.
The lightest neutralino contours reveal aspects of its mixing angles.
For large $\Lambda_G$, and much larger than $\Lambda_S$,
the lightest neutralino is mostly Bino and 
its mass is determined by $\Lambda_{G}$,
while getting closer to the dark grey region the lightest neutralino 
becomes mostly Higgsino and its mass contours have a shape analogous to the $\mu$ term.
In particular, the neutralino is mostly Higgsino in all the region where it is below the TeV scale.

The region of light stop, 
with $\Lambda_{G} \simeq 10^{5.85}$ GeV and $\Lambda_{S_{1,2}} < 10^{5.5}$ GeV 
shows an interesting spectrum.
The lightest stop is the lightest of the squarks, with squark masses around the TeV scale and a quite heavy gluino
at $5$ TeV, and the Wino is at $1.8$ TeV.
The more common NLSP is the stau, whose mass varies between $500$ and $600$ GeV.
The NNLSP is the lightest Higgsino which is almost degenerate with the other neutral and charged Higgsino. The three Higgsinos can be arbitrarily close in mass to the stau as can be observed 
from the Figure \ref{S3negdiffbis}, eventually becoming even 
lighter in the region very close to the non-convergence of SoftSUSY.

This pattern is an usual one when accidental cancellations occur, and we already discussed it 
in the section 4.1 for long running. 
The only new ingredient here is having a stop mass around 2 TeV which, 
however, would not affect so much the collider phenomenology.

%

\section{More General Scenarios}
In the previous section we have studied the possibility of disentangling the 
supersymmetry breaking scales associated to the $SU(3)$ gauge group from the 
others. 
Here we comment on the possibility of enlarging further the parameter space
and on the resulting scenarios that we can envisage, leaving a detailed study for future works.

As already observed, the Higgs mass corrections are more sensitive to the 
colored sector. For this reason 
we do not expect that by considering independent supersymmetry breaking
scales for the $U(1)$ and $SU(2)$ gauge group will change deeply 
our conclusions about the effect of the Higgs mass bound on the sparticle spectrum.
However, such extension of the parameter space could lead to 
modifications in the pattern of the uncolored soft terms.

Possible extensions of our analysis in the gaugino scales 
$\Lambda_{G_i}$ 
consist in taking $\Lambda_{G_1} \neq \Lambda_{G_2}$
and/or considering sign differences among these scales, that we 
have instead taken all positive for simplicity.
For what concerns the scalar mass scales the only
relevant extension is in
taking $\Lambda_{S_1}^2 \neq \Lambda_{S_2}^2$,
since we have considered also UV 
tachionic boundary conditions.

Within our assumptions, we have already realized 
low energy spectra with a very large set of different possible 
un-colored NLSP: Bino, Higgsino, stau, selectron, tau-sneutrino.
We have described the main features of the spectra and we have discussed
their possible collider signatures.
The extensions could 
lead to new classes of un-colored NLSP, or
realize the same NLSP scenario
but through different mechanism and possibly different spectrum structures.

For instance, implementing a hierarchy between the
electroweak gaugino mass scales 
as $\Lambda_{G_1} > \Lambda_{G_2}$, it is possible to
obtain low energy spectra with sneutrino NLSP. 
Among the sneutrino, 
the lightest is typically the $\nu_{\tau}$ because of the large Yukawa coupling.
In our analysis, we have obtained a $\nu_{\tau}$ NLSP
driven by extra contribution to the Higgs soft terms or messenger-parity violations.
The common aspects of our two cases is that the effect is 
induced by the violation of the GGM sum rule $\Tr (Y m^2)$ at the messenger scale. 
This would not be the case for a set up with $\Lambda_{G_1} > \Lambda_{G_2}$.
It would be interesting to compare the two mechanisms that lead to sneutrino NLSP
and find characteristics of the sparticle spectrum which are peculiar
of a specific realization.

An interesting case for collider physics that we have not covered is the one of
the chargino NLSP.
In \cite{Kribs:2008hq} it has been shown that this possibility can be realized in the MSSM in a 
small portion of the parameter space taking $\text{sign}(M_1) \neq \text{sign}(M_2)$.
It would be interesting to explore also this possibility in GGM 
and find the consequences of the Higgs mass bound on the resulting spectrum.

In our low energy spectra, the other scalar Higgses 
were always very heavy, 
realizing the decoupling limit besides when accidental cancellations were driving them light as in section 4.3.
Generically, 
very large values of $\tan \beta$ can suppress the mass of the pseudo-scalar Higgs boson
$A$ and lower the entire set of scalar Higgses as well. 
We did not discuss those region of the parameter space where the MSSM is not in the
decoupling limit, that we leave for future studies.
These cases could be relevant if sizable deviations in the Higgs coupling to the 
Standard Model particles will be discovered which seems unlikely at the present stage.

With the classification we have provided along the paper
and the few extensions that we have
discussed here, essentially all the possibilities for uncolored sparticles,
plus the gluino, have been covered.
Without relaxing the GGM assumption on the vanishing $A$-terms, the Higgs mass bound
forces the squarks to be very heavy, among the heaviest of the sparticle spectrum.
We do not expect that modifying hierarchies among the UV supersymmetry breaking
scales can
substantially change this conclusion.
So within our initial hypothesis, 
the squarks cannot be the lightest sparticles, and
hence we have eventually discussed all possible NLSP species.
For this reason,
enlarging the parameter space in the scalar mass scales $\Lambda_{S_i}$ 
does not seem particularly promising.

\section{Conclusions and Perspectives}
In this paper we have extensively investigated the parameter space of General Gauge Mediation
and some extension of it, in the light of the $125$ GeV Higgs mass discovery.
We have simulated the RG flow evolution of the UV parameter with SoftSUSY and we 
have identified the possible low energy sparticle spectra compatible
with the Higgs mass bound.
We have imposed flavour constraints and
 mild direct bounds on the sparticle masses.
Nevertheless, the $125$ Higgs constrains tremendously 
the mass spectrum at low energy and as a consequence restricts 
seriously the allowed parameter space.
We discussed the main phenomenological features of the possible
low energy spectra and the corresponding collider signatures
and discovery prospects.

We have shown that generically the $125$ Higgs mass 
forces the colored sector of the sparticles to be
quite heavy, typically beyond the reach of the LHC.
Our investigation suggests that viable collider scenarios in GGM 
should probably abandon the assumption of
universality among the supersymmetry breaking scales for the 
gauge group factors of the MSSM.
This possibility can make the low energy 
spectrum more accessible at LHC
in two directions: on one side we can obtain lighter un-colored
sparticles, boosting EW production; on the other side we can lower 
almost arbitrarly the gluino mass, increasing the rate of the gluino
pair production.

During our study, 
we have identified several interesting regions of the GGM parameter space,
characterized by exotic NLSP species and/or by unusual signals for collider
physics.
We believe that many of these scenarios deserve further dedicated studies.
Our 
results
provides the connection between the UV realizations 
and their low energy sparticle signature, and
support the conviction that GGM is still a powerful
generator for supersymmetric simplified models for LHC searches.

A natural extension of our work is to include in the GGM definition extra contributions
to the $A$-terms. The possibility that direct couplings between the Higgs sector
and the supersymmetry breaking sector induce new corrections to the $A$-terms,
without affecting the appealing properties of gauge mediation,
has been recently studied in the literature \cite{Chacko:2001km,Chacko:2002et,Evans:2011bea, Evans:2012hg, Kang:2012ra, Craig:2012xp, Craig:2013wga}.
This would motivate a deep modification of our analysis, which we hope to 
address in the near future. 

%
%

\section*{Acknowledgements}
We would like to thank Steve Abel, Jorgen D'Hondt, Matt Dolan, Benjamin Fuks, Lorenzo Calibbi, Federico Urban
for useful discussions. 
We are especially grateful to Riccardo Argurio, Gabriele Ferretti,
Christoffer Petersson, Kentarou Mawatari and Chris Wymant for numerous helpful discussions 
and for comments on the manuscript.
We thank the IIHE for the possibility of having access to their computational facilities,
and especially Olivier Devroede, St\'{e}phane Gerard, and Shkelzen Rugovac, for assistance with
the cluster. 
The research of D.R. is supported in part by IISN-Belgium (conventions
4.4511.06, 4.4505.86 and 4.4514.08), by the ``Communaut\'e
Fran\c{c}aise de Belgique" through the ARC program and by a ``Mandat d'Impulsion Scientifique" of the F.R.S.-FNRS.
A.M. thanks the Vrije Universiteit Brussel and the FWO PostDoctoral fellowship 
for support during the first part of this project.
A.M. acknowledges funding by the Durham International Junior Research Fellowship.

\let\oldbibitem=\bibitem\renewcommand{\bibitem}{\filbreak\oldbibitem}
\bibliographystyle{JHEP}
\fussy
\bibliography{GGMpheno}
\end{document}